\documentstyle[11pt]{article}

\def\Bagd{B_\al\,^{\ga\de}}
\def\Det{\mbox{Det}}

\def\Habg{H_{\al\be}\,^\ga}

\def\intx{\int\! d^{4}x}
\def\inty{\int\! d^{4}y}
\def\Rabgd{R_{\al\be}\,^{\ga\de}}
\def\Rab{R_{\al\be}}
\def\rvec{\!\!\!\!^{^\rightarrow}}
\def\rrvec{\!\!\!\!^{^{^\rightarrow}}}

\def\Tabg{T_{\al\be}\,^\ga}

\def\ar{\rightarrow}

\def\bib{\bibitem}

\def\dem{\det e^{-1}}

\def\intf{\int d^{4}x\,}

\def\lar{\longrightarrow}
\def\lbr{\lbrack}

\def\pa{\partial}

\def\rbr{\rbrack}

\def\Tr{\,\mbox{Tr}\,}
\def\al{\alpha}
\def\be{\beta}
\def\ga{\gamma}
\def\Ga{{\it\Gamma}}
\def\de{\delta}
\def\ep{\varepsilon}
\def\ze{\zeta}
\def\th{\vartheta}
\def\ka{\kappa\,}

\def\La{{\it\Lambda}}

\def\si{\sigma}
\def\Si{\Sigma}

\def\om{\omega}

\def\beq{\begin{equation}}
\def\eeq{\end{equation}}
\def\bed{\begin{displaymath}}
\def\eed{\end{displaymath}}
\def\beqq{\begin{eqnarray}}
\def\eeqq{\end{eqnarray}}
\def\bedd{\begin{eqnarray*}}
\def\eedd{\end{eqnarray*}}

\title{\huge{\bf{The Quantum Theory of Gravitation \\
\vspace{0.5cm} II -- Renormalizability Proof}}}
\author{\vspace{0.2cm}\\
C. Wiesendanger\\
Aurorastr. 24\\
8032 Zurich, Switzerland\\
{\it christian.wiesendanger@ubs.com}}
\date{May 14, 2019}

\begin{document}

\maketitle

\noindent {\sl Le second (precepte estoit) de diuiser chascune des difficultez que i' exami-nerois en autant de Parcelles qu' il se pourroit \& qu'il seroit requis pour les mieux resoudre}\footnote{Ren\'e Descartes, Discours de la M\'ethode, Deuxi\`eme Partie, 1637, Imprimerie Ian Maire (Leyde)}

\begin{abstract}
A new {\bf SO(1,3)\/} gauge field theory classically equivalent to General Relativity is quantized and the gauge-fixed path integral representation of the quantum effective action (QEA) is derived. Both the gauge-fixed classical action and the QEA are shown to be invariant under nilpotent BRST variations of the gauge, matter, ghost, antighost and Nakanishi-Lautrup fields defining the theory and a Zinn-Justin equation constraining the QEA is derived. Dimensional analysis and the various linear constraints put on the QEA plus the ones from the non-linear Zinn-Justin equation are deployed to demonstrate full renormalizability such that all infinities appearing in a perturbative expansion of the QEA can be absorbed into the gauge-fixed classical action solely by field renormalizations and coupling redefinitions -- alltogether providing the third step in consistently quantizing gravitation\\
\end{abstract}

\clearpage

\section{Introduction}

\paragraph{}
This is the third in a series of papers on the classical and quantum theory of gravitation \cite{chw1, chw2} in which we step-by-step develop a programme aimed at consistently quantizing gravity.

So far we have taken two steps.

The first step has been to formulate a theory for classical gravity which is not equal to GR in its outset, yet equivalent to it in its predictions, and which allows for renormalizable actions in its fundamental dynamical field \cite{chw1}. Technically it is a new gauge field theory of the Lorentz group {\bf SO(1,3)\/} which a) contains as the only dynamical field the dimension-one Lorentz gauge field in terms of which all else can be expressed, which b) allows for actions at most quadratic in the first derivatives of the gauge field and renormalizable by power-counting, and which c) is equivalent to GR in its predictions in a limiting case. In other words it is a candidate theory of gravitation viable at the classical level which is not plagued by the well-known flaws preventing consistent quantization in the usual approaches.

The second step has been to establish that the canonical quantization of the non-interacting gauge field in the {\bf SO(1,3)\/} gauge field theory allows for the definition of positive-norm, positive-energy states and a corresponding relativistically-invariant physical Fock space for the quantum theory in spite of the non-compactness of the gauge group {\bf SO(1,3)\/} and the corresponding indefinite Cartan metric on the gauge algebra \cite{chw2}. This has been achieved by intertwining relativistic covariance with positivity of the norm and energy expectation values for physical states, and consequentially putting restrictions needed in establishing a physical Fock space on state vectors, and not on the algebra of creation and annihilation operators.

The third and current step of our programme is the renormalizability proof for the full quantum theory including the demonstration that unphysical ghosts decouple which appear in the gauge-fixed path integral quantization of the classical theory, and establishing the pseudo-unitarity of the $S$-matrix on the na\"\i ve Fock space containing negative-norm, negative-energy states besides the physical ones.

The final step of our programme will be to establish the unitarity of the $S$-matrix on the physical Fock space constructed in \cite{chw2}.

To effectively establish renormalizability of the full quantum theory in this paper we start in section two revisiting the gauge field theory of the Lorentz group {\bf SO(1,3)\/} at the classical level to then derive gauge-fixed path integral expressions for the expectation values of physical observables and the quantum effective action (QEA). In section three we rewrite these expressions in terms of path integrals over additional ghost, antighost and Nakanishi-Lautrup fields allowing in section four in an elegant way to introduce nilpotent BRST field variations and to demonstrate the invariance of the gauge-fixed classical action and the QEA under these variations. In section five we derive the Zinn-Justin equation which puts crucial constraints on the QEA and its loop-wise expansion. In section six we finally demonstrate the perturbative renormalizability of the {\bf SO(1,3)\/} gauge field theory which marks a further key step towards a consistent quantum theory of gravitation.

All fields in this paper are defined on Minkowski spacetime {\bf M\/}$^{4}$ $\equiv$ ({\bf R\/}$^{4}$,$\eta$) with points $x \in$ {\bf M\/}$^{4}$ given in Cartesian coordinates. $\eta=\mbox{diag}(-1,1,1,1)$ is the flat spacetime metric with which indices $\al,\be,\ga,\dots$ are raised and lowered. They appear in quantities defined on {\bf M\/}$^{4}$ which transform covariantly. All other notations deployed in the paper are explained wherever they appear first.

\section{Path Integral Quantization of the {\bf SO(1,3)\/} Gauge Field Theory}

\paragraph{}
In this section we review some key elements of the {\bf SO(1,3)\/} gauge field theory equivalent to General Relativity developed in \cite{chw1}. We then quantize the theory and derive gauge-fixed path-integral representations for gauge-invariant physical quantities applying the Faddeev-Popov-deWitt approach.

Let us start with the gauge-invariant path integral representing the expectation value of a physical observable ${\cal O} [B]$
\beq \label{1} \int\!\Pi_{\!\!\!\!\!\!_{_{_{x;\al,\ga,\de}}}} \!\!\!\!\!\!\!\!d\Bagd (x)\,
{\cal O} [B]\, \exp\,i\left\{S_G [B] + \ep \mbox{-terms} \right\}.
\eeq

Above $\Bagd$ denotes the {\bf SO(1,3)\/} gauge field antisymmetric in the indices $\ga, \de$ which we have introduced in \cite{chw1}, and ${\cal O} [B]$ a gauge-invariant observable which is a functional of $\Bagd$. $\Pi_{\!\!\!\!\!\!_{_{_{x;\al,\ga,\de}}}} \!\!\!\!\!\!\!\!d\Bagd (x)$ is the integration measure over gauge field space which is invariant under the gauge transformations Eqn.(\ref{17}) below as demonstrated in Appendix B.

The dynamics of the gauge field $\Bagd$ is governed classically by the most general action of dimension four or less for the $\Bagd$ \cite{chw1}
\beq \label{2} S_G [B] = S^{(0)}_G [B] + S^{(2)}_G [B] + S^{(4)}_G [B].
\eeq
Here
\beq S^{(0)}_G [B] = \La\, \intf \dem [B] \eeq is the most general dimension-zero contribution
with $ \La $ a constant of dimension $[\La] = 4$ and $ \dem [B] $ the determinant of a matrix $ e_\al\,^\th[B] $ which will be properly introduced in Eqn.(\ref{7}) below.

The most general dimension-two contribution reads
\beqq S^{(2)}_G [B]
&=& \frac{1}{\ka}\, \intf \dem [B]\,
\Big\{\al_1\, R_{\al\be}\,^{\al\be} [B] \nonumber \\
&+& \al_2\, T_{\al\be\ga} [B]\, T^{\al\be\ga} [B]
+ \al_3\, T_{\al\be\ga} [B]\, T^{\ga\be\al} [B] \\
&+&  \al_4\, T_\al \,^{\ga\al} [B]\, T_{\be\ga} \,^\be [B]
+  \al_5\, \nabla^B_\al T_\be \,^{\al\be} [B] \Big\}. \nonumber
\eeqq
The constant $\frac{1}{\ka} = \frac{1}{16\pi \it\Ga}$ has mass-dimension $[ \frac{1}{\ka} ] = 2$ with ${\it\Ga} $ denoting the Newtonian gravitational constant. $T [B]$ and $R [B]$ are tensors properly introduced in Eqns.(\ref{15}) and (\ref{16}) below and the $ \al_i $ above are constants of dimension $[\al_i] = 0$.

Finally the most general dimension-four contribution reads
\beqq S^{(4)}_G [B] &=& \intf \dem [B]\,
\Big\{ \be_1\, R_{\al\be}\,^{\ga\de} [B]\, R^{\al\be}\,_{\ga\de} [B] \nonumber \\
&+&  \be_2\, R_{\al\ga}\,^{\al\de} [B]\, R^{\be\ga}\,_{\be\de} [B] 
+ \be_3\, R_{\al\be}\,^{\al\be} [B] \,R_{\ga\de}\,^{\ga\de} [B] \nonumber \\
&+& \be_4\, \nabla_B^\ga \nabla^B_\de R_{\al\ga}\,^{\al\de} [B]
+ \be_5\, \nabla_B^\ga \nabla^B_\ga R_{\al\be}\,^{\al\be} [B] \nonumber \\
&+& \dots \nonumber \\
&+& \ga_1\, \nabla^B_\ga T_{\al\be\de} [B]\,
\nabla_B^\ga T^{\al\be\de} [B]
+ \ga_2\, \nabla^B_\ga T_{\al\be\de} [B]\, 
\nabla_B^\ga T^{\de\be\al} [B] \\
&+& \dots \nonumber \\
&+& \ga_j\, T^4-\mbox{terms} \nonumber \\
&+& \dots \nonumber \\
&+& \de_k\, R\,T^2-\mbox{terms},\, R\,\nabla^B\, T-\mbox{terms} \nonumber \\
&+& \dots \Big\} \nonumber 
\eeqq 
with $ \be_i $, $ \ga_j $, $ \de_k $ constants of dimension $[\be_i] = [\ga_j] = [\de_k] = 0$.

Above
\beqq \label{6} \nabla^B_\al &\equiv& \pa_\al +\frac{i}{2}\Bagd {L\rvec}_{\ga\de}
+\frac{i}{2}\Bagd \Si_{\ga\de} \nonumber \\
&=& \left( \eta_\al\,^\ga - \Bagd x_\de \right) \pa_\ga
+\frac{i}{2}\Bagd \Si_{\ga\de} \\
&\equiv& d^B_\al + {\bar B}_\al \nonumber
\eeqq
denotes the covariant derivative w.r.t to the gauge group {\bf SO(1,3)\/} as introduced in \cite{chw1}. ${L\rvec}_{\ga\de} = -i ( x_\ga \pa_\de - x_\de \pa_\ga )$ are the generators of the {\bf so(1,3)\/} Lorentz algebra acting on spacetime coordinates and $\Si_{\ga\de}$ generic generators of the Lorentz algebra acting on spin degrees of freedom.

To simplify notations we have defined the matrix
\beq \label{7} e_\al\,^\th[B] \equiv \eta_\al\,^\th - B_\al\,^{\th\ze}x_\ze
\eeq
resembling a Vierbein which, however, is solely a functional of the fundamental dynamical variable $\Bagd$ in our theory, and have introduced
\beq \label{8} d^B_\al \equiv e_\al\,^\th[B]\, \pa_\th,
\quad {\bar B}_\al \equiv \frac{i}{2}\Bagd \Si_{\ga\de}.
\eeq

We have elaborated in depth in \cite{chw1} why $e_\al\,^\th[B]$ not being a fundamental dynamical field in our approach is so crucial for the further development of the theory to be both equivalent to General Relativity and renormalizable.

To define the covariant objects of the theory we next look at the field strength operator $G$ acting on fields
\beq G_{\al\be} [B]\equiv [ \nabla^B_\al,\nabla^B_\be]
\eeq
and express it in terms of the gauge field $B$
\beqq G_{\al\be} [B]&=& [d^B_\al ,d^B_\be ] + d^B_\al {\bar B}_\be - d^B_\be {\bar B}_\al \\
&+& [{\bar B}_\al , {\bar B}_\be] + ( B_{\al\be}\,^\eta - B_{\be\al}\,^\eta ) \nabla^B_\eta. \nonumber
\eeqq
To re-express
\beq [d^B_\al ,d^B_\be ] = \left( e_\al\,^\ze [B]\,\pa_\ze e_\be\,^\eta [B]-
e_\be\,^\ze [B]\,\pa_\ze e_\al\,^\eta [B] \right)\pa_\eta
\eeq
we assume that the matrix $e_\al\,^\ze [B]$ is non-singular, i.e. $\det e[B]\neq 0$. Hence there is an inverse $e^\ga\,_\eta [B]$ with
$e^\ga\,_\eta [B]\, e_\ga\,^\ze [B]=\de_\eta\,^\ze$, and we can write
\beq [d^B_\al ,d^B_\be ] = \Habg [B]\, d^B_\ga
\eeq
introducing
\beq \Habg [B] \equiv e^\ga\,_\eta [B] \left( e_\al\,^\ze [B]\,\pa_\ze e_\be\,^\eta [B]-
e_\be\,^\ze [B]\,\pa_\ze e_\al\,^\eta [B] \right).
\eeq
As a result we can rewrite
\beqq G_{\al\be} [B] &=& (\Habg [B]+ B_{\al\be}\,^\ga 
- B_{\be\al}\,^\ga ) \nabla^B_\ga  \nonumber \\
&+& d^B_\al {\bar B}_\be - d^B_\be {\bar B}_\al + [{\bar B}_\al ,{\bar B}_\be] - \Habg [B]\, {\bar B}_\ga \\
&\equiv& - \Tabg [B]\, \nabla^B_\ga + \Rab [B]  \nonumber 
\eeqq
in terms of the covariant field strength components $T$
\beq \label{15} \Tabg [B] \equiv -(B_{\al\be}\,^\ga - B_{\be\al}\,^\ga) - \Habg [B]
\eeq
and $R$
\beqq \label{16} \Rab [B] &\equiv& \frac{i}{2} \Rabgd [B] \, \Si_{\ga\de} \nonumber \\
\Rabgd [B] &=& d^B_\al  B_\be\,^{\ga\de} - d^B_\be B_\al\,^{\ga\de}
+\, B_\al\,^{\ga\eta}\, B_{\be\eta}\,^\de \\ 
&-& B_\be\,^{\ga\eta}\, B_{\al\eta}\,^\de 
- H_{\al\be}\,^\eta [B]\, B_\eta\,^{\ga\de}. \nonumber
\eeqq

Under a local variation
\beq \label{17} \de_\om\Bagd = -\om^{\eta\ze}x_\ze \pa_\eta \Bagd 
- d^B_\al \om^{\ga\de} + \om_\al\,^\be B_\be\,^{\ga\de} 
+ \om^\ga\,_\eta B_\al\,^{\eta\de} + \om^\de\,_\eta B_\al\,^{\ga\eta}
\eeq
of the gauge field $\Bagd$ assuring covariance of the derivative Eqn.(\ref{6}) as established in \cite{chw1} we find the field strength components to display the homogenous variations
\beqq \de_\om T_{\al\be}\,^\ga [B]&=& 
- \om^{\eta\ze}x_\ze\,\pa_\eta T_{\al\be}\,^\ga [B]
+ \om_\al\,^\eta T_{\eta\be}\,^\ga [B]\\
&+& \om_\be\,^\eta T_{\al\eta}\,^\ga [B]
+\om^\ga\,_\eta T_{\al\be}\,^\eta [B] \nonumber \eeqq
and
\beqq \de_\om \Rabgd [B] &=& - \om^{\eta\ze}x_\ze\,\pa_\eta \Rabgd [B]
+ \om_\al\,^\eta R_{\eta\be}\,^{\ga\de} [B]\\
&+& \om_\be\,^\eta R_{\al\eta}\,^{\ga\de} [B]
+ \om^\ga\,_\eta R_{\al\be}\,^{\eta\de} [B] 
+ \om^\de\,_\eta R_{\al\be}\,^{\ga\eta} [B],
\nonumber \eeqq
where $\de_\om$ denotes the variation under an infinitesimal gauge transformation. Note that the first terms $- \om^{\eta\ze}x_\ze\,\pa_\eta \dots$ in all the variations above account for the coordinate change related to a local Lorentz transformation in our approach whilst $\de_\om x^\al = 0$ \cite{chw1}.

By construction the action $S_G [B]$ in Eqn.(\ref{2}) is the most general action of dimension $\leq 4$ in the gauge fields $\Bagd$ and their first and second derivatives $\pa_\be \Bagd, \pa_\eta \pa_\be \Bagd$ which is locally Lorentz invariant and renormalizable by power-counting.

The actual proof of renormalizability delivered in this paper requires the much more involved demonstration that counterterms needed to absorb infinite contributions to the perturbative expansion of the effective action of the full quantum theory are again of the form Eqn.(\ref{2}) plus gauge-fixing and ghost terms with possibly renormalized fields and coupling constants.

We finally note that for the choice
\beq \al_1 = 1,\quad \al_2 = -\frac{1}{4},\quad \al_3 = -\frac{1}{2},\quad
\al_4 = -1,\quad \al_5 = 2
\eeq
$S^{(0)}_G [B] + S^{(2)}_G [B]$ coupled to scalar matter is equivalent to General Relativity with a cosmological constant term as demonstrated in \cite{chw1}.

Let us go back to the path integral in Eqn.(\ref{1}). It runs over all possible gauge-equivalent field configurations thence counting a physically relevant field configuration multiple times in the integration. In order to separate the part of the integration related to gauge-invariance from the physically relevant integration over gauge-non-equivalent field configurations we divide the configuration space $\{\Bagd \}$ into equivalence classes $[B^g_\al\,^{\ga\de} ]$ of fields which are gauge-equivalent under the gauge transformation Eqn.(\ref{17}). The integrand in Eqn.(\ref{1}) is then constant over a given equivalence class, and the integral itself proportional to the infinite volume of the Lorentz gauge group. In itself this poses no insurmountable problem when calculating Eqn.(\ref{1}) non-perturbatively. However, the quadratic part of the action  $S_G$ in Eqn.(\ref{2}) as defined in \cite{chw2} is not invertible due to zero eigenvalues related to the gauge symmetry. So in order to perturbatively deal with calculating integrals of the type of Eqn.(\ref{1}) we have to factor out the volume of the Lorentz gauge group in the integration.

Following the Faddeev-Popov-deWitt approach \cite{wein2, poko} we introduce
\beq \label{21} 1 = \Delta [B] \int\!\Pi_{\!\!\!\!\!\!_{_{_{x}}}} \,\,\, d g (x)\, \de\! \left(f^{\eta\ze}[B^g] (x) \right),
\eeq
where $g$ is an element of the gauge group {\bf SO(1,3)\/}, $\Pi_{\!\!\!\!\!\!_{_{_{x}}}} \,\,\, d g (x)$ is a gauge-invariant measure over the gauge group and $f^{\eta\ze}[B^g] (x) = 0$ has exactly one solution and hence fixes a gauge. Note that $\Delta [B]$ is gauge-invariant.

Let us next insert the expression above into the path integral Eqn.(\ref{1}) and change the order of integration
\beqq & & \int\!\Pi_{\!\!\!\!\!\!_{_{_{x}}}} \,\,\, d g (x)\! 
\int\!\Pi_{\!\!\!\!\!\!_{_{_{x;\al,\ga,\de}}}} \!\!\!\!\!\!\!\!d\Bagd (x)\,
\Delta [B]\, \de\! \left(f^{\eta\ze}[B^g] (x) \right) \\
& & \quad\quad\quad \cdot{\cal O} [B]\, \exp\,i\left\{S_G [B] + \ep \mbox{-terms} \right\}. \nonumber
\eeqq
The expression 
\beqq \label{23} & & \int\!\Pi_{\!\!\!\!\!\!_{_{_{x;\al,\ga,\de}}}} \!\!\!\!\!\!\!\!d B^g_\al\,^{\ga\de} (x)\,
\Delta [B^g]\, \de\! \left(f^{\eta\ze}[B^g] (x) \right) \\
& & \quad \cdot{\cal O} [B^g]\, \exp\,i\left\{S_G [B^g] + \ep \mbox{-terms} \right\} \nonumber
\eeqq
turns out to be gauge-invariant which allows us to separate the group volume from the gauge-fixed remainder of the integral in the $ f^{\eta\ze}[B^g] = 0 $ gauge
\beqq \label{24} & & \left( \int\!\Pi_{\!\!\!\!\!\!_{_{_{x}}}} \,\,\, d g (x)\right)
\int\!\Pi_{\!\!\!\!\!\!_{_{_{x;\al,\ga,\de}}}} \!\!\!\!\!\!\!\!d\Bagd (x)\,
\Delta [B]\, \de\! \left(f^{\eta\ze}[B^g] (x) \right) \\
& & \quad\quad\quad\quad \cdot{\cal O} [B]\, \exp\,i\left\{S_G [B] + \ep \mbox{-terms} \right\}. \nonumber
\eeqq

Next we calculate $ \Delta [B] $ by changing variables
\beq \Delta^{-1} [B] = \int\!\Pi_{\!\!\!\!\!\!_{_{_{x;\eta,\ze}}}} \!\!\!\! d\!\, f^{\eta\ze} (x)\, 
\left( \Det\frac{\de f^{\eta\ze}[B^g]}{\de g} \right)^{-1}
\de\! \left(f^{\eta\ze} (x) \right)
\eeq
and find
\beq \Delta [B] = \Det\frac{\de f^{\eta\ze}[B^g]}{\de g}_{\mid_{f^{\eta\ze}[B^g]=0}}
=\; \Det\frac{\de f^{\eta\ze} [B^\om]}{\de \om^{\iota\ka}}_{\mid_{_{\om=0}}},
\eeq
where the last equality relates to it being sufficient to calculate the value of the Jacobian related to infinitesimal variations.

The Faddeev-Popov-deWitt operator is defined as
\beq {\cal F}^{\eta\ze}\,_{\iota\ka} \left[B; x,y \right] \equiv
\frac{\de f^{\eta\ze} [B^\om(x)]}{\de \om^{\iota\ka}(y)}_{\mid_{_{\om=0}}}.
\eeq
Choosing the axial gauge with the gauge fixing functional
\beq f^{\ga\de}[B] = n^\al \Bagd,
\eeq
where $n^\al$ is a constant vector in tangent space with $\de_\om n^\al = - n^\be\, \om_\be\,^\al$, we find
\beqq & & \inty\, {\cal F}^{\ga\de}\,_{\al\be} \left[B; x, y \right] \om^{\al\be} (y)
= n^\al \Big(\! -\om^{\eta\ze}x_\ze \pa_\eta \Bagd \nonumber \\
& & \quad\quad\,\, -\, d^B_\al \om^{\ga\de} + \om_\al\,^\be B_\be\,^{\ga\de} 
+ \om^\ga\,_\eta B_\al\,^{\eta\de} + \om^\de\,_\eta B_\al\,^{\ga\eta} \Big) \\
& & \quad\quad\,\, -\, n^\be\, \om_\be\,^\al \Bagd = - n^\al \pa_\al \om^{\ga\de}. \nonumber
\eeqq
Note that we have used $n^\al \Bagd = 0$. In this case the Faddeev-Popov-deWitt determinant 
$ \Det\frac{\de f^{\eta\ze} [B^\om]}{\de \om^{\iota\ka}}_{\mid_{_{\om=0}}} =  \Det (- n^\al \pa_\al ) $ is field-independent and can be taken in front of the integral Eqn.(\ref{24}) which is generally not the case.

The existence of a gauge with this property guarantees the decoupling of ghosts and anti-ghosts from the real physics in our theory and the pseudo-unitarity of the $S$-matrix on the na\"\i ve Fock space of both positive-norm, positive-energy and negative-norm, negative-energy states related to the gauge field as introduced in \cite{chw2}.

To demonstrate the actual renormalizability we however choose the Lorentz gauge condition
\beq \label{31} f^{\ga\de}[B] = \pa^\al \Bagd = 0
\eeq
with $\de_\om {\pa\rvec}^\al = - {\pa\rvec}^\be\! \om_\be\,^\al$. Here we find the field-dependent Faddeev-Popov-deWitt operator
\beqq \label{32} & & \inty\, {\cal F}^{\ga\de}\,_{\al\be} \left[B; x, y \right] \om^{\al\be} (y)
= \pa^\al \Big(\! -\om^{\eta\ze}x_\ze \pa_\eta \Bagd \nonumber \\
& & \quad\quad\,\, -\, d^B_\al \om^{\ga\de} + \om_\al\,^\be B_\be\,^{\ga\de} 
+ \om^\ga\,_\eta B_\al\,^{\eta\de} + \om^\de\,_\eta B_\al\,^{\ga\eta} \Big) \\
& & \quad\quad\,\, -\, {\pa\rvec}^\be\! \om_\be\,^\al \Bagd = \pa^\al \Big(\! -\om^{\eta\ze}x_\ze \pa_\eta \Bagd \nonumber \\
& & \quad\quad\,\, -\, d^B_\al \om^{\ga\de}
+ \frac{1}{2}\, C^{\ga\de}\,_{\iota\ka\,\eta\ze}\, B_\al\,^{\iota\ka} \om^{\eta\ze}\Big) \nonumber
\eeqq
with the expression in brackets on the last line being the covariant derivative $ \nabla^B_\al \om^{\ga\de} $ of the infinitesimal gauge parameter $\om^{\ga\de}$ as expected.

Note that in both cases above the term $\dots \Big(\! -\om^{\eta\ze}x_\ze \pa_\eta \Bagd - \dots \Big)$ relates to taking into account both the spacetime and spin degrees of freedom of the gauge group {\bf SO(1,3)\/} in the Faddeev-Popov-deWitt approach. As the expressions in Eqns.(\ref{1}), (\ref{21}) and (\ref{23}) are separately invariant under inner {\bf SO(1,3)\/} gauge transformations acting on spin degrees of freedom only we could also work with Faddeev-Popov-deWitt operators without the term $\dots \Big(\! -\om^{\eta\ze}x_\ze \pa_\eta \Bagd - \dots \Big)$ which we will when determining the scaling behaviour of couplings in a separate paper.

Next we note that we can change the gauge fixing condition $f^{\eta\ze}[B^g] = 0$ to $f^{\eta\ze}[B^g] - C^{\eta\ze} = 0$ in
\beqq & & \left( \int\!\Pi_{\!\!\!\!\!\!_{_{_{x}}}} \,\,\, d g (x) \right)
\int\!\Pi_{\!\!\!\!\!\!_{_{_{x;\al,\ga,\de}}}} \!\!\!\!\!\!\!\!d\Bagd (x)\,
\de\! \left(f^{\eta\ze}[B^g] (x) - C^{\eta\ze} (x) \right) \\
& & \quad\quad\,\, \cdot{\cal O} [B]\, \Det {\cal F} [B]\, \exp\,i\left\{S_G [B] + \ep \mbox{-terms} \right\}, \nonumber
\eeqq
and integrate over a field-independent weight function ${\cal G} [C]$
\beqq & & \int\!\Pi_{\!\!\!\!\!\!_{_{_{x;\al,\ga,\de}}}} \!\!\!\!\!\!\!\!d\Bagd (x)\,
{\cal O} [B]\, \Det {\cal F} [B]\, \exp\,i\left\{S_G [B] + \ep \mbox{-terms} \right\} \\
& & \quad\quad\quad\,\, \cdot\int\!\Pi_{\!\!\!\!\!\!_{_{_{x}}}} \,\,\, d C (x)
\de\! \left(f^{\eta\ze}[B^g] (x) - C^{\eta\ze} (x) \right)\, {\cal G} [C] \nonumber
\eeqq
without altering the physics involved \cite{wein2, poko}. A familiar choice compatible with renormalizability is
\beq {\cal G} [C] = \exp - \frac{i}{2\xi} \int\! C_{\ga\de}\, C^{\ga\de}.
\eeq

Leaving aside the infinite gauge group volume $\int\!\Pi_{\!\!\!\!\!\!_{_{_{x}}}} \,\,\, d g (x)$ this amounts to adding a gauge-fixing term
\beqq & & \quad\quad\,\, S_G [B] \lar S_G [B] + S_{GF} [B] \\
& & S_{GF} [B] \equiv - \frac{i}{2\xi} \int\! f_{\ga\de}[B]\, f^{\ga\de}[B] \nonumber
\eeqq
to the gauge field action, destroying gauge invariance in the process as it better should if we want to use the combined action to perturbatively evaluate our path integrals. So finally we get the gauge-fixed expression for the path integral representing the expectation value of an observable ${\cal O} [B]$
\beq \int\Pi_{\!\!\!\!\!\!_{_{_{x;\al,\ga,\de}}}} \!\!\!\!\!\!\!\!d\Bagd (x)\,
{\cal O} [B]\, \Det {\cal F} [B]\, \exp\,i\left\{S_G [B] + S_{GF} [B] + \ep \mbox{-terms} \right\}.
\eeq

\section{Ghosts, Antighosts and Nakanishi-Lautrup Fields}

\paragraph{}
In this section we recast the Faddeev-Popov-deWitt determinant as a fermionic path integral over ghost and antighost fields and introduce the Nakanishi-Lautrup fields in preparation of the demonstration of BRST invariance of the gauge-fixed action.

Using the fact that Gaussian path integrals yield determinants we can re-express the Faddeev-Popov-deWitt determinant as a fermionic Gaussian path integral over anti-commuting ghost and antighost fields $\om^{\iota\ka}$ and $\om^*_{\eta\ze}$
\beq \Det {\cal F} \left[B \right] \propto \int\!\Pi_{\!\!\!\!\!\!_{_{_{x;\eta,\ze}}}}\!\!\!\!d\om^*_{\eta\ze}(x)\,
\int\!\Pi_{\!\!\!\!\!\!_{_{_{x;\iota,\ka}}}}\!\!\!\!d\om^{\iota\ka}(x)\, \exp\,i\,S_{GH}.
\eeq
Above $\om^{\iota\ka}$ and $\om^*_{\eta\ze}$ are antisymmetric tensors of integer spin and the ghost action $S_{GH}$ is given by
\beqq S_{GH} &\equiv& \intx \inty\,\,
\om^*_{\eta\ze} (x) {\cal F}^{\eta\ze}\,_{\iota\ka} \left[B; x,y \right] \om^{\iota\ka} (y) \\
&=& \intx\,\, \om^*_{\eta\ze} (x) {\it\Delta}^{\eta\ze} \left[B; x \right], \nonumber
\eeqq
where we have introduced the shorthand notation
\beq {\it\Delta}^{\eta\ze} \left[B; x \right] \equiv \inty\,\, {\cal F}^{\eta\ze}\,_{\iota\ka} \left[B; x,y \right] \om^{\iota\ka} (y)
\eeq
for later use.

Finally we re-express $\exp - \frac{i}{2\xi} \int\! f_{\ga\de} [B] \, f^{\ga\de} [B]$ as a bosonic Gaussian path integral over the Nakanishi-Lautrup fields $h^{\eta\ze}$
\beqq & &  \exp - \frac{i}{2\xi} \int\! f_{\ga\de} [B] \, f^{\ga\de} [B] \propto \\
& &\!\!\!\!\!\!\!\!\!\!\!\!\!\!\!\!\!\!\!\!\!\!\!\!\!\!\!\!\!\!\!\!\!\!\!\! \int\,\Pi_{\!\!\!\!\!\!_{_{_{x;\eta,\ze}}}}\!\!\!\!dh^{\eta\ze}(x)\,
\exp\,i \left\{\frac{\xi}{2} \int\! h_{\eta\ze} \, h^{\eta\ze}
+ \int\! h_{\eta\ze} \, f^{\eta\ze} [B] \right\} \nonumber
\eeqq
to arrive at the form of the gauge-fixed expression for the path integral representing the expectation value of an observable ${\cal O} [B]$ which is most convenient for our purpose to demonstrate renormalizability
\beqq & & \!\!\!\!\!\!\!\!\!\!\!\!\!\!\!\!\!\!\!\! \int\!\Pi_{\!\!\!\!\!\!_{_{_{x;\al,\ga,\de}}}} \!\!\!\!\!\!\!\!d\Bagd (x)\,
\int\!\Pi_{\!\!\!\!\!\!_{_{_{x;\eta,\ze}}}}\!\!\!\!d\om^*_{\eta\ze}(x)\,
\int\!\Pi_{\!\!\!\!\!\!_{_{_{x;\iota,\ka}}}}\!\!\!\!d\om^{\iota\ka}(x)\,
\int\!\Pi_{\!\!\!\!\!\!_{_{_{x;\rho,\si}}}}\!\!\!\!dh^{\rho\si}(x) \\
& & \quad\quad \cdot{\cal O} [B]\, \exp\,i\left\{S_{NEW} + \ep \mbox{-terms} \right\}. \nonumber
\eeqq
Here
\beq \label{43} S_{NEW} \equiv S_{G} + \int\! \om^*_{\eta\ze} {\it\Delta}^{\eta\ze} [B]
+ \int\! h_{\eta\ze}\, f^{\eta\ze} [B]
+ \frac{\xi}{2} \int\! h_{\eta\ze}\, h^{\eta\ze}
\eeq
is the gauge-fixed action for the gauge, ghost, antighost and Nakanishi-Lautrup fields which we will use as the starting point for the actual renormalizability proof.

Note the absence of the determinant $\dem$ in all contributions to $S_{NEW}$ apart from $S_{G}$ which will prove crucial to rewrite $S_{NEW} - S_{G}$ as a BRST transform in the next section.

The modified action above is not gauge invariant -- indeed, it had better not be, if we want to be able to use it in perturbative calculations.

\section{BRST Invariance}

\paragraph{}
In this section we introduce fermionic BRST field variations, demonstrate their nilpotence and based on this establish the invariance of $S_{NEW}$ under those BRST transformations.

Let us write down the various BRST variations starting with the one for a generic matter field $\psi$
\beq
\de_\theta \psi = \theta s \psi = \frac{i}{2} \theta\, \om^{\ga\de} ({L\rvec}_{\ga\de}\, \psi )
+ \frac{i}{2} \theta\, \om^{\ga\de}  \Si_{\ga\de}\, \psi,
\eeq
where $\theta$ is a fermionic parameter and assures the right statistics for the various field variations above and $s...$ indicates the infinitesimal variation of a given field without the factor $\theta$. We recall that ${L\rvec}_{\ga\de} = -i ( x_\ga \pa_\de - x_\de \pa_\ga )$ denotes the generators of the {\bf so(1,3)\/} Lorentz algebra acting on spacetime coordinates and $\Si_{\ga\de}$ generators of the Lorentz algebra acting on spin degrees of freedom. Note that for a generic matter field the BRST variation is nothing but an infinitesimal gauge variation with gauge parameter $\theta\, \om^{\ga\de}$.

The gauge field variation reads
\beqq \label{45}
\de_\theta \Bagd &=& \theta s \Bagd = \frac{i}{2} \theta\, \om^{\eta\ze} ({L\rvec}_{\eta\ze}\, \Bagd )
- \,\theta\, \pa_\al \om^{\ga\de} \nonumber \\
&-& \frac{i}{2} \theta\, B_\al\,^{\eta\ze} ({L\rvec}_{\eta\ze}\, \om^{\ga\de}) 
+ \frac{i}{2} \theta\, \om^{\eta\ze} \left(\Si^V_{\eta\ze}\right)_\al\,^\be\, B_\be\,^{\ga\de} \\
&+& \frac{1}{2} \theta\, C^{\ga\de}\,_{\iota\ka\,\eta\ze}\, B_\al\,^{\iota\ka} \om^{\eta\ze} \nonumber
\eeqq
which is an infinitesimal gauge variation with gauge parameter $\theta\, \om^{\ga\de}$. Above $\left( \Si^A_{\eta\ze} \right)^{\ga\de}\,_{\iota\ka} = i\, C^{\ga\de}\,_{\eta\ze\,\iota\ka}$ denotes the generators of the Lorentz algebra in the adjoint representation.

Next the ghost field variation is defined by
\beq
\de_\theta \om^{\ga\de} = \theta s \om^{\ga\de} = \frac{i}{2} \theta\, \om^{\eta\ze} ({L\rvec}_{\eta\ze}\, \om^{\ga\de})
- \frac{1}{4} \theta\, C^{\ga\de}\,_{\al\be\,\eta\ze}\, \om^{\al\be} \om^{\eta\ze}
\eeq
and the antighost variation
\beq
\de_\theta \om^*_{\ga\de} = \theta s \om^*_{\ga\de} = - \theta\, h_{\ga\de}
\eeq
with both being perspicously distinct from a regular infinitesimal gauge transformation. Finally the Nakanishi-Lautrup field is taken to be invariant
\beq
\de_\theta h_{\ga\de} = \theta s h_{\ga\de} = 0
\eeq
under BRST variations. Note the absence of the spacetime-related part $\frac{i}{2} \theta\, \om^{\eta\ze} ({L\rvec}_{\eta\ze}\, \dots)$ in both the antighost and Nakanishi-Lautrup field variations.

For later use we also write down the BRST variation of $\dem [B]$
\beq
\de_\theta \dem [B] = - \theta\, \pa_\eta \left( \om^{\eta\ze} x_\ze \dem [B] \right).
\eeq

It is crucial for the sequel that all the BRST variations above are nilpotent, or $ss... = 0$, as some quite tedious algebra in Appendix A demonstrates. This also holds true for any functional $F$ of the fields above, or $ss F= 0$ \cite{wein2}.

Note that we have written the BRST variations above in terms of the Lorentz algebra generators which proves to be of enormous help to organize the lengthy algebra involved in proving nilpotence.

Let us turn to evaluate
\beq \de_\theta \left( 
\om^*_{\eta\ze}\, f^{\eta\ze} [B] + \frac{\xi}{2} \om^*_{\eta\ze}\, h^{\eta\ze}
\right).
\eeq
For the variation of $f^{\eta\ze} [B]$ we find
\beq \de_\theta f^{\eta\ze} [B] 
= \int \frac{\de f^{\eta\ze} [B]}{\de B_\al\,^{\iota\ka}} \left( \de_\theta B_\al\,^{\iota\ka} \right)
+ \int \left( \de_\theta \frac{\de f^{\eta\ze} [B]}{\de B_\al\,^{\iota\ka}} \right) B_\al\,^{\iota\ka}
= \theta{\it\Delta}^{\eta\ze} [B], \nonumber
\eeq
where the second term accounts for the non-trivial transformation of the $\al$-index in $\frac{\de f^{\eta\ze} [B]}{\de B_\al\,^{\iota\ka}}$. Using this and taking into account that $\theta$ and $\om^*$ anticommute we get
\beq \de_\theta\! \left(\! \om^*_{\eta\ze}\, f^{\eta\ze} [B] \!+\! \frac{\xi}{2} \om^*_{\eta\ze}\, h^{\eta\ze} \!\right)
= - \theta\! \left(\! \om^*_{\eta\ze}\, {\it\Delta}^{\eta\ze} [B]
\!+\! h_{\eta\ze}\, f^{\eta\ze} [B]
\!+\! \frac{\xi}{2} h_{\eta\ze}\, h^{\eta\ze} \!\right)
\eeq
which allows us to rewrite
\beq S_{NEW} = S_{G}
- s \left( \om^*_{\eta\ze}\, f^{\eta\ze} [B] + \frac{\xi}{2} \om^*_{\eta\ze}\, h^{\eta\ze} \right).
\eeq
Evoking nilpotence  for the term $s (...)$ in brackets, or  $ss (...) = 0$, and the fact that $S_{G}$ is gauge-invariant we find that
\beq \de_\theta S_{NEW} = 0
\eeq
or that $S_{NEW}$ is indeed BRST invariant -- and so is the gauge-fixed expression for the path integral representing the expectation value of an observable ${\cal O} [B]$
\beqq & & \!\!\!\!\!\!\!\!\!\!\!\!\!\!\!\!\!\!\!\! \int\!\Pi_{\!\!\!\!\!\!_{_{_{x}}}}\,\,\, d\psi (x)\,
\int\!\Pi_{\!\!\!\!\!\!_{_{_{x;\al,\ga,\de}}}} \!\!\!\!\!\!\!\!d\Bagd (x)\,
\int\!\Pi_{\!\!\!\!\!\!_{_{_{x;\eta,\ze}}}}\!\!\!\!d\om^*_{\eta\ze}(x)\,
\int\!\Pi_{\!\!\!\!\!\!_{_{_{x;\iota,\ka}}}}\!\!\!\!d\om^{\iota\ka}(x)\,
\int\!\Pi_{\!\!\!\!\!\!_{_{_{x;\rho,\si}}}}\!\!\!\!dh^{\rho\si}(x) \\
& & \quad\quad\quad \cdot{\cal O} [B]\, \exp\,i\left\{S_{NEW} + S_M 
+ \ep \mbox{-terms} \right\} \nonumber
\eeqq
if the action $S_M [\psi, B]$ for a matter field $\psi$ is gauge-invariant. We note that all the integration measures over field space are BRST invariant as demonstrated in Appendix B.

\section{Zinn-Justin Equation}

\paragraph{}
In this section we derive a fundamental property of the theory, the Zinn-Justin equation for the quantum effective action related to the connected vacuum persistence amplitude ${\cal W}\left[J, K \right]$ in the presence of external currents $J$ and $K$ for the fundamental fields $\chi^n$ and their BRST variations $s \chi^n$ respectively \cite{wein2}.

Let us introduce the shorthand notation $\chi^n$ for the fundamental fields
\beq \chi^n \sim \psi, B_\al, \om, \om^*, h
\eeq
The BRST transformations in this notation read
\beqq \chi^n (x) \ar {\chi^n} ' (x) &=& \chi^n (x) + \de_\theta \chi^n [\chi^t; x] \\
\de_\theta \chi^n [\chi^t; x] &=& \theta\, s \chi^n [\chi^t; x] \equiv \theta\, \Delta^n [\chi^t; x]  \nonumber
\eeqq
with
\beqq \label{58} & & \Delta^\psi = \frac{i}{2} \om^{\ga\de} {L\rvec}_{\ga\de}\, \psi
+ \frac{i}{2} \om^{\ga\de}  \Si_{\ga\de}\, \psi \nonumber \\
& & \Delta^B_\al\,^{\ga\de} = \frac{i}{2} \om^{\eta\ze} {L\rvec}_{\eta\ze}\, \Bagd 
- \, \pa_\al \om^{\ga\de} - \frac{i}{2} B_\al\,^{\eta\ze} {L\rvec}_{\eta\ze}\, \om^{\ga\de} \nonumber \\
& &\quad\quad\quad +\, \frac{i}{2} \om^{\eta\ze} \left(\Si^V_{\eta\ze}\right)_\al\,^\be\, B_\be\,^{\ga\de} 
+ \frac{1}{2}\, C^{\ga\de}\,_{\iota\ka\,\eta\ze}\, B_\al\,^{\iota\ka} \om^{\eta\ze} \\
& & \Delta^{\om\,\ga\de} = \frac{i}{2} \om^{\eta\ze} {L\rvec}_{\eta\ze}\, \om^{\ga\de}
- \frac{1}{4}\, C^{\ga\de}\,_{\al\be\,\eta\ze}\, \om^{\al\be} \om^{\eta\ze}  \nonumber \\
& & \Delta^{\om^*}\,\!_{\ga\de} = - h_{\ga\de}
 \nonumber \\
& & \Delta^h\,_{\ga\de} = 0. \nonumber
\eeqq

As demonstrated above we have
\beqq \label{59} & & S_{TOT} [{\chi^n} '] = S_{TOT} \left[\chi^n + \theta\, \Delta^n [\chi^t] \right] \\
& &\quad = S_{TOT} [\chi^n] +  \de_\theta S_{TOT} [\chi^n ]\, \theta\, \Delta^n [\chi^t]= S_{TOT} [\chi^n] \nonumber
\eeqq
with $S_{TOT} = S_{NEW} + S_M$. In addition we have
\beq \label{60} \Pi_{\!\!\!\!\!\!_{_{_{x;n}}}} \!d \left( \chi^n (x) + \de_\theta \chi^n [\chi^t; x] \right)
= \Pi_{\!\!\!\!\!\!_{_{_{x;m}}}} \!d\chi^m (x) \, {\cal J}
\eeq
with the Berezinian
\beq
{\cal J} = \Det \left( \frac{\de {\chi^n}' }{\de \chi^m } \right) = 1 + \Tr \log \left( \frac{\de {\chi^n}' }{\de \chi^m } \right) = 1
\eeq
being trivial as demonstrated in Appendix B.

Next we introduce the connected vacuum persistence amplitude ${\cal W}\left[J, K \right]$ in the presence of external currents $J$ and $K$ for the fundamental fields $\chi^n$ and their BRST variations $s \chi^n$ respectively
\beqq & & {\cal Z}\left[J, K \right] \equiv \exp\,i\, {\cal W}\left[J, K \right]
\equiv \int\Pi_{\!\!\!\!\!\!_{_{_{x;n}}}} \!d\chi^n (x) \\
& & \quad \cdot \exp\,i\,\left\{S_{TOT} + S_M + \intx\, \Delta^n K_n 
+ \intx\, \chi^n J_n + \ep \mbox{-terms} \right\}. \nonumber
\eeqq
This allows us to derive a condition on the quantum effective action
\beq \Ga\left[\chi, K \right] \equiv {\cal W}\left[J_{_{\chi,K}}, K \right]
- \int \chi^n \, J_{n\,_{\chi,K}}
\eeq
belonging to the connected vacuum persistence amplitude ${\cal W}\left[J, K \right]$. 

Note that for $K = 0$ the functional ${\cal Z}\left[J, 0 \right]$ reduces to the usual gene- rating functional for the Green functions of the interacting theory which are equal to the vacuum expectation values of time-ordered products of interacting field operators from which the $S$-matrix is derived via the LSZ approach. Also, $\Ga\left[\chi, 0 \right]$ is the usual quantum effective action which contains all connected one-particle irreducible graphs of the interacting theory in the presence of the current $J_{_{\chi,0}}$.

The condition referred to above, a Slavnov-Taylor identity, follows from the BRST invariance of ${\cal W}\left[0, 0 \right]$ for vanishing currents $J, K$ which is easy to demonstrate on the basis of Eqns.(\ref{59}) and (\ref{60}). To derive the Slavnov-Taylor identity we calculate
\beqq \label{64} & & {\cal Z}\left[J, K \right] 
= \int\Pi_{\!\!\!\!\!\!_{_{_{x;n}}}} \!d\left(\chi^n + \theta \Delta^n \left[\chi \right] \right) (x)\nonumber \\
& & \quad \cdot \exp\,i\,\bigg\{ S_{TOT} \left[\chi^n + \theta \Delta^n \left[\chi \right] \right] 
+ \intx\, \Delta^n \left[\chi^m + \theta \Delta^m \left[\chi \right] \right] K_n \nonumber \\
& & \quad\quad\quad\quad\quad\quad +\, \intx\, \left(\chi^n + \theta \Delta^n \left[\chi \right] \right) J_n \bigg\} \\
& & = {\cal Z}\left[J, K \right]
+\, i\, \theta \int\Pi_{\!\!\!\!\!\!_{_{_{x;n}}}}\! d\chi^n (x)\,
\left( \inty\, \Delta^m \left[\chi^t; y \right] \, J_m (y) \right) \nonumber \\
& & \quad \cdot \exp\,i\,\left\{S_{TOT} + \intx\, \Delta^n K_n + \intx\, \chi^n J_n \right\}, \nonumber
\eeqq 
where we have taken into account the nilpotence of the BRST variations.

Defining the quantum average
\beqq & & \left\langle \Delta^n \left[\chi^t; y \right] \right\rangle_{J_{_{\chi,K}},K} 
= \int\Pi_{\!\!\!\!\!\!_{_{_{x;n}}}}\! d\chi^n (x)\, \Delta^m \left[\chi^t; y \right] \\
& & \quad \cdot \exp\,i\,\left\{S_{TOT} + \intx\, \Delta^n K_n + \intx\, \chi^n J_n \right\}, \nonumber
\eeqq
in the presence of currents $J$ and $K$ we get
\beq \label{65} \intx\, \left\langle \Delta^n
\left[\chi^t; x \right] \right\rangle_{J_{_{\chi,K}},K}\, J_n (x) = 0.
\eeq
Noting that
\beq \frac{\de_L \Ga\left[\chi, K \right]}{\de \chi^n (x)} = - J_{n\,_{\chi,K}} (x)
\eeq
we can recast Eqn.(\ref{65}) in the more perspicuous form
\beq \label{67} \intx\, \left\langle \Delta^n
\left[\chi^t; x \right] \right\rangle_{J_{_{\chi,K}},K}\, \frac{\de_L \Ga\left[\chi, K \right]}{\de \chi^n (x)} = 0.
\eeq
In other words $\Ga\left[\chi, K \right]$ is invariant under the infinitesimal transformations
\beq \chi^n (x) \lar \chi^n (x)\, + \theta\, \left\langle \Delta^n
\left[\chi^t; x \right] \right\rangle_{J_{_{\chi,K}},K}
\eeq
establishing a Slavnov-Taylor identity which is the basis for the Zinn-Justin equation we derive next.

Noting that
\beq \frac{\de_R \Ga\left[\chi, K \right]}{\de K_n (x)}
= \left\langle \Delta^n
\left[\chi^t; x \right] \right\rangle_{J_{_{\chi,K}},K},
\eeq
where we have introduced the left and right derivatives $\de_L$ and $\de_R$ respectively taking  the (anti-)commuting properties of the various fields into proper account, Eqn.(\ref{67}) can finally be rewritten as the Zinn-Justin equation
\beq \intx\, \frac{\de_R \Ga\left[\chi, K \right]}{\de K_n (x)}\, 
\frac{\de_L \Ga\left[\chi, K \right]}{\de \chi^n (x)} = 0.
\eeq

Defining the antibracket of two functionals $F\left[\chi, K \right]$ and $G\left[\chi, K \right]$ w.r.t to the fields $\chi^n$ and the currents $K_n$
\beq \left( F,\, G \right) \equiv \intx\, \left\{ 
\frac{\de_R F\left[\chi, K \right]}{\de \chi^n (x)}\, 
\frac{\de_L G\left[\chi, K \right]}{\de K_n (x)} -\, \frac{\de_R F\left[\chi, K \right]}{\de K_n (x)}\,
\frac{\de_L G\left[\chi, K \right]}{\de \chi^n (x)} \right\}
\eeq
we can recast the Zinn-Justin equation in its final form as
\beq \left( \Ga,\, \Ga \right) = 0
\eeq
which is the starting point for the renormalizability proof for our theory in the next and final section of the paper.

\section{Perturbative Renormalizability of the Quantum Effective Action $\Ga [\chi, K]$ }

\paragraph{}
In this section we prove the renormalizability of our theory closely following the approach outlined in \cite{wein2}. First, we use renormalizability in the Dyson sense to derive the explicit $K$-dependence of $\Ga_{N,\infty} [\chi, K]$ which contains the infinite contributions of order $N$ to the loop expansion of the effective action $\Ga [\chi, K] = \sum_{N = 0}^\infty \Ga_N [\chi, K]$. Second, evaluating the Zinn-Justin equation we find the combination $\Delta_N^{(\ep) n} (x) \equiv \Delta^n (x) + \ep\, {\cal D}_N^n (x)$ to be nilpotent with ${\cal D}_N^n$ properly defined below, and the combination $\Ga_N^{(\ep)} [\chi] \equiv S_R [\chi] + \ep\, \Ga_{N,\infty} [\chi, 0]$ to be invariant under the renormalized BRST field variations $\de_{\theta^{(\ep)}} \chi^n (x) = \theta\, \Delta_N^{(\ep) n} (x)$. This will finally allow us to prove the renormalizablity of our theory.

\subsection{$K$-dependence of $\Ga_{N,\infty} [\chi, K]$ }

\paragraph{}
In this subsection we use the renormalizability of our theory in the Dyson sense to derive the explicit $K$-dependence of $\Ga_{N,\infty} [\chi, K]$ which contains the infinite contributions of order $N$ to the loop expansion of the effective action $\Ga [\chi, K] = \sum_{N = 0}^\infty \Ga_N [\chi, K]$.

We start by noting that $S_{TOT} [\chi]$ is by construction a sum of integrals over Lagrangians of dimension four or less expressed in the fundamental fields $\chi^n$ -- in fact it is the most general BRST-invariant action of dimension four or less in those fields. As a consequence power-counting allows to show that the corresponding quantum effective action of the quantized theory only contains divergent contributions of dimension four or less in those fields -- or is renormalizable in the Dyson sense \cite{wein1}. They then can be cancelled by counterterms of dimensionality four or less.

However, there is more to full renormalizability. The action used in the path integral or canonical quantization of our gauge field theory is constrained by BRST invariance -- in fact it is the most general BRST-invariant action of dimension four or less in all the dynamical fields. For the quantum theory to be renormalizable, i.e. all infinities to be absorbable solely by field renormalizations and coupling redefinitions, the infinite contributions to the quantum effective action and the counterterms needed to cancel them have to satisfy the same BRST constraints up to such renormalizations of fields and couplings -- which guarantees that the counterterms must be of the same form as the terms in the original action. In other words BRST invariance and the resulting Zinn-Justin equation are enough of an algebraic "straightjacket" to assure renormalizability.

The first of a sequence of steps to prove full renormalizability is to determine the $K$-dependence of the infinite contributions to the effective action $\Ga [\chi, K]$ deploying dimensional analysis.

Based on the Dyson renormalizability we can rewrite the action $S [\chi, K]$ in the presence of sources $K$ as
\beqq S [\chi, K] &=& S_{TOT} [\chi] + \int \Big\{ \Delta^\psi\, K_\psi
+ \Delta^B_\al\,^{\ga\de}\, K_B^\al\,_{\ga\de} \nonumber \\
& &\quad +\, \Delta^{\om\,\ga\de}\, K_{\om\,\ga\de}
+ \Delta^{\om^*}\,\!_{\ga\de}\, K_{\om^*}\,\!^{\ga\de} \Big\} \\
&=& S_R [\chi, K] + S_\infty [\chi, K], \nonumber
\eeqq
where masses and coupling constants in $S_R [\chi, K]$ are set to their renormalized values plus the correction $S_\infty [\chi, K]$ containing all the counterterms needed to cancel infinities from loop graphs in the perturbative loop expansion of the effective action $\Ga [\chi, K]$
\beq \label{74} \Ga [\chi, K] = \sum_{N = 0}^\infty \Ga_N [\chi, K].
\eeq
Above $\Ga_N$ contains all the diagrams with $N$ loops, plus contributions from graphs with $N-M$ loops, $1 \leq M \leq N$, involving the counterterms in $S_\infty$ introduced to cancel infinities in graphs with $M$ loops. Note that no source term $K_h\,\!^{\ga\de}$ for $\Delta^h\,\!_{\ga\de} = 0$ appears.

The power-counting rules of renormalization theory imply that after all infinities in subgraphs of $\Ga_N$ have been cancelled the infinite part $\Ga_{N,\infty} [\chi, K]$ of $\Ga_N [\chi, K]$ must be an integral over a sum of local products of fields $\chi, K$ and their derivatives of dimension four or less \cite{wein1}.

Now it is possible to determine the $K$-dependence of the infinite contributions to the effective action $\Ga [\chi, K]$. To that end we first establish the dimensions of the various fields. If the fields $\chi^n$ have dimensionality $[\chi^n] \equiv d_n$ then inspection of Eqns.(\ref{58}) shows that the dimensionality of $\Delta^n$ is $[\Delta^n] = d_n + 1$ and $K_n$ has dimensionality $[K_n] = 4 - [\Delta^n] = 3 - d_n$.

So we find the dimensionalities for the various fields to be
\beqq & & [ B ] = [ \om ] = [ \om^* ] = 1,
\nonumber \\
& & [ h ] = [ K_B ] = [ K_\om ] = [ K_{\om^*} ] = 2, \\
& & [ \psi ] = 3/2,\, [ K_\psi ] = 3/2, \nonumber
\eeqq
where we assume the matter field $\psi$ to be a spin-$\frac{1}{2}$ Dirac fermion. The dimension four quantity $\Ga_{N,\infty} [\chi, K]$ can then be at most quadratic in any of the $K_n$, and terms quadratic in any of the $K_n$ cannot involve any other fields with the exception of a term quadratic in $K_\psi$ which may contain one additional field of dimension one.

Using ghost number conservation we next demonstrate $\Ga_{N,\infty} [\chi, K]$ to be at most linear in $K_n$. If the fields $\chi^n$ have ghost number $|\chi^n| \equiv g_n$ then inspection of Eqns.(\ref{58}) shows that the ghost number of $\Delta^n$ is $|\Delta^n| = g_n + 1$ and $K_n$ has ghost number $|K_n| = - |\Delta^n| = -g_n - 1$.

So we find the ghost numbers for the various fields to be
\beqq & & | B | = | \psi | = | h | = 0,\quad | K_B | = | K_\psi | = -1, \nonumber \\
& & | \om | = 1,\quad | K_\om | = -2, \\
& & | \om^* | = -1,\quad | K_{\om^*} | = 0. \nonumber
\eeqq
This rules out all potential contributions to $\Ga_{N,\infty} [\chi, K]$ of second order in $K_n$ with the exception of a potential term of second order in $K_{\om^*}$. Now as
\beq \label{77} \frac{\de_R \Ga\left[\chi, K \right]}{\de K_{\om^*}\,\!^{\ga\de}}
= \left\langle \Delta^{\om^*}\,\!_{\ga\de} \right\rangle_{J_{_{\chi,K}},K}
= -\left\langle h_{\ga\de} \right\rangle_{J_{_{\chi,K}},K}
= - h_{\ga\de}
\eeq
is independent of $K_{\om^*}$ the effective action $\Ga\left[\chi, K \right]$ is linear in $K_{\om^*}$ through a term $-\intf\, K_{\om^*}\,\!^{\ga\de} h_{\ga\de}$ and the $\Ga_{N,\infty} [\chi, K]$ are independent of $K_{\om^*}$ for $N\geq 1$.

Note that the last equality in Eqn.(\ref{77}) above follows from the fact that for transformations
\beq \chi^n (x) \ar \chi^n (x) + \ep F^n [\chi^m; x]
\eeq 
which are linear in the fields
\beq \label{79} F^n \left[\chi^m; x \right] = s^n (x) + \int t^n\,_m (x,y)\,
\chi^m (y)
\eeq
with $\ep$ infinitesimal and $s$ and $t$ field-independent, the quantum average of the field variation $\langle F^n \left[\chi^m \right] \rangle_{J_{_{\chi,K}},K}$ equals its classical value $F^n \left[\chi^m \right]$ as is easily shown in a calculation analogous to the one in Eqn.(\ref{64}).

In fact, if the effective action $\Ga [\chi, K]$ is invariant under a variation with a general $\left\langle F^n \left[\chi^m \right] \right\rangle$
\beq \de_{\!_F} \Ga = \int\, \left\langle F^n \left[\chi^m \right] \right\rangle_{J_{_{\chi,K}},K} 
\, \frac{\de \Ga [\chi, K]}{\de \chi^n} 
= \int\, \left\langle F^n \left[\chi^m \right] \right\rangle_{J_{_{\chi,K}},K}\, J_n = 0
\eeq
then for linear transformations Eqn.(\ref{79}) we have
 \beqq & & \langle F^n \left[\chi^m \right] \rangle_{J_{_{\chi,K}},K} 
= s^n + \int t^n\,_m\, \langle \chi^m \rangle_{J_{_{\chi,K}},K} \\
& &\quad = s^n + \int t^n\,_m\, \chi^m
= F^n \left[\chi^m \right] \nonumber
\eeqq
so the invariance becomes
\beq \de_{\!_F} \Ga = \int F^n \left[\chi^m \right] 
\, \frac{\de \Ga [\chi, K]}{\de \chi^n} = 0
\eeq
and the full quantum effective action is invariant under the same linear transformation under which the classical action is, assuming the integration measure is invariant as well.

Hence, finally we find the desired $K$-dependence of $\Ga_{N,\infty} [\chi, K]$ to be
\beq \Ga_{N,\infty} [\chi, K] = \Ga_{N,\infty} [\chi, 0]
+ \intx\, {\cal D}_N^n [\chi; x]\, K_n (x)
\eeq
defining ${\cal D}_N^n [\chi; x]$ in the process which is a functional of the fields $\chi^n$ only.

\subsection{Invariance of $\Ga_N^{(\ep)} [\chi]$ under Nilpotent Transformations $\Delta_N^{(\ep) n}$ }

\paragraph{}
In this subsection we evaluate the Zinn-Justin equation perturbatively allowing us to demonstrate the combination $\Ga_N^{(\ep)} [\chi] \equiv S_R [\chi] + \ep\, \Ga_{N,\infty} [\chi, 0]$ to be invariant under nilpotent renormalized BRST field variations $\de_{\theta^{(\ep)}} \chi^n (x) = \theta\, \Delta_N^{(\ep) n} (x)$ with $\Delta_N^{(\ep) n} (x) \equiv \Delta^n (x) + \ep\, {\cal D}_N^n (x)$ and $\ep$ infinitesimal.

Taking the Zinn-Justin equation $\left( \Ga,\, \Ga \right) = 0$ and inserting the perturbative expansion $\Ga [\chi, K] = \sum_{N = 0}^\infty \Ga_N [\chi, K]$
we get for fixed $N$
\beq \label{84} \sum_{M = 0}^N \left(\Ga_{M},\, \Ga_{N - M} \right) = 0.
\eeq
The leading term in the expansion Eqn.(\ref{74}) is
\beq \Ga_0 [\chi, K] = S_R [\chi, K]
\eeq
which is finite. Supposing that all infinities from loops for $M\leq N - 1$ have been absorbed by respective counterterms in $S_\infty [\chi, K]$ new infinities can only appear in the $M = 0$ and $M = N$ terms which are equal. Now the infinite part of the condition Eqn.(\ref{84}) is
\beq \left(S_R,\, \Ga_{N,\infty} \right) = 0.
\eeq

Recalling that $S_R [\chi, K] = S_R [\chi] + \intx\, \Delta^n [\chi; x] K_n (x)$ and inserting this together with $\Ga_{N,\infty} [\chi, K] = \Ga_{N,\infty} [\chi, 0]
+ \intx\, {\cal D}_N^n [\chi; x]\, K_n (x)$ into the infinite part of the $N$-th order contribution to the Zinn-Justin equation above we get to zeroth order in $K$
\beq \label{87} \intx\, \left\{ \Delta^n [\chi; x]\,
\frac{\de_L \Ga_{N,\infty} [\chi, 0]}{\de \chi^n (x)}
+ {\cal D}_N^n [\chi; x]\,
\frac{\de_L S_R [\chi]}{\de \chi^n (x)} \right\} = 0
\eeq
whilst terms linear in $K$ yield
\beq \label{88} \inty\, \left\{ \Delta^n [\chi; x]\,
\frac{\de_L {\cal D}_N^m [\chi; y]}{\de \chi^n (x)}
+ {\cal D}_N^n [\chi; x]\,
\frac{\de_L \Delta^m [\chi; y]}{\de \chi^n (x)} \right\} = 0.
\eeq 

To bring these two results into their most perspicous form we define the $N$-th order contribution to the corrected quantum effective action
\beq \Ga_N^{(\ep)} [\chi] \equiv S_R [\chi]
+ \ep\, \Ga_{N,\infty} [\chi, 0]
\eeq 
and
\beq \Delta_N^{(\ep) n} [\chi; x] \equiv \Delta^n [\chi; x]
+ \ep\, {\cal D}_N^n [\chi; x]
\eeq
with $ \ep$ infinitesimal. Then Eqn.(\ref{87}) in conjunction with the BRST invariance of $S_R$ tells us that to leading order in $\ep$  the expression $\Ga_N^{(\ep)} [\chi]$ is invariant under the field variations $\de_{\theta^{(\ep)}} \chi^n (x)$
\beq \chi^n (x) \lar \chi^n (x) + \theta\, \Delta_N^{(\ep) n} [\chi; x]
\eeq
or
\beq  \de_{\theta^{(\ep)}} \Ga_N^{(\ep)} [\chi] = \intx\, \Delta_N^{(\ep) n} [\chi; x]\,
\frac{\de_L \Ga_N^{(\ep)} [\chi]}{\de \chi^n (x)} = 0.
\eeq
In addition Eqn.(\ref{88}) in conjunction with the BRST invariance of $\Delta^n [\chi; x]$ tells us that to leading order in $\ep$ the variations $\Delta_N^{(\ep) n} [\chi; x]$ are nilpotent
\beq \de_{\theta^{(\ep)}} \Delta_N^{(\ep) n} (x) = 0.
\eeq

\subsection{Nilpotence forcing the $\Delta_N^{(\ep) n}$ to be Renormalized BRST Transformations}

\paragraph{}
In this subsection we determine the most general form of the nilpotent field variations $\Delta_N^{(\ep) n} [\chi; x]$.

Noting that $\Ga_N^{(\ep)} [\chi]$ is of dimensionality four or less, ${\cal D}_N^n [\chi; x]$ and hence $\Delta_N^{(\ep) n} [\chi; x]$ have at most the dimension of the original BRST transformations $\Delta^n [\chi; x]$. In addition the ${\cal D}_N^n [\chi; x]$ must share their Lorentz transformation behaviours with those of the $\Delta^n [\chi; x]$.

The most general renormalized nilpotent BRST variations are then found to be for the Dirac field
\beq \label{94}
\de_{\theta^{(\ep)}} \psi = \theta \Delta_N^{(\ep) \psi} 
= \frac{i}{2} \theta\, \om^{\ga\de} {\cal Z}_N^{(\ep)} ({L\rvec}_{\ga\de}\, \psi )
+ \frac{i}{2} \theta\, \om^{\ga\de} {\cal Z}_N^{(\ep)} \Si_{\ga\de}\, \psi,
\eeq
for the gauge field
\beqq \label{95} \!\!\!\!\!\!\!\!
\de_{\theta^{(\ep)}} \Bagd &=& \theta \Delta_N^{(\ep) B\, \ga\de} 
= \frac{i}{2} \theta\, \om^{\eta\ze} {\cal Z}_N^{(\ep)} ({L\rvec}_{\eta\ze}\, \Bagd )
- \,\theta\, {\cal Z}_N^{(\ep)} {\cal N}_N^{(\ep)} \pa_\al \om^{\ga\de} \nonumber \\
&-& \frac{i}{2} \theta\, B_\al\,^{\eta\ze} {\cal Z}_N^{(\ep)} ({L\rvec}_{\eta\ze}\, \om^{\ga\de}) 
+ \frac{i}{2} \theta\, \om^{\eta\ze} {\cal Z}_N^{(\ep)} \left(\Si^V_{\eta\ze}\right)_\al\,^\be\, B_\be\,^{\ga\de} \\
&+& \frac{1}{2} \theta\, {\cal Z}_N^{(\ep)} C^{\ga\de}\,_{\iota\ka\,\eta\ze}\, B_\al\,^{\iota\ka} \om^{\eta\ze} \nonumber
\eeqq
and for the ghost field
\beq
\de_{\theta^{(\ep)}} \om^{\ga\de} = \theta \Delta_N^{(\ep) \om\, \ga\de} 
= \frac{i}{2} \theta\, \om^{\eta\ze} {\cal Z}_N^{(\ep)} ({L\rvec}_{\eta\ze}\, \om^{\ga\de})
- \frac{1}{4} \theta\, {\cal Z}_N^{(\ep)} C^{\ga\de}\,_{\al\be\,\eta\ze}\, \om^{\al\be} \om^{\eta\ze}.
\eeq
In comparison to the original BRST variations $\Delta^n [\chi; x]$ we find the various generators of the Lorentz algebra to be renormalized by a factor ${\cal Z}_N^{(\ep)}$ whilst the derivative term in the gauge field transformation picks up a separate factor ${\cal Z}_N^{(\ep)} {\cal N}_N^{(\ep)}$. The renormalized BRST variations above are easily shown to be nilpotent by repetition of the calculations in Appendix A.

Noting that both the BRST transformations for the antighost and Nakani-shi-Lautrup fields are linear we get the original BRST variations back for the antighost field 
\beq
\de_{\theta^{(\ep)}} \om^*_{\ga\de} = \theta \Delta_{N\, \ga\de}^{(\ep) \om^*} = - \theta\, h_{\ga\de}
\eeq
and for the Nakanishi-Lautrup field
\beq \label{98}
\de_{\theta^{(\ep)}} h_{\ga\de} = \theta \Delta_{N\, \ga\de}^{(\ep) h} = 0.
\eeq

\subsection{Renormalized BRST Invariance forcing $\Ga_N^{(\ep)} [\chi]$ to be of the Form of the original Action $S_{TOT}$ up to possible Field Renormalizations and Coupling Redefinitions}

\paragraph{}
In this subsection we determine the most general form of the renormalized action $\Ga_N^{(\ep)} [\chi]$ invariant A) under all the linear symmetry operations under which the original action is invariant and B) under the renormalized BRST variations $\Delta_N^{(\ep) n} [\chi; x]$ determined in the preceding subsection.

Let us turn to the final step in our renormalization proof: the demonstration that the most general form of the renormalized action invariant under i) all the linear symmetry operations under which the original action is invariant as well as under ii) the renormalized BRST variations $\Delta_N^{(\ep) n} [\chi; x]$ is of the form of our original BRST-invariant action $S_{TOT}$ up to potential field and coupling constant renormalizations.

We start with the $N$-th order contribution to the corrected renormalized action $\Ga_N^{(\ep)} [\chi] \equiv S_R [\chi]
+ \ep\, \Ga_{N,\infty} [\chi, 0]$ which contains the original renormalized action plus the infinite part of the $N$-loop contributions to the quantum effective action. According to the general rules of renormalization theory it must be the integral over local terms in the dynamical fields and their derivatives of dimensionality equal or less than four \cite{wein1}
\beq \Ga_N^{(\ep)} [\chi] = \int {\cal L}_N^{(\ep)} [\chi].
\eeq 
The expression $\Ga_N^{(\ep)} [\chi]$ is invariant under all linearly realized symmetries of the original action as argued above. To identify them we explicitly write down the original action in the Lorentz gauge as given by Eqns.(\ref{43}), (\ref{31}), (\ref{32}), (\ref{8}) and (\ref{45})
\beqq \label{100} & & S_{TOT} = S_M + S_{NEW} = S_M + S_{G} \nonumber \\
& & \quad + \int\! \om^*_{\ga\de}\, \pa^\al \Big( \frac{i}{2} \om^{\eta\ze} {L\rvec}_{\eta\ze}\, \Bagd 
- \, \pa_\al \om^{\ga\de} \\
& &\quad\quad\quad  - \frac{i}{2} B_\al\,^{\eta\ze} {L\rvec}_{\eta\ze}\, \om^{\ga\de}
+ \frac{1}{2}\, C^{\ga\de}\,_{\iota\ka\,\eta\ze}\, B_\al\,^{\iota\ka} \om^{\eta\ze}
\Big) \nonumber \\
& & \quad + \int\! h_{\ga\de}\, \pa^\al B_\al\,^{\ga\de}
+ \frac{\xi}{2} \int\! h_{\ga\de}\, h^{\ga\de}. \nonumber
\eeqq

By inspection $S_{TOT}$ is invariant under all the linearly realized symmetry operations which in particular are A) global Lorentz transformations equalling global gauge transformations parametrized by the constant infinitesimal gauge parameter $\rho$. 

Under the latter the Dirac field living in the spin-$\frac{1}{2}$ representation varies as
\beq \label{101} \de_\rho \psi = \frac{i}{2} \rho^{\ga\de} ({L\rvec}_{\ga\de}\, \psi )
+ \frac{i}{2} \rho^{\ga\de} \Si_{\ga\de}\, \psi,
\eeq
the gauge field living in the vector-cum-adjoint representation varies as
\beqq \!\!\!\!\!\!\!\! \de_\rho \Bagd &=& \frac{i}{2} \rho^{\eta\ze} ({L\rvec}_{\eta\ze}\, \Bagd )
+ \frac{i}{2} \rho^{\eta\ze} \left(\Si^V_{\eta\ze}\right)_\al\,^\be\, B_\be\,^{\ga\de} \\
&+& \frac{1}{2} C^{\ga\de}\,_{\iota\ka\,\eta\ze}\, B_\al\,^{\iota\ka} \rho^{\eta\ze}, \nonumber
\eeqq
the ghost field living in the adjoint representation varies as
\beq \de_\rho \om^{\ga\de}
= \frac{i}{2} \rho^{\eta\ze} ({L\rvec}_{\eta\ze}\, \om^{\ga\de})
+ \frac{1}{2} C^{\ga\de}\,_{\al\be\,\eta\ze}\, \om^{\al\be} \rho^{\eta\ze}, \nonumber
\eeq
the antighost field living in the adjoint representation varies as
\beq \de_\rho \om^*_{\ga\de}
= \frac{i}{2} \rho^{\eta\ze} ({L\rvec}_{\eta\ze}\, \om^*_{\ga\de})
+ \frac{1}{2} C^{\ga\de}\,_{\al\be\,\eta\ze}\, \om^*_{\al\be}\, \rho^{\eta\ze} \nonumber
\eeq
and the Nakanishi-Lautrup field living in the adjoint representation varies as
\beq \label{105} \de_\rho h_{\ga\de}
= \frac{i}{2} \rho^{\eta\ze} ({L\rvec}_{\eta\ze}\, h_{\ga\de})
+ \frac{1}{2} C^{\ga\de}\,_{\al\be\,\eta\ze}\, h_{\al\be}\, \rho^{\eta\ze}.
\eeq

In addition $S_{TOT}$ is invariant under B) the linearly realized antighost translations
\beq \om^*_{\ga\de} \ar \om^*_{\ga\de} + c_{\ga\de}
\eeq 
with $c_{\ga\de}$ an arbitrary constant antisymmetric Lorentz tensor -- which is a particular feature of the Lorentz gauge condition, and is subject to C) ghost number conservation. The invariance under B) is obvious as the $c_{\ga\de}$-term adds nothing but a total divergence.

Next we turn to determine the most general form of the renormalized action invariant under all the linear symmetry operations A) to C) under which the original action is invariant. Recalling the dimensionalities and ghost numbers of the various fields from subsection 6.1 we first note that ghost number conservation requires that $\om$ and $\om^*$ come in pairs, and antighost invariance that $\om^*$ comes together with a derivative $\om^* \pa$ which we can always shuffle to the left of any other expression in the fields. Alltogether any pair of $\om$ and $\om^* \pa$ carries dimension three, so there cannot be more than one such pair in $\Ga_N^{(\ep)} [\chi]$ and such a pair can come with only one more derivative $\pa$ or gauge field $B$. As a result the only remaining possibilities respecting the invariances under A) above are linear combinations of $\int\! \om^*_{\ga\de}\, \pa^\al \left( \om^{\eta\ze} ({L\rvec}_{\eta\ze}\, \Bagd) \right)$,
$\int\! \om^*_{\ga\de}\, \pa^\al \left(\pa_\al \om^{\ga\de} \right)$,
$\int\! \om^*_{\ga\de}\, \pa^\al \left( B_\al\,^{\eta\ze} ({L\rvec}_{\eta\ze}\, \om^{\ga\de}) \right)$,\\ 
$\int\! \om^*_{\ga\de}\, \pa^\al \left( \om^{\eta\ze} \left(\Si^V_{\eta\ze}\right)_\al\,^\be\, B_\be\,^{\ga\de} \right)$
and $\int\! \om^*_{\ga\de}\, \pa^\al \left(C^{\ga\de}\,_{\iota\ka\,\eta\ze}\, B_\al\,^{\iota\ka} \om^{\eta\ze} \right)$.

We turn to terms containing $h$ and other fields but neither $\om$ nor $\om^*$. As $h$ has dimension two the condition A) only allow for linear combinations of $\int\! h_{\ga\de}\, \pa^\al B_\al\,^{\ga\de}$,
$\int\! h^{\ga\de} \left( B_\al\,^{\eta\ze} ({L\rvec}_{\eta\ze}\, B^\al\,_{\ga\de}) \right)$, 
$\int\! h^{\ga\de} \left( B_\al\,^{\eta\ze} \left(\Si^V_{\eta\ze}\right)^\al\,_\be\, B^\be\,_{\ga\de} \right)$,\\
$\int\! h^{\ga\de} \left( C^{\ga\de\,\iota\ka}\,_{\eta\ze}\, B^\al\,_{\iota\ka}\, B^\al\,_{\eta\ze} \right)$
and $\int\! h_{\ga\de}\, h^{\ga\de}$.

Finally $\Ga_N^{(\ep)} [\chi]$ contains terms involving gauge and matter fields only of dimension four or less which we collect in the expression $S_{B,\psi} [\chi]$.

As a result the most general form of the renormalized action invariant under all the linear symmetry operations under which the original action is invariant -- most notably under the global gauge transformations Eqns.(\ref{101}) to (\ref{105}) -- takes the form
\beqq & & \Ga_N^{(\ep)} [\chi] = S_R [\chi]
+ \ep\, \Ga_{N,\infty} [\chi, 0] = S_{B,\psi} [\chi] \nonumber \\
& & \quad +\, Z^{{N (\ep)}}_{\, \om}\! \int\! \om^*_{\ga\de}\, \pa^\al\, \Bigg(
\frac{i}{2} a^{{N (\ep)}}_{\, 1} \om^{\eta\ze} ({L\rvec}_{\eta\ze}\, \Bagd )
+ \, \pa_\al \om^{\ga\de} \nonumber \\
& & \quad\quad +\, \frac{i}{2} a^{{N (\ep)}}_{\, 2} B_\al\,^{\eta\ze} ({L\rvec}_{\eta\ze}\, \om^{\ga\de}) 
+ \frac{i}{2} a^{{N (\ep)}}_{\, 3} \om^{\eta\ze} \left(\Si^V_{\eta\ze}\right)_\al\,^\be\, B_\be\,^{\ga\de}  \nonumber\\
& & \quad\quad +\, \frac{1}{2} a^{{N (\ep)}}_{\, 4} C^{\ga\de}\,_{\iota\ka\,\eta\ze}\, B_\al\,^{\iota\ka} \om^{\eta\ze}
\Bigg) \\
& & \quad +\, b^{{N (\ep)}}_{\, 1}\! \int\! h_{\ga\de}\,  \Bigg(
\pa^\al B_\al\,^{\ga\de}
+ \frac{i}{2} b^{{N (\ep)}}_{\, 2} B_\al\,^{\eta\ze} ({L\rvec}_{\eta\ze}\, B^\al\,_{\ga\de}) \nonumber \\
& & \quad\quad +\, \frac{i}{2} b^{{N (\ep)}}_{\, 3} B_\al\,^{\eta\ze} \left(\Si^V_{\eta\ze}\right)^\al\,_\be\, B^\be\,_{\ga\de}
+ \frac{1}{2} b^{{N (\ep)}}_{\, 4} C^{\ga\de\,\iota\ka}\,_{\eta\ze}\, B^\al\,_{\iota\ka}\, B^\al\,_{\eta\ze}
\Bigg) \nonumber \\
& & \quad +\, \frac{\xi_N^{(\ep)}}{2}\! \int\! h_{\ga\de}\, h^{\ga\de}  \nonumber
\eeqq
with $Z^{{N (\ep)}}_{\, \om}, a^{{N (\ep)}}_{\, 1}, a^{{N (\ep)}}_{\, 2}, a^{{N (\ep)}}_{\, 3}, a^{{N (\ep)}}_{\, 4}, b^{{N (\ep)}}_{\, 1}, b^{{N (\ep)}}_{\, 2}, b^{{N (\ep)}}_{\, 3}, b^{{N (\ep)}}_{\, 4}$ and $\xi_N^{(\ep)}$ unknown constants.

Imposing invariance under the renormalized BRST variations Eqns.(\ref{94}) to (\ref{98})
\beqq & & \de_{\theta^{(\ep)}} \Ga_N^{(\ep)} [\chi] = \de_{\theta^{(\ep)}} S_{B,\psi} [\chi] \nonumber \\
& & \quad -\, Z^{{N (\ep)}}_{\, \om}\, \theta\! \int\! h_{\ga\de}\, \pa^\al\, \Bigg(
\frac{i}{2} a^{{N (\ep)}}_{\, 1} \om^{\eta\ze} ({L\rvec}_{\eta\ze}\, \Bagd )
+ \, \pa_\al \om^{\ga\de} \nonumber \\
& & \quad\quad +\, \frac{i}{2} a^{{N (\ep)}}_{\, 2} B_\al\,^{\eta\ze} ({L\rvec}_{\eta\ze}\, \om^{\ga\de}) 
+ \frac{i}{2} a^{{N (\ep)}}_{\, 3} \om^{\eta\ze} \left(\Si^V_{\eta\ze}\right)_\al\,^\be\, B_\be\,^{\ga\de} \nonumber \\
& & \quad\quad +\, \frac{1}{2} a^{{N (\ep)}}_{\, 4} C^{\ga\de}\,_{\iota\ka\,\eta\ze}\, B_\al\,^{\iota\ka} \om^{\eta\ze}
\Bigg) \nonumber \\
& & \quad +\, Z^{{N (\ep)}}_{\, \om}\! \int\! \om^*_{\ga\de}\, \de_{\theta^{(\ep)}} {\pa\rvec}^\al\, \Bigg(
\frac{i}{2} a^{{N (\ep)}}_{\, 1} \om^{\eta\ze} ({L\rvec}_{\eta\ze}\, \Bagd )
+ \, \pa_\al \om^{\ga\de} \nonumber \\
& & \quad\quad +\, \frac{i}{2} a^{{N (\ep)}}_{\, 2} B_\al\,^{\eta\ze} ({L\rvec}_{\eta\ze}\, \om^{\ga\de}) 
+ \frac{i}{2} a^{{N (\ep)}}_{\, 3} \om^{\eta\ze} \left(\Si^V_{\eta\ze}\right)_\al\,^\be\, B_\be\,^{\ga\de} \nonumber \\
& & \quad\quad +\, \frac{1}{2} a^{{N (\ep)}}_{\, 4} C^{\ga\de}\,_{\iota\ka\,\eta\ze}\, B_\al\,^{\iota\ka} \om^{\eta\ze}
\Bigg) \\
& & \quad +\, b^{{N (\ep)}}_{\, 1}\, \theta\! \int\! h_{\ga\de}\, \pa^\al\, \Bigg(
\frac{i}{2} \om^{\eta\ze} {\cal Z}_N^{(\ep)} ({L\rvec}_{\eta\ze}\, \Bagd )
- \, {\cal Z}_N^{(\ep)} {\cal N}_N^{(\ep)} \pa_\al \om^{\ga\de} \nonumber \\
& & \quad\quad -\, \frac{i}{2} B_\al\,^{\eta\ze} {\cal Z}_N^{(\ep)} ({L\rvec}_{\eta\ze}\, \om^{\ga\de}) 
+ \frac{i}{2} \om^{\eta\ze} {\cal Z}_N^{(\ep)} \left(\Si^V_{\eta\ze}\right)_\al\,^\be\, B_\be\,^{\ga\de} \nonumber \\
& & \quad\quad +\, \frac{1}{2} {\cal Z}_N^{(\ep)} C^{\ga\de}\,_{\iota\ka\,\eta\ze}\, B_\al\,^{\iota\ka} \om^{\eta\ze}
\Bigg) \nonumber \\
& & \quad -\, b^{{N (\ep)}}_{\, 1}\, \theta\! \int\! h_{\ga\de}\, \pa^\be\, 
\Bigg( \frac{i}{2} \om^{\eta\ze} {\cal Z}_N^{(\ep)} \left(\Si^V_{\eta\ze}\right)_\be\,^\al\, B_\al\,^{\ga\de}
\Bigg) \nonumber \\
& & \quad +\, b^{{N (\ep)}}_{\, 1}\! \int\! h_{\ga\de}\, \de_{\theta^{(\ep)}} \Bigg(
\frac{i}{2} b^{{N (\ep)}}_{\, 2} B_\al\,^{\eta\ze} ({L\rvec}_{\eta\ze}\, B^\al\,_{\ga\de}) \nonumber \\
& & \quad\quad +\, \frac{i}{2} b^{{N (\ep)}}_{\, 3} B_\al\,^{\eta\ze} \left(\Si^V_{\eta\ze}\right)^\al\,_\be\, B^\be\,_{\ga\de}
+ \frac{1}{2} b^{{N (\ep)}}_{\, 4} C^{\ga\de\,\iota\ka}\,_{\eta\ze}\, B^\al\,_{\iota\ka}\, B^\al\,_{\eta\ze}
\Bigg) \nonumber \\
& & \quad =^{\!\!\!\! !}\,\, 0 \nonumber
\eeqq
forces the various constants to take the following values
\beqq \label{109} b^{{N (\ep)}}_{\, 1} &=& -\frac{Z^{{N (\ep)}}_{\, \om}}{{\cal Z}_N^{(\ep)} {\cal N}_N^{(\ep)}} \nonumber \\
a^{{N (\ep)}}_{\, 1} &=& -a^{{N (\ep)}}_{\, 2}
= a^{{N (\ep)}}_{\, 4} = -\frac{1}{{\cal N}_N^{(\ep)}} \\
a^{{N (\ep)}}_{\, 3} &=& b^{{N (\ep)}}_{\, 2} = b^{{N (\ep)}}_{\, 3}
= b^{{N (\ep)}}_{\, 4} = 0. \nonumber
\eeqq
Setting the constants to the values as in Eqns.(\ref{109}) the variation of the term $\de_{\theta^{(\ep)}} {\pa\rvec}^\al\, \Big(
\frac{i}{2} a^{{N (\ep)}}_{\, 1} \om^{\eta\ze} ({L\rvec}_{\eta\ze}\, \Bagd ) +\dots \Big) =  - {\cal Z}_N^{(\ep)} {\cal N}_N^{(\ep)}
\de_{\theta^{(\ep)}} \left( s^{(\ep)} \pa^\al \Bagd \right) = 0$ must -- and indeed does -- vanish due to the nilpotence of the renormalized BRST transformation as the expression in brackets is nothing but the renormalized BRST variation $s^{(\ep)}$ of $\pa^\al \Bagd$, and the variation of the term $\de_{\theta^{(\ep)}} \Big(
\frac{i}{2} b^{{N (\ep)}}_{\, 2} B_\al\,^{\eta\ze} ({L\rvec}_{\eta\ze}\, B^\al\,_{\ga\de}) +\dots \Big) = 0$ vanishes as per the values of the constants $b^{{N (\ep)}}_{\, 2} = b^{{N (\ep)}}_{\, 3} = b^{{N (\ep)}}_{\, 4} = 0$.

As a result we are just left with the new constants $Z^{{N (\ep)}}_{\, \om}$ and $\xi_N^{(\ep)}$ and we get
\beqq & & \Ga_N^{(\ep)} [\chi] = S_R [\chi]
+ \ep\, \Ga_{N,\infty} [\chi, 0] = S_{B,\psi} [\chi] \nonumber \\
& & \quad -\, \frac{Z^{{N (\ep)}}_{\, \om}}{{\cal Z}_N^{(\ep)} {\cal N}_N^{(\ep)}}\! \int\! \om^*_{\ga\de}\, \pa^\al\, \Bigg(
\frac{i}{2} \om^{\eta\ze} {\cal Z}_N^{(\ep)} ({L\rvec}_{\eta\ze}\, \Bagd ) \nonumber \\
& & \quad\quad -\, {\cal Z}_N^{(\ep)} {\cal N}_N^{(\ep)} \pa_\al \om^{\ga\de}
-\, \frac{i}{2} B_\al\,^{\eta\ze} {\cal Z}_N^{(\ep)} ({L\rvec}_{\eta\ze}\, \om^{\ga\de}) \\
& & \quad\quad +\, \frac{1}{2} {\cal Z}_N^{(\ep)} C^{\ga\de}\,_{\iota\ka\,\eta\ze}\, B_\al\,^{\iota\ka} \om^{\eta\ze}
\Bigg) \nonumber \\
& & \quad -\, \frac{Z^{{N (\ep)}}_{\, \om}}{{\cal Z}_N^{(\ep)} {\cal N}_N^{(\ep)}}\! \int\! h_{\ga\de}\, \pa^\al B_\al\,^{\ga\de}
+\, \frac{\xi_N^{(\ep)}}{2}\! \int\! h_{\ga\de}\, h^{\ga\de}  \nonumber
\eeqq

Turning to $\de_{\theta^{(\ep)}} S_{B,\psi} [\chi]$ we first note that the renormalized BRST variations for the gauge and matter fields Eqns.(\ref{95}) and (\ref{94}) are nothing but local gauge transformations with gauge parameter
\beq \label{111} \rho^{\ga\de} (x) = {\cal Z}_N^{(\ep)} {\cal N}_N^{(\ep)} \theta\, \om^{\ga\de} (x)
\eeq
and with the generators $J_{\ga\de}$ of the {\bf so(1,3)\/} gauge algebra replaced by the rescaled generators ${\tilde J}_{\ga\de}$  
\beq \label{112} J_{\ga\de} \ar {\tilde J}_{\ga\de} = \frac{1}{{\cal N}_N^{(\ep)}}\, J_{\ga\de}.
\eeq

Because terms from $\de_{\theta^{(\ep)}} S_{B,\psi} [\chi]$ will not contain any $h$ or $\om^*$ they cannot mix with the terms discussed above and we separately have to have
\beq \de_{\theta^{(\ep)}} S_{B,\psi} [\chi] = 0
\eeq
which means that $S_{B,\psi} [\chi]$ must be locally gauge-invariant under the renormalized gauge transformations defined by the BRST variations Eqns.(\ref{94}) to (\ref{98}) for $B$ and $\psi$ with renormalized gauge parameter and gauge algebra generators as in Eqns.(\ref{111}) and (\ref{112}).

As the gauge field action $S_G [B]$ in Eqn.(\ref{2}) is by construction the most general gauge-invariant action of dimension four or less we conclude that the most general $\Ga_N^{(\ep)} [\chi]$ compatible with renormalized BRST invariance is
\beqq \label{114} & & \Ga_N^{(\ep)} [\chi] = {\tilde S}_G [B]+ {\tilde S}_M [B, \psi] \nonumber \\
& & \quad -\, Z^{{N (\ep)}}_{\, \om}\! \int\! \om^*_{\ga\de}\, \pa^\al\, \Bigg(
\frac{i}{2} \om^{\eta\ze} ({{\tilde L}\rrvec}_{\eta\ze}\, \Bagd ) -\, \pa_\al \om^{\ga\de} \\
& & \quad\quad -\, \frac{i}{2} B_\al\,^{\eta\ze} ({{\tilde L}\rrvec}_{\eta\ze}\, \om^{\ga\de})
+\, \frac{1}{2} {\tilde C}^{\ga\de}\,_{\iota\ka\,\eta\ze}\, B_\al\,^{\iota\ka} \om^{\eta\ze}
\Bigg) \nonumber \\
& & \quad -\, \frac{Z^{{N (\ep)}}_{\, \om}}{{\cal Z}_N^{(\ep)} {\cal N}_N^{(\ep)}}\! \int\! h_{\ga\de}\, \pa^\al B_\al\,^{\ga\de}
+\, \frac{\xi_N^{(\ep)}}{2}\! \int\! h_{\ga\de}\, h^{\ga\de},  \nonumber
\eeqq
where ${\tilde S}_{..}$ indicates that all gauge algebra generators $J$ in $S_{..}$ have been replaced by the rescaled generators ${\tilde J}$ as in Eqn.(\ref{112}). Above we have assumed that $S_M$ and as a consequence ${\tilde S}_M$ is the most general renormalizable Dirac matter action coupled to the gauge field.

Inspection of Eqn.(\ref{114}) shows that apart from the appearance of new constants $\Ga_N^{(\ep)} [\chi]$ is functionally the same expression in the dynamical fields as is the action $S_{TOT}$ given by Eqn.(\ref{100}) with which we have started.

By adjusting the $N$-th order terms in the corresponding constants in the original unrenormalized action all the new constants may be absorbed in $S_R$ so that finally
\beq \Ga_N^{(\ep)} [\chi] = S_R [\chi] + \ep\, \Ga_{N,\infty} [\chi, 0] = S_R [\chi].
\eeq
For this particular choice of renormalized constants in $S_R [\chi]$ we then have $\Ga_{N,\infty} [\chi, 0] = 0$. \\

\noindent Q.E.D.

\section{Conclusions}

\paragraph{}
In two preceding papers we have developed a classical theory of gravitation equivalent to GR, yet free from the well-known flaws of GR when it comes to quantization \cite{chw1}, and we have canonically quantized the non-interacting gauge field of that theory and defined the corresponding relativistically-invariant physical Fock space of positive-norm, positive-energy particle states \cite{chw2}.

In this paper we have given full proof of the renormalizability of the quantized theory. In fact we have proven the renormalizability of the perturba-tively defined QEA in essence following the steps usually taken to prove the renormalizability of the QEA of the Standard Model of particle physics. As in that case the ghosts and antighosts appearing as a byproduct of gauge-fixing the path integral expressions for the Green functions of the theory decouple, and the $S$-matrix is as a result unitary -- in our case on the na\"\i ve Fock space of both positive-norm, positive-energy and negative-norm, negative-energy states related to only the gauge field, and on the physical Fock spaces related to possible other physical fields.

The last step to be taken in consistently quantizing gravitation will be the demonstration of the unitarity of the $S$-Matrix on the physical Fock space for the gauge field.

And then more real work starts: what about asymptotic freedom versus the observability of the gravitational interaction -- or the $\be$-function of the theory determining the running of the gauge coupling? What about instantons which definitely exist in the Euclidean version of the theory given that {\bf SO(4)\/} $=$ {\bf SU(2)\/} $\times$ {\bf SU(2)\/}, and anomalies? And what about the interplay of $S^{(2)}_G [B]$ and $S^{(4)}_G [B]$ whereby the former dominates the gravitational interaction at long distances or in the realm of classical physics and the latter at the short distances governing quantum physics? And what about the gravitational quanta implied by the latter already in the non-interacting theory? Could they be at the origin of dark energy -- forming a cosmological radiation background consisting of gravitational quanta in analogy to the CMB -- and helping to resolve the mystery surrounding 70\% of the observed energy content of the Universe? And in that process nicely feeding back as a sort of cosmological constant into $S^{(0)}_G [B]$ in the current Standard Model of Cosmology at a phenomenological level? And $\dots$?

Finally let us take a step back from the more technical aspects and look at the potentially emerging holistic understanding of the four fundamental interactions.

Also in the case of gravitation it seems fruitful to take the historically superbly successful approach to fundamental physics starting with a set of conservations laws based on observations, then evoking Emmy Noether's miraculous theorem linking such conservation laws to global symmetries of local field theories and finally uncovering a related force field and its dynamics by gauging the global symmetry.

So what the observed conservation of the electric, weak and colour charges have done for the formulation of the Standard Model the observed conservation of angular momentum and the observed uniformity of the speed of light across all Lorentz frames of motion might indeed do for the formulation of a consistent quantum theory of gravitation - and its inclusion in the existing Standard Model of particle physics.

In that case both A) working from the conservation of energy-momentum and B) geometrizing gravitation might ultimately prove to have been optical illusions too close to reality to be easily recognized as such, but not close enough to provide the final keys to quantize gravitation.

Hence, there might be no "World Formula" finally, but there might be very well a consistent framing of all physics across the four observed fundamental interactions and it seems that a programme started long ago resulting in the Standard Model of particle physics might eventually come to its ultimate fruition by seamlessly including Gravitation.

\appendix

\section{The {\bf so(1,3)\/} Lorentz Gauge Algebra}

\paragraph{}
In this section we introduce notations and normalizations for the {\bf so(1,3)\/} Lorentz gauge algebra central to this work.

The {\bf so(1,3)\/} Lie or Lorentz gauge algebra is defined by the commutation relations
\beqq \label{116}
\lbr J_{\al\be},J_{\ga\de}\rbr
&=& i\{\eta_{\al\ga}J_{\be\de}-\eta_{\be\ga}J_{\al\de}
+\eta_{\be\de}J_{\al\ga}-\eta_{\al\de}J_{\be\ga}\} \\
&\equiv& i \,C^{\eta\ze}\,_{\al\be\,\ga\de} \,J_{\eta\ze}, \nonumber
\eeqq
where $J_{\al\be}$ denotes a generic set of the six Lie algebra generators and $C_{\al\be\,\ga\de\,\eta\ze}$ its structure constants
\beqq \!\!\!\!\!\!\!\! C_{\al\be\,\ga\de\,\eta\ze}  &=&
\frac{1}{2} \Big\{ \eta_{\ga\eta}\, (\eta_{\de\al}\eta_{\ze\be} - \eta_{\ze\al}\eta_{\de\be})
- \eta_{\de\eta}\, (\eta_{\ga\al}\eta_{\ze\be} - \eta_{\ze\al}\eta_{\ga\be}) \nonumber \\
& &\quad +\, \eta_{\de\ze}\, (\eta_{\ga\al}\eta_{\eta\be} - \eta_{\eta\al}\eta_{\ga\be})
- \eta_{\al\ze}\, (\eta_{\de\al}\eta_{\eta\be} - \eta_{\eta\al}\eta_{\de\be}) \Big\} \\
&=& C_{\eta\ze\,\al\be\,\ga\de}\, =\, C_{\ga\de\,\eta\ze\,\al\be}. \nonumber
\eeqq

The $C_{\al\be\,\ga\de\,\eta\ze}$ are antisymmetric in all the pairs of indices $C_{\be\al\,\ga\de\,\eta\ze} = - C_{\al\be\,\ga\de\,\eta\ze} = \dots$, antisymmetric in exchanging two adjacent pairs of indices $C_{\ga\de\,\al\be\,\eta\ze} = - C_{\al\be\,\ga\de\,\eta\ze} = \dots$ and subject to the Jacobi identity
\beq C^{\ga\de}\,_{\al\be\,\eta\ze}\, C^{\eta\ze}\,_{\rho\si\,\tau\chi}
+ C^{\ga\de}\,_{\rho\si\,\eta\ze}\, C^{\eta\ze}\,_{\tau\chi\,\al\be}
+ C^{\ga\de}\,_{\tau\chi\,\eta\ze}\, C^{\eta\ze}\,_{\al\be\,\rho\si} = 0
\eeq
which follows from
\beqq & & \!\!\!\!\!\!\!\!\!\!\!\!\!\!\!\!\!\!\!\!  \lbr J_{\al\be}, \lbr J_{\rho\si}, J_{\tau\chi} \rbr \rbr + \mbox{cycl. perm.}
= \lbr J_{\al\be}, i\,  C^{\eta\ze}\,_{\rho\si\,\tau\chi} J_{\eta\ze} \rbr + \mbox{cycl. perm.} \\
& & \quad\quad\quad = i^2\, C^{\eta\ze}\,_{\rho\si\,\tau\chi}\, C^{\ga\de}\,_{\al\be\,\eta\ze}\, J_{\ga\de} + \mbox{cycl. perm.} 
= 0. \nonumber
\eeqq

Let us next we display three sets of {\bf so(1,3)\/} generators regularly appearing throughout the paper, namely:

\noindent A) the generators of infinitesimal gauge transformations acting on spacetime coordinates
\beq {L\rvec}_{\eta\ze} = -i ( x_\eta \pa_\ze - x_\ze \pa_\eta ),
\eeq
B) the generators of infinitesimal gauge transformations in the vector representation
\beq \left( \Si^V_{\eta\ze} \right)^\ga\,_\de = -i\, \left(\eta_\eta\,^\ga \eta_\ze\,^\de - \eta_\ze\,^\de  \eta_\eta\,^\ga \right)
\eeq
from which the generators of tensor representations are built and

\noindent C) the generators of infinitesimal gauge transformations in the adjoint representation
\beq \left( \Si^A_{\eta\ze} \right)^{\ga\de}\,_{\iota\ka} = i\, C^{\ga\de}\,_{\eta\ze\,\iota\ka}.
\eeq
It is easy to show that they all obey the commutation relations Eqns.(\ref{116}).

\section{Nilpotence of BRST Transformations}

\paragraph{}
In this section we demonstrate the nilpotence of the BRST transformations introduced in Section 4.

\subsection{Ghosts}

Ghost BRST variation:
\beq s \om^{\ga\de} =  \frac{i}{2}\, \om^{\eta\ze} {L\rvec}_{\eta\ze}\, \om^{\ga\de}
- \frac{1}{4}\, C^{\ga\de}\,_{\al\be\,\eta\ze}\, \om^{\al\be} \om^{\eta\ze}
\eeq

\noindent Nilpotence of ghost BRST variation:
\beqq \de_\theta s \om^{\ga\de} &=& \frac{i}{2}\, \de_\theta \om^{\eta\ze} {L\rvec}_{\eta\ze}\, \om^{\ga\de}
+ \frac{i}{2}\, \om^{\eta\ze} {L\rvec}_{\eta\ze}\, \de_\theta \om^{\ga\de} \\
&-& \frac{1}{4}\, C^{\ga\de}\,_{\al\be\,\eta\ze}\,
(\de_\theta \om^{\al\be} \om^{\eta\ze}  + \om^{\al\be} \de_\theta\om^{\eta\ze}) \nonumber \\
&=& \frac{i}{2}\, \Big\{
\frac{i}{2}\, \theta\, \om^{\rho\si} ({L\rvec}_{\rho\si}\, \om^{\eta\ze}) ({L\rvec}_{\eta\ze}\, \om^{\ga\de}) \nonumber \\
& & -\, \frac{1}{4}\, \theta\, C^{\eta\ze}\,_{\al\be\,\rho\si}\, \om^{\al\be} \om^{\rho\si} ({L\rvec}_{\eta\ze}\, \om^{\ga\de}) 
\Big\} \nonumber \\
&+& \frac{i}{2}\, \om^{\eta\ze} {L\rvec}_{\eta\ze}\, \Big\{ \frac{i}{2}\, \theta\, \om^{\rho\si} ({L\rvec}_{\rho\si}\, \om^{\ga\de})
\nonumber \\
& & -\, \frac{1}{4}\, \theta\, C^{\ga\de}\,_{\al\be\,\rho\si}\, \om^{\al\be} \om^{\rho\si} 
\Big\} \nonumber \\
&-& \frac{1}{2}\, C^{\ga\de}\,_{\al\be\,\eta\ze}\, \om^{\al\be}\, \Big\{
\frac{i}{2} \theta\, \om^{\rho\si} ({L\rvec}_{\rho\si}\, \om^{\eta\ze}) \nonumber \\
& & -\, \frac{1}{4}\, \theta\, C^{\eta\ze}\,_{\rho\si\,\tau\chi}\, \om^{\rho\si} \om^{\tau\chi} 
\Big\} \nonumber \\
&=& \left( \frac{i}{2} \right)^2 \theta\, \om^{\rho\si} 
({L\rvec}_{\rho\si}\, \om^{\eta\ze}) ({L\rvec}_{\eta\ze}\, \om^{\ga\de}) \nonumber \\
&-& \frac{i}{8}\, \theta\, C^{\eta\ze}\,_{\al\be\,\rho\si}\, \om^{\al\be} \om^{\rho\si} ({L\rvec}_{\eta\ze}\, \om^{\ga\de}) 
\nonumber \\
&-& \left(\frac{i}{2}\right)^2 \theta\, \om^{\eta\ze} 
({L\rvec}_{\eta\ze}\, \om^{\rho\si}) ({L\rvec}_{\rho\si}\, \om^{\ga\de}) \nonumber \\
&-& \left(\frac{i}{2}\right)^2 \theta\, \om^{\eta\ze} \om^{\rho\si}
({L\rvec}_{\eta\ze}\, {L\rvec}_{\rho\si}\, \om^{\ga\de}) \nonumber \\
&+& \frac{i}{8}\, \theta\, C^{\ga\de}\,_{\al\be\,\rho\si}\, \om^{\eta\ze} 
\Big\{ ({L\rvec}_{\eta\ze}\, \om^{\al\be}) \om^{\rho\si}
+  \om^{\al\be} ({L\rvec}_{\eta\ze}\, \om^{\rho\si}) \Big\} \nonumber \\
&+& \frac{i}{4}\, \theta\, C^{\ga\de}\,_{\al\be\,\eta\ze}\, \om^{\al\be} \om^{\rho\si} ({L\rvec}_{\rho\si}\, \om^{\eta\ze}) 
\nonumber \\
&-& \frac{1}{8}\, \theta\, C^{\ga\de}\,_{\al\be\,\eta\ze}\, C^{\eta\ze}\,_{\rho\si\,\tau\chi}\, 
\om^{\al\be} \om^{\rho\si} \om^{\tau\chi} \nonumber \\
&=& \left( \frac{i}{2} \right)^2 \theta\, \om^{\rho\si} 
({L\rvec}_{\rho\si}\, \om^{\eta\ze}) ({L\rvec}_{\eta\ze}\, \om^{\ga\de}) \nonumber \\
&-& \left(\frac{i}{2}\right)^2 \theta\, \om^{\eta\ze} 
({L\rvec}_{\eta\ze}\, \om^{\rho\si}) ({L\rvec}_{\rho\si}\, \om^{\ga\de}) \nonumber \\
&-& \frac{i}{8}\, \theta\, C^{\eta\ze}\,_{\al\be\,\rho\si}\, \om^{\al\be} \om^{\rho\si} ({L\rvec}_{\eta\ze}\, \om^{\ga\de}) 
\nonumber \\
&+& \frac{1}{4}\, \theta\, \om^{\al\be} \om^{\rho\si}
\frac{1}{2}\,\lbr {L\rvec}_{\al\be}, {L\rvec}_{\rho\si} \rbr \, \om^{\ga\de} \nonumber \\
&+& \frac{i}{4}\, \theta\, C^{\ga\de}\,_{\al\be\,\eta\ze}\, \om^{\rho\si} \om^{\al\be} ({L\rvec}_{\rho\si}\, \om^{\eta\ze}) 
\nonumber \\
&+& \frac{i}{4}\, \theta\, C^{\ga\de}\,_{\al\be\,\eta\ze}\, \om^{\al\be} \om^{\rho\si} ({L\rvec}_{\rho\si}\, \om^{\eta\ze}) 
\nonumber \\
&-& \frac{1}{24}\, \theta\, \Big\{ C^{\ga\de}\,_{\al\be\,\eta\ze}\, C^{\eta\ze}\,_{\rho\si\,\tau\chi}
+ C^{\ga\de}\,_{\rho\si\,\eta\ze}\, C^{\eta\ze}\,_{\tau\chi\,\al\be} \nonumber \\
& & +\, C^{\ga\de}\,_{\tau\chi\,\eta\ze}\, C^{\eta\ze}\,_{\al\be\,\rho\si} \Big\}
\om^{\al\be} \om^{\rho\si} \om^{\tau\chi} \nonumber \\
&=& 0 \nonumber
\eeqq

\subsection{Matter}

Matter BRST variation:
\beq s \psi = \frac{i}{2}\, \om^{\ga\de} {L\rvec}_{\ga\de}\, \psi
+ \frac{i}{2}\, \om^{\ga\de} \Si_{\ga\de}\, \psi
\eeq

\noindent Nilpotence of matter BRST variation:
\beqq \de_\theta s \psi &=& \frac{i}{2}\, \de_\theta \om^{\ga\de} {L\rvec}_{\ga\de}\, \psi
+ \frac{i}{2}\, \om^{\ga\de} {L\rvec}_{\ga\de}\, \de_\theta \psi \\
&+& \frac{i}{2}\, \de_\theta \om^{\ga\de} \Si_{\ga\de}\, \psi
+ \frac{i}{2}\, \om^{\ga\de} \Si_{\ga\de}\, \de_\theta \psi \nonumber \\
&=& \dots = \left( \frac{i}{2} \right)^2 \theta\, \om^{\eta\ze} 
({L\rvec}_{\eta\ze}\, \om^{\ga\de}) ({L\rvec}_{\ga\de}\, \psi) \nonumber \\
&-& \frac{i}{8}\, \theta\, C^{\ga\de}\,_{\al\be\,\eta\ze}\, \om^{\al\be} \om^{\eta\ze} ({L\rvec}_{\ga\de}\, \psi) 
\nonumber \\
&-& \left(\frac{i}{2}\right)^2 \theta\, \om^{\ga\de} 
({L\rvec}_{\ga\de}\, \om^{\eta\ze}) ({L\rvec}_{\eta\ze}\, \psi) \nonumber \\
&-& \left(\frac{i}{2}\right)^2 \theta\, \om^{\ga\de} \om^{\eta\ze}
({L\rvec}_{\ga\de}\, {L\rvec}_{\eta\ze}\, \psi) \nonumber \\
&-& \left(\frac{i}{2}\right)^2 \theta\, \om^{\ga\de} 
({L\rvec}_{\ga\de}\, \om^{\eta\ze}) (\Si_{\eta\ze}\, \psi) \nonumber \\
&-& \left(\frac{i}{2}\right)^2 \theta\, \om^{\ga\de} \om^{\eta\ze}
({L\rvec}_{\ga\de}\, \Si_{\eta\ze}\, \psi) \nonumber \\
&+& \left(\frac{i}{2}\right)^2 \theta\, \om^{\eta\ze} 
({L\rvec}_{\eta\ze}\, \om^{\ga\de}) (\Si_{\ga\de}\, \psi) \nonumber \\
&-& \frac{i}{8}\, \theta\, C^{\ga\de}\,_{\al\be\,\eta\ze}\, \om^{\al\be} \om^{\eta\ze} (\Si_{\ga\de}\, \psi) 
\nonumber \\
&-& \left(\frac{i}{2}\right)^2 \theta\, \om^{\ga\de} 
(\Si_{\ga\de}\, \om^{\eta\ze}) ({L\rvec}_{\eta\ze}\, \psi) \nonumber \\
&-& \left(\frac{i}{2}\right)^2 \theta\, \om^{\ga\de} \om^{\eta\ze}
(\Si_{\ga\de}\, \Si_{\eta\ze}\, \psi) \nonumber \\
&=& \dots = 0 \nonumber
\eeqq

\subsection{Gauge Fields}

Gauge field BRST variation:
\beqq s B_\mu\,^{\ga\de} &=& \frac{i}{2}\, \om^{\eta\ze} {L\rvec}_{\eta\ze}\, B_\mu\,^{\ga\de}
- \pa_\mu  \om^{\ga\de}
- \frac{i}{2}\, B_\mu\,^{\eta\ze}\, {L\rvec}_{\eta\ze}\, \om^{\ga\de} \\
&+& \frac{i}{2}\, \om^{\eta\ze}\, (\Si_{\eta\ze})_\mu\,^\nu B_\nu\,^{\ga\de}
+ \frac{1}{2}\, C^{\ga\de}\,_{\al\be\,\eta\ze}\, B_\mu\,^{\al\be} \om^{\eta\ze} \nonumber
\eeqq

\noindent Nilpotence of gauge field BRST variation:
\beqq \de_\theta s B_\mu\,^{\ga\de} &=& \frac{i}{2}\, \de_\theta \om^{\eta\ze} {L\rvec}_{\eta\ze}\, B_\mu\,^{\ga\de}
+ \frac{i}{2}\, \om^{\eta\ze} {L\rvec}_{\eta\ze}\, \de_\theta B_\mu\,^{\ga\de} \\
&-& \pa_\mu  \de_\theta \om^{\ga\de}
- \frac{i}{2}\, \de_\theta B_\mu\,^{\eta\ze}\, {L\rvec}_{\eta\ze}\, \om^{\ga\de}
- \frac{i}{2}\, B_\mu\,^{\eta\ze}\, {L\rvec}_{\eta\ze}\, \de_\theta \om^{\ga\de} \nonumber \\
&+& \frac{i}{2}\, \de_\theta \om^{\eta\ze}\, (\Si_{\eta\ze})_\mu\,^\nu B_\nu\,^{\ga\de}
+ \frac{i}{2}\, \om^{\eta\ze}\, (\Si_{\eta\ze})_\mu\,^\nu \de_\theta B_\nu\,^{\ga\de} \nonumber \\
&+& \frac{1}{2}\, C^{\ga\de}\,_{\al\be\,\eta\ze}\, \de_\theta B_\mu\,^{\al\be} \om^{\eta\ze}
+ \frac{1}{2}\, C^{\ga\de}\,_{\al\be\,\eta\ze}\, B_\mu\,^{\al\be} \de_\theta \om^{\eta\ze} \nonumber \\
&=& \dots = \left( \frac{i}{2} \right)^2 \theta\, \om^{\rho\si} 
({L\rvec}_{\rho\si}\, \om^{\eta\ze}) ({L\rvec}_{\eta\ze}\, B_\mu\,^{\ga\de}) \nonumber \\
&-& \frac{i}{8}\, \theta\, C^{\eta\ze}\,_{\rho\si\,\tau\chi}\, \om^{\rho\si} \om^{\tau\chi} 
({L\rvec}_{\eta\ze}\, B_\mu\,^{\ga\de}) \nonumber \\
&-& \left(\frac{i}{2}\right)^2 \theta\, \om^{\eta\ze} 
({L\rvec}_{\eta\ze}\, \om^{\rho\si}) ({L\rvec}_{\rho\si}\, B_\mu\,^{\ga\de}) \nonumber \\
&-& \left(\frac{i}{2}\right)^2 \theta\, \om^{\eta\ze} \om^{\rho\si}
({L\rvec}_{\eta\ze}\, {L\rvec}_{\rho\si}\, B_\mu\,^{\ga\de}) \nonumber \\
&+& \frac{i}{2}\, \theta\, \om^{\eta\ze} ({L\rvec}_{\eta\ze}\, \pa_\mu \om^{\ga\de}) \nonumber \\
&+& \left( \frac{i}{2} \right)^2 \theta\, \om^{\eta\ze} 
({L\rvec}_{\eta\ze}\, B_\mu\,^{\rho\si} ) ({L\rvec}_{\rho\si}\, \om^{\ga\de}) \nonumber \\
&+& \left(\frac{i}{2}\right)^2 \theta\, \om^{\eta\ze} B_\mu\,^{\rho\si}
({L\rvec}_{\eta\ze}\, {L\rvec}_{\rho\si}\, \om^{\ga\de}) \nonumber \\
&-& \left(\frac{i}{2}\right)^2 \theta\, \om^{\eta\ze} 
({L\rvec}_{\eta\ze}\, \om^{\rho\si}) (\Si_{\rho\si})_\mu\,^\nu B_\nu\,^{\ga\de} \nonumber \\
&-& \left(\frac{i}{2}\right)^2 \theta\, \om^{\eta\ze} \om^{\rho\si}
(\Si_{\rho\si})_\mu\,^\nu\, ({L\rvec}_{\eta\ze}\, B_\nu\,^{\ga\de}) \nonumber \\
&-& \frac{i}{4}\, \theta\, C^{\ga\de}\,_{\rho\si\,\tau\chi}\, \om^{\eta\ze} 
({L\rvec}_{\eta\ze}\, B_\mu\,^{\rho\si})\, \om^{\tau\chi} \nonumber \\
&-& \frac{i}{4}\, \theta\, C^{\ga\de}\,_{\rho\si\,\tau\chi}\, \om^{\eta\ze} B_\mu\,^{\rho\si}
({L\rvec}_{\eta\ze}\, \om^{\tau\chi}) \nonumber \\
&-& \frac{i}{2}\, \theta\, \pa_\mu \om^{\eta\ze}\, ({L\rvec}_{\eta\ze}\, \om^{\ga\de}) \nonumber \\
&-& \frac{i}{2}\, \theta\, \om^{\eta\ze}\, \pa_\mu ({L\rvec}_{\eta\ze}\, \om^{\ga\de}) \nonumber \\
&+& \frac{1}{4}\, \theta\, C^{\ga\de}\,_{\al\be\,\eta\ze}\, 
( \pa_\mu \om^{\al\be}\, \om^{\eta\ze} +  \om^{\al\be}\, \pa_\mu \om^{\eta\ze} ) \nonumber \\
&-& \left( \frac{i}{2} \right)^2 \theta\, \om^{\rho\si} 
({L\rvec}_{\rho\si}\, B_\mu\,^{\eta\ze} ) ({L\rvec}_{\eta\ze}\, \om^{\ga\de}) \nonumber \\
&+& \frac{i}{2}\, \theta\, \pa_\mu \om^{\eta\ze}\, ({L\rvec}_{\eta\ze}\, \om^{\ga\de}) \nonumber \\
&+& \left( \frac{i}{2} \right)^2 \theta\, B_\mu\,^{\rho\si} 
({L\rvec}_{\rho\si}\, \om^{\eta\ze} ) ({L\rvec}_{\eta\ze}\, \om^{\ga\de}) \nonumber \\
&-& \left( \frac{i}{2} \right)^2 \theta\, \om^{\rho\si} 
(\Si_{\rho\si})_\mu\,^\nu B_\nu\,^{\eta\ze} ({L\rvec}_{\eta\ze}\, \om^{\ga\de}) \nonumber \\
&-& \frac{i}{4}\, \theta\, C^{\eta\ze}\,_{\rho\si\,\tau\chi}\, B_\mu\,^{\rho\si} \om^{\tau\chi}
({L\rvec}_{\eta\ze}\, \om^{\ga\de}) \nonumber \\
&-& \left( \frac{i}{2} \right)^2 \theta\, B_\mu\,^{\rho\si} 
({L\rvec}_{\rho\si}\, \om^{\eta\ze} ) ({L\rvec}_{\eta\ze}\, \om^{\ga\de}) \nonumber \\
&-& \left( \frac{i}{2} \right)^2 \theta\, B_\mu\,^{\rho\si} \om^{\eta\ze}
({L\rvec}_{\rho\si}\, {L\rvec}_{\eta\ze}\, \om^{\ga\de}) \nonumber \\
&+& \frac{i}{4}\, \theta\, C^{\ga\de}\,_{\al\be\,\eta\ze}\, B_\mu\,^{\rho\si} \om^{\al\be}
({L\rvec}_{\rho\si}\, \om^{\eta\ze}) \nonumber \\
&+& \left( \frac{i}{2} \right)^2 \theta\, \om^{\rho\si} 
 ({L\rvec}_{\rho\si}\, \om^{\eta\ze}) (\Si_{\eta\ze})_\mu\,^\nu B_\nu\,^{\ga\de} \nonumber \\
&-& \frac{i}{8}\, \theta\, C^{\eta\ze}\,_{\rho\si\,\tau\chi}\, \om^{\rho\si} \om^{\tau\chi}
(\Si_{\eta\ze})_\mu\,^\nu B_\nu\,^{\ga\de} \nonumber \\
&-& \left( \frac{i}{2} \right)^2 \theta\, \om^{\rho\si} 
(\Si_{\rho\si})_\mu\,^\nu \om^{\eta\ze} ({L\rvec}_{\eta\ze}\, B_\nu\,^{\ga\de}) \nonumber \\
&+& \frac{i}{2}\, \theta\, \om^{\rho\si}\, (\Si_{\rho\si})_\mu\,^\nu \pa_\nu \om^{\ga\de} \nonumber \\
&+& \left( \frac{i}{2} \right)^2 \theta\, \om^{\rho\si} 
(\Si_{\rho\si})_\mu\,^\nu B_\nu\,^{\eta\ze} ({L\rvec}_{\eta\ze}\, \om^{\ga\de}) \nonumber \\
&-& \left(\frac{i}{2}\right)^2 \theta\, \om^{\rho\si} \om^{\eta\ze}
(\Si_{\rho\si})_\mu\,^\ka\, (\Si_{\eta\ze})_\ka\,^\nu\, B_\nu\,^{\ga\de} \nonumber \\
&-& \frac{i}{4}\, \theta\, \om^{\rho\si} (\Si_{\rho\si})_\mu\,^\nu 
C^{\ga\de}\,_{\al\be\,\eta\ze}\, B_\nu\,^{\al\be}\, \om^{\eta\ze} \nonumber \\
&+& \frac{i}{4}\, \theta\, C^{\ga\de}\,_{\al\be\,\rho\si}\, \om^{\eta\ze} 
({L\rvec}_{\eta\ze}\, B_\mu\,^{\al\be})\, \om^{\rho\si} \nonumber \\
&-& \frac{1}{2}\, \theta\, C^{\ga\de}\,_{\al\be\,\rho\si}\, \pa_\mu \om^{\al\be}\, \om^{\rho\si} \nonumber \\
&-& \frac{i}{4}\, \theta\, C^{\ga\de}\,_{\al\be\,\rho\si}\, B_\mu\,^{\eta\ze}
({L\rvec}_{\eta\ze}\, \om^{\al\be}) \om^{\rho\si} \nonumber \\
&+& \frac{i}{4}\, \theta\, C^{\ga\de}\,_{\al\be\,\rho\si}\, \om^{\eta\ze} \om^{\rho\si}
(\Si_{\eta\ze})_\mu\,^\nu B_\nu\,^{\al\be} \nonumber \\
&+& \frac{1}{4}\, \theta\, C^{\ga\de}\,_{\al\be\,\rho\si}\, C^{\al\be}\,_{\tau\chi\,\eta\ze}\,
B_\mu\,^{\tau\chi}\, \om^{\eta\ze} \om^{\rho\si} \nonumber \\
&+& \frac{i}{4}\, \theta\, C^{\ga\de}\,_{\al\be\,\eta\ze}\, B_\mu\,^{\al\be} \om^{\rho\si}
({L\rvec}_{\rho\si}\, \om^{\eta\ze}) \nonumber \\
&-& \frac{1}{8}\, \theta\, C^{\ga\de}\,_{\al\be\,\eta\ze}\, C^{\eta\ze}\,_{\rho\si\,\tau\chi}\,
B_\mu\,^{\al\be}\, \om^{\rho\si} \om^{\tau\chi} \nonumber \\
&=& \dots = \frac{i}{2}\, \theta\, \om^{\eta\ze}\, ({L\rvec}_{\eta\ze}\, \pa_\mu \om^{\ga\de}) \nonumber \\
&-& \frac{i}{2}\, \theta\, \om^{\eta\ze}\, ({L\rvec}_{\eta\ze}\, \pa_\mu \om^{\ga\de}) \nonumber \\
&-& \frac{i}{2}\, \theta\, \om^{\eta\ze}\, (\pa_\mu {L\rvec}_{\eta\ze})\, \om^{\ga\de} \nonumber \\
&+& \frac{i}{2}\, \theta\, \om^{\rho\si} (\Si_{\rho\si})_\mu\,^\nu \pa_\nu \om^{\ga\de} \nonumber \\
&-& \frac{1}{8}\, \theta\, \Big\{ C^{\ga\de}\,_{\al\be\,\eta\ze}\, C^{\eta\ze}\,_{\rho\si\,\tau\chi}
+ C^{\ga\de}\,_{\tau\chi\,\eta\ze}\, C^{\eta\ze}\,_{\al\be\,\rho\si} \nonumber \\
& & +\, C^{\ga\de}\,_{\rho\si\,\eta\ze}\, C^{\eta\ze}\,_{\tau\chi\,\al\be}  \Big\}
B_\mu\,^{\al\be} \om^{\rho\si} \om^{\tau\chi} \nonumber \\
&=& 0 \nonumber
\eeqq
using
\beqq & & \frac{i}{2}\, \theta\, \om^{\eta\ze}\, (\pa_\mu {L\rvec}_{\eta\ze})\, \om^{\ga\de}
+ \frac{i}{2}\, \theta\, \om^{\rho\si} (\Si_{\rho\si})_\mu\,^\nu \pa_\nu \om^{\ga\de} \\
& &\quad\quad =\, \theta\, \om^\eta\,_\mu\, \pa_\eta \om^{\ga\de}
+ \theta\, \om_\mu\,^\nu\, \pa_\nu \om^{\ga\de} = 0 \nonumber
\eeqq

\subsection{Antighosts}

Antighost BRST variation:
\beq s \om^*_{\ga\de} =  -h_{\ga\de}
\eeq

\noindent Nilpotence of antighost BRST variation:
\beq \de_\theta s \om^*_{\ga\de} =  -\de_\theta h_{\ga\de} =  0
\eeq

\subsection{Nakanishi-Lautrup Fields}

Nakanishi-Lautrup fields BRST variation:
\beq s h_{\ga\de} =  0
\eeq

\noindent Nilpotence of Nakanishi-Lautrup fields BRST variation:
\beq \de_\theta s h_{\ga\de} =  0
\eeq

\section{Berezinian Determinant}

\paragraph{}
In this section we demonstrate the triviality of the Berezinian determinant introduced in Section 5.

\subsection{Ghosts}

Jacobian matrix for ghosts:
\beqq & &  \!\!\!\!\!\!\! \frac{\de \om'^{\ga\de} (x)}{\de \om^{\iota\ka} (y)} =
\Big( \eta^\ga\,_{[ \iota,} \eta^\de\,_{\ka ]} 
+ \frac{i}{2}\, \theta\, \om^{\eta\ze} (x) {L\rvec}_{\eta\ze}\, \eta^\ga\,_{[ \iota,} \eta^\de\,_{\ka ]} \\
& & -\, \frac{1}{4}\, \theta\, C^{\ga\de}\,_{\al\be\,\eta\ze}\,
\left( \om^{\eta\ze} (x)\, \eta^\al\,_{[ \iota,} \eta^\be\,_{\ka ]}
- \om^{\al\be} (x)\, \eta^\eta\,_{[ \iota,} \eta^\ze\,_{\ka ]} \right)\! \Big)\,
\de(x-y) \nonumber
\eeqq
Above the square brackets with comma $\eta^\eta\,_{[ \iota,} \eta^\ze\,_{\ka ]} \equiv 
\eta^\eta\,_\iota \eta^\ze\,_\ka - \eta^\eta\,_\ka \eta^\ze\,_\iota$ indicate antisymmetrization in the indices concerned

\subsection{Matter}

Jacobian matrix for matter:
\beqq & &  \!\!\!\!\!\!\! \frac{\de \psi' (x)}{\de \psi (y)} = 
\Big( 1 + \frac{i}{2}\, \theta\, \om^{\eta\ze} (x) {L\rvec}_{\eta\ze}\, \\
& & +\, \frac{i}{2}\, \theta\, \om^{\eta\ze} (x) \Si_{\eta\ze} \Big)\,
\de(x-y) \nonumber
\eeqq

\subsection{Gauge Fields}

Jacobian matrix for gauge fields:
\beqq & &  \!\!\!\!\!\!\!  \frac{\de B'_\mu\,^{\ga\de} (x)}{\de B_\rho\,^{\iota\ka} (y)} =
\Big( \eta_\mu\,^\rho\, \eta^\ga\,_{[ \iota,} \eta^\de\,_{\ka ]} 
+ \frac{i}{2}\, \theta\, \om^{\eta\ze} (x) {L\rvec}_{\eta\ze}\,
\eta_\mu\,^\rho\, \eta^\ga\,_{[ \iota,} \eta^\de\,_{\ka ]} \nonumber \\
& & -\, \frac{i}{2}\, \theta\, \left({L\rvec}_{\eta\ze}\, \om^{\ga\de} (x) \right) 
\eta_\mu\,^\rho\, \eta^\ga\,_{[ \iota,} \eta^\de\,_{\ka ]} \\
& & +\, \frac{i}{2}\, \theta\, \om^{\eta\ze} (x) \left( \Si_{\eta\ze}\right)_\mu\,^\nu\,
\eta_\nu\,^\rho\,\eta^\ga\,_{[ \iota,} \eta^\de\,_{\ka ]}  \nonumber \\
& & +\, \frac{1}{2}\, \theta\, C^{\ga\de}\,_{\al\be\,\eta\ze}\,
\om^{\eta\ze} (x)\, \eta_\mu\,^\rho\, \eta^\al\,_{[ \iota,} \eta^\be\,_{\ka ]} \Big)\,
\de(x-y) \nonumber
\eeqq

\subsection{Antighosts}

Jacobian matrix for antighosts:
\beq \frac{\de {\om^*}'_{\ga\de} (x)}{\de \om^*_{\iota\ka} (y)} = 
\eta_\ga\,^{[ \iota,} \eta_\de\,^{\ka ]}\, \de(x-y)
\eeq

\subsection{Nakanishi-Lautrup Fields}

Jacobian matrix for Nakanishi-Lautrup fields:
\beq \frac{\de h'_{\ga\de} (x)}{\de h_{\iota\ka} (y)} = 
\eta_\ga\,^{[ \iota,} \eta_\de\,^{\ka ]}\, \de(x-y)
\eeq

\subsection{Berezinian Determinant}

Berezinian determinant: 
\beq {\cal J} = \Det \left( \frac{\de {\chi^n}' }{\de \chi^m } \right) = 1 + \Tr \log \left( \frac{\de {\chi^n}' }{\de \chi^m } \right)
\eeq
Note that this is an exact expression as all higher terms on the r.h.s. vanish due to the antisymmetric nature of $\theta$ to which all the non-trivial contributions to the Jacobian matrices above are proportional \\

\noindent Trace of the sum of logarithms of the Jacobians:
\beqq \Tr \log \left( \frac{\de {\chi^n}' }{\de \chi^m } \right) &=&
-\, \theta \Tr_\psi \left( \frac{i}{2}\, \theta\, \om^{\eta\ze} {L\rvec}_{\eta\ze}
+ \frac{i}{2}\, \theta\, \om^{\eta\ze} \Si_{\eta\ze} \right)  \nonumber\\
& & +\,\, \theta \Tr_B (\dots) - \theta \Tr_\om (\dots) \\
&=& \theta\, \Tr\! \left( \frac{i}{2}\,  \om^{\eta\ze} {L\rvec}_{\eta\ze} \right)
 (- \dim_\psi + \dim_B - \dim_\om) \nonumber
\eeqq

\noindent Functional trace of the infinitesimal algebra parameter $\om^{\eta\ze} (x) {L\rvec}_{\eta\ze}$:
\beqq & & \!\!\!\!\!\!\!\!\!\!\!\!\!\!\!\!  \Tr\! \left( \frac{i}{2}\,  \om^{\eta\ze} (x) {L\rvec}_{\eta\ze}\, \de(x-y) \right)
= - \intx \inty\, \om^{\eta\ze} (x)\, x_\ze \pa^x_\eta \de(x-y) \nonumber \\
& & \quad\quad\quad\quad =\, \intx \inty\, \de(x-y)\, \pa^x_\eta \left(\om^{\eta\ze} (x)\, x_\ze \pa^x_\eta \right) \\
& & \quad\quad\quad\quad =\, \intx\, \pa^x_\eta \left(\om^{\eta\ze} (x)\, x_\ze \right) = 0 \nonumber
\eeqq

\noindent Berezinian determinant: 
\beq {\cal J} = 1
\eeq

\end{document}


\maketitle

\noindent {\sl Le permier (precepte) estoit de ne receuoir iamais aucune chose pour vraye que ie ne la connusse euidemment estre telle}\footnote{Ren\'e Descartes, Discours de la M\'ethode, Deuxi\`eme Partie, 1637, Imprimerie Ian Maire (Leyde)}
\vspace{0.5cm}

\begin{abstract}
The non-interacting field belonging to a new {\bf SO(1,3)\/} gauge field theory equivalent to General Relativity is canonically quantized in the Lorentz gauge and the physical Fock space ${\cal F}^{\Phys}$ for free gauge particles is constructed. To assure both relativistic covariance and positivity of the norm and energy expectation value for physical states restrictions needed in the construction of the physical Fock space are put consequentially on state vectors, and not on the algebra of creation and annihilation operators as usual -- alltogether providing the second step in consistently quantizing gravitation\\
\end{abstract}

\clearpage

\section{Introduction}

\paragraph{}
Gravity has defied so far all attempts at consistent quantization. In fact, Einstein's  theory of General Relativity (GR) and its generalizations turn out to be either not renormalizable or do not respect unitarity at the quantum level.

GR itself is not renormalizable in essence due to the fundamental field $g$, the metric, or equivalently $e$, the Vierbein, being of mass-dimension zero. Simple power-counting allows to demonstrate that the loop expansion of the quantum effective action contains divergent contributions of ever higher mass-dimension -- destroying renormalizability \cite{buc}.

All attempts to deal with this fundamental difficulty have failed essentially because all generalizations of GR analyzed in the literature offering a seeming way out still contain $g$, or equivalently $e$, as at least one dynamical field due to the local translation or coordinate invariance to which these theories are subject to. As a result in all cases one ends up with either a non-renormalizable or a non-unitary theory.

In \cite{chw1} we have started a programme aiming to overcome those difficulties step-by-step.

Our first step as been to take a point of departure different from the literature at the classical level already: a theory for classical gravity which is not equal, but only equivalent to GR, and which in return allows for renormalizable actions for the fundamental field. In fact in \cite{chw1} we have developed a new classical gauge field theory of the Lorentz group {\bf SO(1,3)\/} which a) contains as the only dynamical field the dimension-one Lorentz gauge field in terms of which all else can be expressed and which allows for actions renormalizable by power-counting, and b) is equivalent to GR in a limiting case. Whilst a renormalizability proof seems within reach for this theory \cite{chw2}, another difficulty has to be resolved first, which arises from the non-compactness of {\bf SO(1,3)\/}: establishing that the canonical quantization of the gauge fields allows for the definition of positive-norm, positive-energy states and a corresponding relativistically-invariant physical Fock space for these fields - and that negative-norm states completely decouple.

The second step of our programme and the goal of this paper is exactly to provide that resolution. We do this by strictly adhering to our guiding principles of keeping to relativistic and {\bf SO(1,3)\/} gauge covariance, to treating the {\bf SO(1,3)\/} gauge symmetry strictly as an inner symmetry, and to renormalizability which were at the origin of developing the {\bf SO(1,3)\/} gauge field theory in the first place \cite{chw1}. In addition in this paper we intertwine relativistic covariance with positivity of the norm and energy expectation value for physical states, and consequentially put restrictions needed in the search of a physical Fock space on state vectors, and not on the algebra of creation and annihilation operators as usual.

The next step of our programme will have to be the renormalizability proof of the full quantum theory \cite{chw2} and the final one to establish the unitarity of the $S$-matrix on the physical Fock space constructed in this paper.

To establish that canonical quantization of the free gauge fields allows for the definition of positive-norm, positive-energy states and a corresponding relativistically invariant physical Fock space we have organized this paper as follows. In section two we revisit the fundamentals of the new Lorentz gauge field theory of gravitation which is renormalizable by power-counting and equivalent to General Relativity. In section three we derive the Lagrangian of the free Lorentz gauge field theory in the Lorentz gauge which serves as starting point for the covariant canonical quantization of the non-interacting theory in terms of the gauge fields and their conjugate canonical momenta in section four. In section five we recast the theory in terms of the creation and annihilation operators corresponding to the gauge field operators. In sections six and seven we establish the one-particle subspace of states  ${\cal F}_1^{\mbox{\footnotesize LG}}$ fulfilling the Lorentz gauge condition and the one-particle subspace of states ${\cal F}_1^{\mbox{\footnotesize P}}$ orthogonal to their dual state with positive semidefinite norm respectively which in combination allow for the definition of the one-particle subspace of states ${\cal F}_1^{\mbox{\footnotesize LG $\!\&\!$ P}}$ with positive semidefinite norm and energy expectation value in section eight. The final three sections are devoted to establish the physical one-particle space of states ${\cal F}_1^{\Phys}$ with positive norm and energy expectation value, to prove the covariance of the quantization approach on ${\cal F}_1^{\Phys}$ and to define the physical n-particle space of states ${\cal F}_n^{\Phys}$ and the physical Fock space ${\cal F}^{\Phys}$ for free gauge particles respectively.

\section{Lorentz gauge field theory of gravitation equi- valent to General Relativity}

\paragraph{}
In this section we revisit local Lorentz invariance and the corresponding classical gauge field theory equivalent to General Relativity as developed in \cite{chw1} to prepare for the derivation and quantization of the free Lorentz quantum gauge field theory.

Let us start with some notations around the Lorentz gauge group {\bf SO(1,3)\/} and algebra {\bf so(1,3)\/} at the core of the theory. We will mainly work with infinitesimal group elements ${\bf 1} + \Theta_\om (x)$, where
\beq \Theta_\om (x) = \frac{i}{2}\om^{\ga\de} (x) (L_{\ga\de} + \Si_{\ga\de}) \in {\bf so(1,3)\/}
\eeq
is a local element of the gauge algebra {\bf so(1,3)\/}, $x$ a point in Minkowski spacetime {\bf M\/}$^{4}$ $\equiv$ ({\bf R\/}$^{4}$,$\eta$) given in Cartesian coordinates and $\eta=\mbox{diag}(-1,1,1,1)$ is the flat spacetime metric with which indices are raised and lowered throughout this paper. Indices $\al,\be,\ga,...$ denote quantities defined on {\bf M\/}$^{4}$ which transform covariantly.

Above $\om^{\ga\de} = - \om^{\de\ga}$ are the six infinitesimal gauge parameters parametrizing the Lorentz algebra. Treating Lorentz transformations as "inner" transformations \cite{chw1} the spacetime-related algebra generators acting on field space are given by
\beq L_{\ga\de} = -L_{\de\ga} = -i(x_{\ga}\pa_{\de}-x_{\de}\pa_{\ga})
\eeq
whilst $\Si_{\ga\de}$ denote the generators of an arbitrary finite-dimensional spin representation of the Lorentz algebra.

Generators of any representation obey the commutation relations
\beq \lbr J_{\al\be},J_{\ga\de}\rbr
= i\{\eta_{\al\ga}J_{\be\de}-\eta_{\be\ga}J_{\al\de}
+\eta_{\be\de}J_{\al\ga}-\eta_{\al\de}J_{\be\ga}\}
\eeq
which in effect defines the Lorentz algebra {\bf so(1,3)\/}.

The action of a local infinitesimal group element ${\bf 1} + \Theta_\om (x)$ on a field $\va(x)$ living in an arbitrary finite-dimensional spin representation of the Lorentz group is given by
\beqq x^\al \lar x'^\al&=&x^\al \\
\va(x)\lar \va'(x)&=&\va(x) + \de_\om \va(x), \nonumber
\eeqq
where
\beqq & & \de_\om \va (x) = (\Theta_\om (x) \va)(x) \\
&=& -\om^{\ga\de}(x)  x_\de \pa_\ga \va (x) 
+\frac{i}{2}\om^{\ga\de}(x)  \Si_{\ga\de} \va (x). \nonumber
\eeqq

In order to define locally gauge-covariant expressions we need to introduce a covariant derivative $\nabla_\al(x)$ as usual which obeys
\beq (\nabla_\al(x) \va)' = \nabla'_\al(x) \va'
\eeq
and a related gauge field $B_\al(x)$
\beqq \nabla^B_\al &=& \pa_\al + B_\al(x) \nonumber \\
B_\al &=& \frac{i}{2}\Bagd(x) (L_{\ga\de} + \Si_{\ga\de}) \\
&=& -\Bagd(x) x_\de \pa_\ga
+\frac{i}{2}\Bagd(x) \Si_{\ga\de} \nonumber
\eeqq
living in the Lorentz algebra {\bf so(1,3)\/}. Above primed quantities refer to gauge-transformed quantities. For further details we refer to \cite{chw1}.

To deal with the ever more involved algebraic expressions in the further development of the theory we rewrite the covariant derivative $\nabla_\al(x)$
\beqq \nabla^B_\al &=& \pa_\al - \Bagd x_\de \pa_\ga
+\frac{i}{2}\Bagd \Si_{\ga\de} \nonumber \\
&=& \left( \eta_\al\,^\ga - \Bagd x_\de \right) \pa_\ga
+\frac{i}{2}\Bagd \Si_{\ga\de} \\
&=& d^B_\al + {\bar B}_\al, \nonumber
\eeqq
and introduce the expression
\beq e_\al\,^\th[B] \equiv \eta_\al\,^\th - B_\al\,^{\th\ze}x_\ze
\eeq
resembling a Vierbein which, however, is a functional of the dynamical gauge field $\Bagd(x)$ in our theory, and the short-hand notations
\beq d^B_\al \equiv e_\al\,^\th[B]\, \pa_\th,
\quad {\bar B}_\al \equiv \frac{i}{2}\Bagd \Si_{\ga\de}.
\eeq

We have elaborated in depth why $e_\al\,^\th[B]$ not being a fundamental field is so crucial for the further development of the theory which turns out to be both equivalent to General Relativity and renormalizable  \cite{chw1}.

As usual we next define the field strength $G$ operator acting on field space
\beq G_{\al\be} [B]\equiv [ \nabla^B_\al,\nabla^B_\be]
\eeq
and express it in terms of the gauge field $B$
\beqq G_{\al\be} [B]&=& [d^B_\al ,d^B_\be ] + d^B_\al {\bar B}_\be - d^B_\be {\bar B}_\al \\
&+& [{\bar B}_\al , {\bar B}_\be] + ( B_{\al\be}\,^\eta - B_{\be\al}\,^\eta ) \nabla^B_\eta. \nonumber
\eeqq

To calculate
\beq [d^B_\al ,d^B_\be ] = \left( e_\al\,^\ze [B]\,\pa_\ze e_\be\,^\eta [B]-
e_\be\,^\ze [B]\,\pa_\ze e_\al\,^\eta [B] \right)\pa_\eta
\eeq
we assume that $e_\al\,^\ze [B]$ is non-singular, i.e. $\det e[B]\neq 0$. Hence there is an inverse $e^\ga\,_\eta [B]$ with
$e^\ga\,_\eta [B]\, e_\ga\,^\ze [B]=\de_\eta\,^\ze$, and we can write
\beq [d^B_\al ,d^B_\be ] = \Habg [B]\, d^B_\ga
\eeq
with
\beq \Habg [B] \equiv e^\ga\,_\eta [B] \left( e_\al\,^\ze [B]\,\pa_\ze e_\be\,^\eta [B]-
e_\be\,^\ze [B]\,\pa_\ze e_\al\,^\eta [B] \right).
\eeq

As a result we can express
\beqq G_{\al\be} [B] &=& (\Habg [B]+ B_{\al\be}\,^\ga 
- B_{\be\al}\,^\ga ) \nabla^B_\ga  \nonumber \\
&+& d^B_\al {\bar B}_\be - d^B_\be {\bar B}_\al + [{\bar B}_\al ,{\bar B}_\be] - \Habg [B]\, {\bar B}_\ga \\
&=& - \Tabg [B]\, \nabla^B_\ga + \Rab [B]  \nonumber 
\eeqq
in terms of the covariant quantities
\beq \Tabg [B] \equiv -(B_{\al\be}\,^\ga - B_{\be\al}\,^\ga) - \Habg [B]
\eeq
and
\beqq \Rab [B] &\equiv& \frac{i}{2} \Rabgd [B] \, \Si_{\ga\de} \nonumber \\
\Rabgd [B] &=& d^B_\al  B_\be\,^{\ga\de} - d^B_\be B_\al\,^{\ga\de}
+\, B_\al\,^{\ga\eta}\, B_{\be\eta}\,^\de \\ 
&-& B_\be\,^{\ga\eta}\, B_{\al\eta}\,^\de 
- H_{\al\be}\,^\eta [B]\, B_\eta\,^{\ga\de}. \nonumber
\eeqq
The geometrical significance of all: the gauge field $B$, field strength $G$ and $T$ as well as $R$ has been further illuminated in terms of a Banach fibre bundle structure with trivial base manifold {\bf M\/}$^{4}$ and infinite-dimensional fibres for the various fields in \cite{chw1}.

There we also have given the transformation behaviour of the various quantities under local Lorentz gauge transformations.

Next we write down the most general gauge-invariant action for the gauge field $B$ which is renormalizable by power-counting. It contains dimension-zero, -two and -four contributions $S^{(0)}_G [B]$, $S^{(2)}_G [B]$ and $S^{(4)}_G [B]$ respectively which we cite from \cite{chw1} without further details here starting with
\beq S^{(0)}_G [B] = \La\, \intf \dem [B], \eeq
where $ \La $ is a constant of dimension $[\La] = 4$

The most general dimension-two contribution reads
\beqq S^{(2)}_G [B]
&=& \frac{1}{\ka}\, \intf \dem [B]
\Big\{\al_1\, R_{\al\be}\,^{\al\be} [B] \nonumber \\
&+& \al_2\, T_{\al\be\ga} [B]\, T^{\al\be\ga} [B]
+ \al_3\, T_{\al\be\ga} [B]\, T^{\ga\be\al} [B] \\
&+&  \al_4\, T_\al \,^{\ga\al} [B]\, T_{\be\ga} \,^\be [B]
+  \al_5\, \nabla^B_\al T_\be \,^{\al\be} [B] \Big\}. \nonumber
\eeqq
$\frac{1}{\ka} = \frac{1}{16\pi \it\Ga}$ has mass-dimension $[ \frac{1}{\ka} ] = 2$ with ${\it\Ga} $ denoting the Newtonian gravitational constant. The $ \al_i $ above are constants of dimension $[\al_i] = 0$.

Finally, the most general dimension-four contribution reads
\beqq S^{(4)}_G [B] &=& \intf \dem [B]
\Big\{ \be_1\, R_{\al\be}\,^{\ga\de} [B]\, R^{\al\be}\,_{\ga\de} [B] \nonumber \\
&+&  \be_2\, R_{\al\ga}\,^{\al\de} [B]\, R^{\be\ga}\,_{\be\de} [B] 
+ \be_3\, R_{\al\be}\,^{\al\be} [B] \,R_{\ga\de}\,^{\ga\de} [B] \nonumber \\
&+& \be_4\, \nabla_B^\ga \nabla^B_\de R_{\al\ga}\,^{\al\de} [B]
+ \be_5\, \nabla_B^\ga \nabla^B_\ga R_{\al\be}\,^{\al\be} [B] \nonumber \\
&+& \dots \nonumber \\
&+& \ga_1\, \nabla^B_\ga T_{\al\be\de} [B]\,
\nabla_B^\ga T^{\al\be\de} [B]
+ \ga_2\, \nabla^B_\ga T_{\al\be\de} [B]\, 
\nabla_B^\ga T^{\de\be\al} [B] \\
&+& \dots \nonumber \\
&+& \ga_j\, T^4-\mbox{terms} \nonumber \\
&+& \dots \nonumber \\
&+& \de_k\, R\,T^2-\mbox{terms},\, R\,\nabla^B\, T-\mbox{terms} \nonumber \\
&+& \dots \Big\} \nonumber 
\eeqq 
with $ \be_i $, $ \ga_j $, $ \de_k $ constants of dimension $[\be_i] = [\ga_j] = [\de_k] = 0$.

By construction
\beq \label{22} S_G [B] = S^{(0)}_G [B] + S^{(2)}_G [B] + S^{(4)}_G [B]
\eeq
is the most general action of dimension $\leq 4$ in the gauge fields $\Bagd$ and their first and second derivatives $\pa_\be \Bagd, \pa_\eta \pa_\be \Bagd$ which is locally Lorentz invariant and -- having consistent field quantization in mind -- renormalizable by power-counting. The actual proof of renormalizability requires the much more involved demonstration that counterterms needed to absorb infinite contributions to the perturbative expansion of the effective action of the full quantum theory are again of the form Eqn.(\ref{22}) with possibly renormalized constants.

We finally note that for the choice
\beq \al_1 = 1,\quad \al_2 = -\frac{1}{4},\quad \al_3 = -\frac{1}{2},\quad
\al_4 = -1,\quad \al_5 = 2
\eeq
$S^{(0)}_G [B] + S^{(2)}_G [B]$ coupled to scalar matter is equivalent to General Relativity with a cosmological constant term \cite{chw1}.

\section{Non-interacting Lagrangian for the {\bf SO(1,3)\/} gauge field theory in the Lorentz gauge}

\paragraph{}
In this section, starting with the most general gauge-invariant Lagrangian $S_G [B]$ renormalizable by power-counting, we introduce a coupling constant $g$ and derive the free action $S^{\!\!\!{^{^o}}}_G [B]$ quadratic in the gauge field $B$ serving as the starting point for the free gauge field quantization in the sections to follow.

Scaling the gauge field $B$ as well as $H$, $T$ and $R$ by a dimension-less constant $g$ as
\beqq \Bagd &\ar& g\, B_\al\,^{\ga\de} \nonumber \\
H_{\al\be}\,^\ga [B] &\ar& g\, H_{\al\be}\,^\ga [g B] \\
T_{\al\be}\,^\ga [B] &\ar& g\, T_{\al\be}\,^\ga [g B] \nonumber \\
R_{\al\be}\,^{\ga\de} [B] &\ar& g\, R_{\al\be}\,^{\ga\de} [g B], \nonumber
\eeqq
and the action $S_G [B]$ as
\beq S_G [B] \ar S_G [B, g] = S^{(0)}_G [B, g] + \frac{1}{g} S^{(2)}_G [B, g] 
+ \frac{1}{g^2} S^{(4)}_G [B, g]
\eeq
we obtain
\beqq \label{26} S_G [B, g] &=& \La\, \intf \dem [g B] \nonumber \\
&+& \frac{1}{\ka}\, \intf \dem [g B]
\Big\{\al_1\, R_{\al\be}\,^{\al\be} [g B]  \nonumber \\
&+& g\, \al_2\, T_{\al\be\ga} [g B]\, T^{\al\be\ga} [g B]
+ g\, \al_3\, T_{\al\be\ga} [g B]\, T^{\ga\be\al} [g B]  \nonumber \\
&+&  g\, \al_4\, T_\al \,^{\ga\al} [g B]\, T_{\be\ga} \,^\be [g B]
+  \al_5\, \nabla^{g B}_\al T_\be \,^{\al\be} [g B] \Big\} \nonumber \\
&+& \intf \dem [g B]
\Big\{ \be_1\, R_{\al\be}\,^{\ga\de} [g B]\, R^{\al\be}\,_{\ga\de} [g B] \nonumber \\
&+&  \be_2\, R_{\al\ga}\,^{\al\de} [g B]\, R^{\be\ga}\,_{\be\de} [g B] 
+ \be_3\, R_{\al\be}\,^{\al\be} [g B] \,R_{\ga\de}\,^{\ga\de} [g B] \nonumber \\
&+& \frac{1}{g} \be_4\, \nabla_{g B}^\ga \nabla^{g B}_\de R_{\al\ga}\,^{\al\de} [g B]
+ \frac{1}{g} \be_5\, \nabla_{g B}^\ga \nabla^{g B}_\ga R_{\al\be}\,^{\al\be} [g B] \nonumber \\
&+& \dots \\
&+& \ga_1\, \nabla^{g B}_\ga T_{\al\be\de} [g B]\,
\nabla_{g B}^\ga T^{\al\be\de} [g B] \nonumber \\
&+& \ga_2\, \nabla^{g B}_\ga T_{\al\be\de} [g B]\, 
\nabla_{g B}^\ga T^{\de\be\al} [g B] \nonumber \\
&+& \dots \nonumber \\
&+& \ga_j\, g^2\, T^4-\mbox{terms} \nonumber \\
&+& \dots \nonumber \\
&+& \de_k\, g\, R\,T^2-\mbox{terms},\, R\,\nabla^{g B}\, T-\mbox{terms} \nonumber \\
&+& \dots \Big\}. \nonumber 
\eeqq

Expanding Eqn.(\ref{26}) in orders of $g$ we get
\beqq S_G [B, g] &=& \La\, \intf \nonumber \\
&+& \frac{1}{\ka}\, \intf \Big\{2 \al_1\, \pa_\al B_\be\,^{\al\be}
+  \al_5\, \pa_\al T^{\!\!\!{^{^o}}}_\be \,^{\al\be} [B] \Big\} \nonumber \\
&+& \intf \Big\{ \be_1\, R^{\!\!\!{^{^o}}}_{\al\be}\,^{\ga\de} [B]\, R^{{\!\!\!{^{^o}}}\al\be}\,_{\ga\de} [B] \nonumber \\
&+&  \be_2\, R^{\!\!\!{^{^o}}}_{\al\ga}\,^{\al\de} [B]\, R^{{\!\!\!{^{^o}}}\be\ga}\,_{\be\de} [B] 
+ \be_3\, R^{\!\!\!{^{^o}}}_{\al\be}\,^{\al\be} [B] \,R^{\!\!\!{^{^o}}}_{\ga\de}\,^{\ga\de} [B] \nonumber \\
&+& \frac{1}{g} \be_4\, \pa^\ga \pa_\de R^{\!\!\!{^{^o}}}_{\al\ga}\,^{\al\de} [B]
+ \frac{1}{g} \be_5\, \pa^\ga \pa_\ga R^{\!\!\!{^{^o}}}_{\al\be}\,^{\al\be} [B] \\
&+& \dots \nonumber \\
&+& \ga_1\, \pa_\ga T^{\!\!\!{^{^o}}}_{\al\be\de} [B]\,
\pa^\ga T^{{\!\!\!{^{^o}}}\al\be\de} [B] \nonumber \\
&+& \ga_2\, \pa_\ga T^{\!\!\!{^{^o}}}_{\al\be\de} [B]\, 
\pa^\ga T^{{\!\!\!{^{^o}}}\de\be\al} [B] \nonumber \\
&+& \dots \nonumber \\
&+& O(g)-\mbox{terms}. \nonumber
\eeqq
Taking the limit $g \ar 0$ above the surviving dimension-zero and -one terms can be neglected because the former is an infinite field-independent constant and the latter a total divergence. For the same reason we can also neglect the $O(\frac{1}{g})-$terms in the dimension-four term which is crucial for a consistent free field limit.

Whilst there is more than one surviving term in the limit above we take
\beq S^{\!\!\!{^{^o}}}_G [B] = \intf {\cal L}^{\!\!\!{^{^o}}}_G [B] \eeq
with
\beq \label{29} {\cal L}^{\!\!\!{^{^o}}}_G [B]  = 
-\frac{1}{4} R^{\!\!\!{^{^o}}}_{\al\be}\,^{\ga\de} [B]\, R^{{\!\!\!{^{^o}}}\al\be}\,_{\ga\de} [B]
-\frac{1}{2}\, \pa^\al B_\al\,^{\ga\de}\, \pa_\be B^\be\,_{\ga\de}
\eeq
as the free gauge field action with
\beq R^{\!\!\!{^{^o}}}_{\al\be}\,^{\ga\de} [B] =  \pa_\al B_\be\,^{\ga\de} -\, \pa_\be B_\al\,^{\ga\de}, \eeq
whilst all the other terms are to be treated as perturbations with small coupling constants $\be_i$, $\ga_j$ etc.

Above we have normalized $\be_1 = -\frac{1}{4}$ and introduced a Lorentz-covariant gauge-fixing term $\pa^\al B_\al\,^{\ga\de}\, \pa_\be B^\be\,_{\ga\de}$ as usual.

\section{Covariant Canonical Quantization of the non-interacting theory}

\paragraph{}
In this section, in a fully relativistically covariant way, we canonically quantize the gauge fields $B$ in the Lorentz gauge whose dynamics is governed by the non-interacting Lagrangian given by Eqn.(\ref{29}), and determine the expressions for the energy and momentum field operators in terms of $B$.

A usual we start with the gauge fields $B_\al\,^{\ga\de} (t,{\underline x})$ for a fixed time $t$ and first determine their canonically conjugate momenta
\beqq {\it\Pi}^\ep\,_{\eta\ze} (t,{\underline y}) &=& \frac{\pa {\cal L}^{\!\!\!{^{^o}}}_G [B]}{\pa (\pa_0B_\ep\,^{\eta\ze})} 
(t,{\underline y}) \\
&=& -R^{{\!\!\!{^{^o}}}0\ep}\,_{\eta\ze} [B] (t,{\underline y}) -\,\eta^{0\ep} \pa_\al B^\al\,_{\eta\ze} (t,{\underline y})
\nonumber
\eeqq
which are well-defined due to the gauge-fixing term in Eqn.(\ref{29}).

We now demand the quantum fields and their conjugates to obey the covariant equal-time canonical commutation relations
\beq \label{33} \Big[ B_\al\,^{\ga\de} (t,{\underline x}),  {\it\Pi}^\be\,_{\eta\ze} (t,{\underline y}) \Big]
= i\,  \eta_\al\,^\be \Big( \eta^\ga\,_\eta \eta^\de\,_\ze - \eta^\ga\,_\ze \eta^\de\,_\eta \Big)\,
\de^3({\underline x} - {\underline y})
\eeq
and
\beqq \label{34} \Big[ B_\al\,^{\ga\de} (t,{\underline x}), B^\be\,_{\eta\ze} (t,{\underline y}) \Big]
&=& 0 \\
\Big[ {\it\Pi}_\al\,^{\ga\de} (t,{\underline x}), {\it\Pi}^\be\,_{\eta\ze} (t,{\underline y}) \Big]
&=& 0, \nonumber
\eeqq
where $\eta^\ga\,_\eta \eta^\de\,_\ze - \eta^\ga\,_\ze \eta^\de\,_\eta$ is the projection operator onto the space of tensors $B_{.}\,^{\ga\de}$ antisymmetric in $\ga$ and $\de$.

Using
\beqq \dot B^0\,_{\eta\ze} (t,{\underline y}) &=& {\it\Pi}^0\,_{\eta\ze} (t,{\underline y}) 
- \pa_i B^i\,_{\eta\ze} (t,{\underline y}) \\
\dot B^i\,_{\eta\ze} (t,{\underline y}) &=& {\it\Pi}^i\,_{\eta\ze} (t,{\underline y}) 
- \pa^i B^0\,_{\eta\ze} (t,{\underline y}) \nonumber
\eeqq
a little algebra shows that Eqns.(\ref{33}) and (\ref{34}) are equivalent to
\beq \label{35} \Big[ \dot B_\al\,^{\ga\de} (t,{\underline x}),  B^\be\,_{\eta\ze} (t,{\underline y}) \Big]
= - i\,  \eta_\al\,^\be \Big( \eta^\ga\,_\eta \eta^\de\,_\ze - \eta^\ga\,_\ze \eta^\de\,_\eta \Big)\,
\de^3({\underline x} - {\underline y})
\eeq
and
\beqq \label{36} \Big[ B_\al\,^{\ga\de} (t,{\underline x}), B^\be\,_{\eta\ze} (t,{\underline y}) \Big]
&=& 0 \\
\Big[ \dot B_\al\,^{\ga\de} (t,{\underline x}), \dot B^\be\,_{\eta\ze} (t,{\underline y}) \Big]
&=& 0. \nonumber
\eeqq
Eqns.(\ref{35}) and (\ref{36}) are nothing but canonical commutation relations for 24 non-interacting scalar fields written in compact form -- however with only 12 of them coming with negative sign in Eqn.(\ref{35}) and 12 with positive.

In the remainder of the paper -- generalizing the covariant Gupta-Bleuler treatment of the photon field \cite{itz} -- we will work our way step-by-step to a relativistically invariant physical Fock space based on Eqns.(\ref{35}) and (\ref{36}) which will display positive norm and energy for all its states, and in which only 6 from the original 24 degrees of freedom will survive. 

Finally we give the Noether expression for the energy-momentum four vector
\beq P^\be = \intt T^{0\be}
\eeq
derived from the canonical energy-momentum tensor
\beq T^{\al\be} = \eta^{\al\be} {\cal L}^{\!\!\!{^{^o}}}_G
- \frac{\pa {\cal L}^{\!\!\!{^{^o}}}_G [B]}{\pa (\pa_\al B_\ep\,^{\eta\ze})}\, \pa_\be B_\ep\,^{\eta\ze}
\eeq
in terms of the fields and their first time derivative. We find for the energy operator
\beqq P^0 &=& H = \intt \Big\{ {\it\Pi}^\al\,_{\ga\de} (x) \dot B_\al\,^{\ga\de} (x) - {\cal L}^{\!\!\!{^{^o}}}_G [B] (x) \Big\} \\
&=& \frac{1}{2} \intt \Big\{ \dot B_\al\,^{\ga\de} (x) \dot B^\al\,_{\ga\de} (x)
+ \pa_i B_\al\,^{\ga\de} (x)\, \pa^i B^\al\,_{\ga\de} (x) \Big\} \nonumber
\eeqq
and for the momentum operator
\beq P^i = - \intt {\it\Pi}^\al\,_{\ga\de} (x)\, \pa^i B_\al\,^{\ga\de} (x)
= - \intt \dot B_\al\,^{\ga\de} (x)\, \pa^i B^\al\,_{\ga\de} (x).
\eeq

\section{Creation and annihilation operators of the non-interacting theory}

\paragraph{}
In this section we introduce the covariant creation and annihilation operators $b^\dagger$ and $b$ respectively and calculate all: their commutation relations and the expressions for the energy and the momentum operators in terms of them. We then naively construct the corresponding one-particle pseudo Hilbert space of states, not yet taking into account neither the Lorentz gauge fixing nor the physical requirements of positive probability and positive energy.

As usual we first Fourier-transform the hermitean gauge field $B$ introducing $b^\dagger$ and $b$
\beq \label{41} B_\al\,^{\ga\de} (x) = \intkt
\Big(b_\al\,^{\ga\de} ({\underline k})\, e^{ikx}  + b^\dagger_\al\,^{\ga\de} ({\underline k})\, e^{- ikx} \Big)
= B^\dagger_\al\,^{\ga\de} (x),
\eeq
where
\beqq
k^0 &=& \om_k = +\, \sqrt{{\underline k}^2} = |{\underline k}| \\
d {\tilde k} &=& \frac{d^{3} k}{(2 \pi)^3\, 2\om_k} 
= \frac{d^{4} k}{(2 \pi)^3}\, \theta(k^0)\, \de^4 (k^2) \nonumber
\eeqq
expresses the on-shell condition $k^2 = 0$ assuring that $B$ is a solution of the field equations derived from extremizing Eqn.(\ref{29}) and $d {\tilde k}$ denotes the invariant on-shell volume element in momentum-space.

Next we invert Eqn.(\ref{41}) and find the time-independent expression
\beq b_\al\,^{\ga\de} ({\underline k}) = i \intt e^{- ikx}\, \pa_0\lrvec\,\, B_\al\,^{\ga\de} (t,{\underline x}).
\eeq

It is now easy to show that the covariant creation and annihilation operators obey the covariant commutation relations
\beq \label{44} \Big[ b_\al\,^{\ga\de} ({\underline k}),  b^\dagger_\be\,^{\eta\ze} ({\underline h}) \Big]
= (2 \pi)^3\, 2\om_k\,  \eta_{\al\be} \Big( \eta^{\ga\eta} \eta^{\de\ze} - \eta^{\ga\ze} \eta^{\de\eta} \Big)\,
\de^3({\underline k} - {\underline h})
\eeq
and
\beqq \label{45} \Big[ b_\al\,^{\ga\de} ({\underline k}),  b_\be\,^{\eta\ze} ({\underline h}) \Big]
&=& 0 \\
\Big[ b^\dagger_\al\,^{\ga\de} ({\underline k}),  b^\dagger_\be\,^{\eta\ze} ({\underline h}) \Big]
&=& 0. \nonumber
\eeqq
Again we obtain 24 commutations relations for the creation and annihilation operators similar to those belonging to a scalar field -- again with only 12 of them coming with positive sign in Eqn.(\ref{44}) and 12 with negative.

Next we express the energy operator in terms of the covariant creation and annihilation operators
\beqq H &=& \frac{1}{2} \intt \Big\{ \dot B_\al\,^{\ga\de} (x) \dot B^\al\,_{\ga\de} (x)
+ \pa_i B_\al\,^{\ga\de} (x)\, \pa^i B^\al\,_{\ga\de} (x) \Big\} \\
&=& \frac{1}{2} \intkt\, \om_k\, \Big\{ b^\al\,_{\ga\de} ({\underline k})\, b^\dagger_\al\,^{\ga\de} ({\underline k})
 + b^\dagger_\al\,^{\ga\de} ({\underline k})\, b^\al\,_{\ga\de} ({\underline k}) \Big\}. \nonumber
\eeqq
As usual we introduce normal ordering denoted by $: . :$ -- writing all the creation operators to the left of the annihilation operators -- to normalize the energy of the vacuum to zero
\beqq H &=& \frac{1}{2} :\!\! \intkt\, \om_k\, 
\Big\{ b^\al\,_{\ga\de} ({\underline k})\, b^\dagger_\al\,^{\ga\de} ({\underline k})
+ b^\dagger_\al\,^{\ga\de} ({\underline k})\, b^\al\,_{\ga\de} ({\underline k}) \Big\}\!\! : \\
&=& \intkt\, \om_k\, b^\dagger_\al\,^{\ga\de} ({\underline k})\, b^\al\,_{\ga\de} ({\underline k}). \nonumber 
\eeqq

Similarly we find for the momentum operator
\beqq P^i &=& - \intt \dot B_\al\,^{\ga\de} (x)\, \pa^i B^\al\,_{\ga\de} (x) \\
&=& \frac{1}{2} \intkt\, k^i\, \Big\{ b^\al\,_{\ga\de} ({\underline k})\, b^\dagger_\al\,^{\ga\de} ({\underline k})
+ b^\dagger_\al\,^{\ga\de} ({\underline k})\, b^\al\,_{\ga\de} ({\underline k}) \Big\} \nonumber
\eeqq
which needs no normal ordering.

The 1 + 3 relations above can be combined into the covariant expression for the hermitean energy-momentum operator
\beqq P^\be &=& \frac{1}{2} :\!\! \intkt\, k^\be\, 
\Big\{ b^\al\,_{\ga\de} ({\underline k})\, b^\dagger_\al\,^{\ga\de} ({\underline k})
+ b^\dagger_\al\,^{\ga\de} ({\underline k})\, b^\al\,_{\ga\de} ({\underline k}) \Big\} \!\! : \\
&=& \intkt\, k^\be\, b^\dagger_\al\,^{\ga\de} ({\underline k})\, b^\al\,_{\ga\de} ({\underline k}). \nonumber 
\eeqq
Its commutator with a creation operator $b^\dagger$ reads
\beq
\Big[ P^\be, b^\dagger_\al\,^{\ga\de} ({\underline k}) \Big] = k^\be\, b^\dagger_\al\,^{\ga\de} ({\underline k})
\eeq
and it commutes with the energy operator, i.e. is time-independent
\beq 
\Big[ H, P^\be \Big] = 0.
\eeq

Next we introduce the normalized vacuum state
\beq
\midd 0\rangle \quad\mbox{with}\quad \langle 0 \midd 0\rangle = 1
\eeq
which is annihilated by the annihilation operators $b$
\beq
b_\al\,^{\ga\de} ({\underline k}) \midd 0\rangle = 0 \quad \forall\, {\underline k}; \al, \ga, \de
\eeq
and out of which the creation operators $b^\dagger$ create one-particle states
\beq
b^\dagger_\al\,^{\ga\de} ({\underline k}) \midd 0\rangle = \mbox{one-particle state} 
\eeq
with normalization
\beq \langle 0\midd b_\al\,^{\ga\de} ({\underline k})\,  b^\dagger_\be\,^{\eta\ze} ({\underline h}) \midd 0\rangle
= (2 \pi)^3\, 2\om_k\,  \eta_{\al\be} \Big( \eta^{\ga\eta} \eta^{\de\ze} - \eta^{\ga\ze} \eta^{\de\eta} \Big)\,
\de^3({\underline k} - {\underline h}).
\eeq

Let us look next at a general one-particle state defined by
\beq \midd f\rangle = \sqrt{\frac{1}{2}} \intkt\, f^\al\,_{\ga\de} ({\underline k})\, 
b^\dagger_\al\,^{\ga\de} ({\underline k}) \midd 0\rangle,
\eeq
where $f^\al\,_{\ga\de} ({\underline k}) = -f^\al\,_{\de\ga} ({\underline k})$ is a generally complex polarization tensor antisymmetric in the indices $\ga$ and $\de$. $f^\al\,_{\ga\de} ({\underline k})$ characterizes a one-particle state completely which will be crucial in the further development of our thoughts.

The energy operator acts on $\midd f\rangle$ as
\beq H \midd f\rangle = \sqrt{2} \intkt\, \om_k\, f^\al\,_{\ga\de} ({\underline k})\, 
b^\dagger_\al\,^{\ga\de} ({\underline k}) \midd 0\rangle
\eeq
and the norm of $\midd f\rangle$ is calculated to be
\beq \langle f \midd f\rangle = 
\intkt\, f^*_\al\,^{\ga\de} ({\underline k})\,  f^\al\,_{\ga\de} ({\underline k})
\eeq
whilst the energy expectation value for the state $\midd f\rangle$ is
\beq \langle f \midd H \midd f\rangle = 
\intkt\, \om_k\, f^*_\al\,^{\ga\de} ({\underline k})\,  f^\al\,_{\ga\de} ({\underline k}).
\eeq

We again recognize the problem: whilst the quantization procedure is manifestly covariant it yields 12 positive-norm, positive-energy states and 12 negative-norm, negative-energy states. In the next sections we turn to covariantly cure the flaws step-by-step.

\section{One-particle subspace of states  ${\cal F}_1^{\mbox{\footnotesize LG}}$ fulfilling the Lorentz gauge condition}

\paragraph{}
In this section we take the first step towards the identification of a viable physical Fock space by introducing the one-particle subspace of states  ${\cal F}_1^{\mbox{\footnotesize LG}}$ fulfilling the Lorentz gauge condition which is in essence a condition on the $\al$-index of the polarization tensor $f_\al\,^{\ga\de} ({\underline k})$.

Let us start by introducing the negative-frequency part of the gauge field $B$ by
\beq B^{(-)}_\al\,^{\ga\de} (x) = \intkt\, b_\al\,^{\ga\de} ({\underline k})\, e^{ikx}. 
\eeq

We then define a one-particle state $\midd f\rangle$
\beq \midd f\rangle\, \in {\cal F}_1^{\mbox{\footnotesize LG}}
\eeq
to be in the one-particle subspace of states  ${\cal F}_1^{\mbox{\footnotesize LG}}$ fulfilling the Lorentz gauge condition if $(i)$ holds
\beq (i)\!: \quad \pa^\al B^{(-)}_\al\,^{\ga\de} (x) \midd f\rangle\, = \sqrt{2} \intkt\, i k^\al\, f_\al\,^{\ga\de} ({\underline k}) \, e^{ikx}\,
\midd 0\rangle {=\!\!\!\!\!^{^!}}\,\,\,\, 0
\eeq
which assures that $\midd f\rangle$ fulfills the Lorentz condition in the mean
\beq \langle f \midd \pa^\al B_\al\,^{\ga\de} (x) \midd f\rangle\, = 0.
\eeq

To fulfill the condition above we have to have
\beqq k^\al f_\al\,^{\ga\de} ({\underline k}) &=& k^0 f_0\,^{\ga\de} ({\underline k})
+ k^i f_i\,^{\ga\de} ({\underline k}) \\
&=& \om_k \left( f_0\,^{\ga\de} ({\underline k}) + \frac{k^i}{\om_k}\, f_i\,^{\ga\de} ({\underline k}) \right)
\, {=\!\!\!\!\!^!}\,\,\,\, 0 \nonumber 
\eeqq
for all $k^\al$ which implies
\beq f_0\,^{\ga\de} ({\underline k}) = -\frac{k^i}{\om_k}\, f_i\,^{\ga\de} ({\underline k}) = 
- f^{0\,\ga\de} ({\underline k}) \quad \forall\, k^\al.
\eeq
Above we have implicitly introduced a coordinate system in Minkowski space singling out a  time- or $0$-axis and three space- or $i$-axes to which we will refer to later again. To be specific let one spatial axis be parallel and the two others be perpendicular to ${\underline k}$.
 
Hence a general one-particle state $\midd f\rangle\, \in {\cal F}_1^{\mbox{\footnotesize LG}}$ can always be written as
\beq \midd f\rangle = \sqrt{\frac{1}{2}} \intkt\,  
f^i\,_{\ga\de} ({\underline k}) \left( b^\dagger_i\,^{\ga\de} ({\underline k}) 
+ \frac{k^i}{\om_k}\, b^\dagger_0\,^{\ga\de} ({\underline k}) \right) \midd 0\rangle.
\eeq
On such a state the energy operator acts as
\beq H \midd f\rangle = \sqrt{2} \intkt\, \om_k\, 
f^i\,_{\ga\de} ({\underline k}) \left( b^\dagger_i\,^{\ga\de} ({\underline k}) 
+ \frac{k^i}{\om_k}\, b^\dagger_0\,^{\ga\de} ({\underline k}) \right) \midd 0\rangle.
\eeq

Finally the norm of a general one-particle state $\midd f\rangle\, \in {\cal F}_1^{\mbox{\footnotesize LG}}$  becomes
\beq \langle f \midd f\rangle = \intkt\, f^*_i\,^{\ga\de} ({\underline k})\, f^j\,_{\ga\de} ({\underline k})
\left( \de^i\,_j - \frac{k^i k_j}{{\underline k}^2} \right)
\eeq
and the energy expectation value becomes
\beq \langle f \midd H \midd f\rangle = \intkt\, \om_k\, f^*_i\,^{\ga\de} ({\underline k})\, f^j\,_{\ga\de} ({\underline k})
\left( \de^i\,_j - \frac{k^i k_j}{{\underline k}^2} \right).
\eeq

Above $\de^i\,_j - \frac{k^i k_j}{{\underline k}^2}$ is the projection operator onto the transversal part of $f^j\,_{\ga\de} ({\underline k})$. As a result both the norm and the energy expectation value of a general state $\midd f\rangle\, \in {\cal F}_1^{\mbox{\footnotesize LG}}$ depend only on the transversal parts of the one-particle state vector -- in effect reducing the 24 degrees of freedom we have started with to 4 $\times$ 3 = 12, however only 6 coming with positive-norm, positive-energy and another 6 still coming with negative-norm, negative-energy.

\section{One-particle subspace of states ${\cal F}_1^{\mbox{\footnotesize P}}$ orthogonal to their dual state with positive semidefinite norm}

\paragraph{}
In this section we take the second step towards the identification of a viable physical Fock space by introducing the one-particle subspace of states  ${\cal F}_1^{\mbox{\footnotesize P}}$ orthogonal to their dual state with positive semidefinite norm which will turn out to be in essence a condition on the $\ga, \de$-indices of the polarization tensor $f_\al\,^{\ga\de} ({\underline k})$.

We start by defining a one-particle state $\midd f\rangle$
\beq \midd f\rangle\, \in {\cal F}_1^{\mbox{\footnotesize P}}
\eeq
to be in the one-particle subspace of states  ${\cal F}_1^{\mbox{\footnotesize P}}$ orthogonal to their dual state with positive semidefinite norm if the two conditions $(ii-a)$ and $(ii-b)$ hold
\beqq \label{71} & &(ii-a)\!: \quad \langle f\midd f\rangle\,  {\geq\!\!\!\!\!^{^!}}\,\,\,\,  0 \\
& & (ii-b)\!: \quad \langle \hat f\midd f\rangle\,  {=\!\!\!\!\!^{^!}}\,\,\,\, 0. \nonumber
\eeqq
Above we have introduced the state $\midd \hat f\rangle$ dual to $\midd f\rangle$
\beq \midd \hat f\rangle \equiv \sqrt{\frac{1}{2}} \intkt\, \ep^{\ga\de}\,_{\eta\ze}\, f^\al\,_{\ga\de} ({\underline k})\, 
b^\dagger_\al\,^{\eta\ze} ({\underline k}) \midd 0\rangle
\eeq
which plays a crucial role in the further development of the theory.

In addition to demanding $(ii-a)$ and $(ii-b)$ to hold for individual states we must require linearity
\beq \label{73} \midd f\rangle ,\, \midd g \rangle\, \in {\cal F}_1^{\mbox{\footnotesize P}} \Rightarrow
\quad \midd f \rangle\, + \midd g \rangle\, =\,\,\,\, \midd f + g \rangle\, \in {\cal F}_1^{\mbox{\footnotesize P}}
\eeq
so as to assure ${\cal F}_1^{\mbox{\footnotesize P}}$ is indeed a linear space and prove consistency of the three requirements alltogether.

To do so we note that the two requirements $(ii-a)'$ and $(ii-b)'$
\beqq & &(ii-a)'\!: \quad\quad f^*_{\al}\,^{\ga\de} ({\underline k})\, f_{\al\,\ga\de} ({\underline k})\,  \geq 0, 
\quad {\mbox{no summation over}}\,\, \al ! \nonumber  \\
& & \quad\quad {\mbox{so that}} \quad\quad f^*_{\al}\,^{\ga\de} ({\underline k})\, f^{\al}\,_{\ga\de} ({\underline k})\,  \geq 0 \\
& & (ii-b)'\!: \quad\quad \hat f^*_{\al}\,^{\ga\de} ({\underline k})\, f_{\al\,\ga\de} ({\underline k})\,  = 0 \nonumber 
\eeqq
are sufficient for $(ii-a)$ and $(ii-b)$ to hold.

On the other hand $(ii-a)'$ and $(ii-b)'$ imply for $\al$ fixed the existence of a Lorentz transformation $\La_{_{P}}^\ga\,_\eta (f, \al, {\underline k})$ and an antisymmetric tensor $f^P_{\al}\,^{\eta\ze} ({\underline k})$ such that $f_{\al}\,^{\ga\de} ({\underline k})$ can be written in terms of the three components $f^P_{\al}\,^{ij} ({\underline k})$ only \cite{lan}
\beqq & & f_{\al}\,^{\ga\de} ({\underline k}) = \La_{_{P}}^\ga\,_\eta (f, \al, {\underline k})\,
\La_{_{P}}^\de\,_\ze (f, \al, {\underline k})\, f^P_{\al}\,^{\eta\ze} ({\underline k}) \\
& & \quad\quad {\mbox{with}}\quad\quad f^P_{\al}\,^{0 i} ({\underline k}) = - f^P_{\al}\,^{i 0} ({\underline k}) = 0. \nonumber
\eeqq
We note that both the Lorentz transformation and $\La_{_{P}}$ and $f^P$ are not uniquely defined at this point. To completely fix both $\La_{_{P}}$ and $f^P$ we first perform a Lorentz transformation on $f_{\al}\,^{\ga\de} ({\underline k})$ to an arbitrary $f^{P'}_{\al}\,^{\eta\ze} ({\underline k})$ fulfilling $f^{P'}_{\al}\,^{0 i} ({\underline k}) = - f^{P'}_{\al}\,^{i 0} ({\underline k}) = 0$. We then can still perform {\bf SO(3)\/} rotations which respect $f^{P''}_{\al}\,^{0 i} ({\underline k}) = - f^{P''}_{\al}\,^{i 0} ({\underline k}) = 0$. Let these rotations be parametrized by their rotation vectors ${\underline \al}$, where the direction of ${\underline \al}$ defines the rotation axis and its length the rotation angle. To fix $\La_{_{P}}$ and $f^P$ completely we choose the rotation such that its rotation axis is parallel to ${\underline k}$ and the coordinate system of the plane perpendicular to ${\underline k}$ coincides with the one defined in the preceding section.

We finally note that at this point $\La_{_{P}}^\ga\,_\eta (f, \al, {\underline k})$ may depend on all: $f$, $\al$ and ${\underline k}$.

Let us turn the consideration above on its head and look at the linear space of tensors $f^P_{\al}\,^{\eta\ze} ({\underline k})$ antisymmetric in $\eta, \ze$ with $f^P_{\al}\,^{0 i} ({\underline k}) = - f^P_{\al}\,^{i 0} ({\underline k}) = 0$. It is then possible to demonstrate that having $f_{\al}\,^{\ga\de} ({\underline k})$ of the form
\beq f_{\al}\,^{\ga\de} ({\underline k}) = \La_{_{P}}^\ga\,_\eta ({\underline k})\, \La_{_{P}}^\de\,_\ze ({\underline k})\, f^P_{\al}\,^{\eta\ze} ({\underline k}) \eeq
with $\La^\ga\,_\eta ({\underline k})$ depending only on ${\underline k}$, but neither on $f$ nor $\al$, is necessary and sufficient for $(ii-a)'$, $(ii-b)'$ and the linearity condition to hold which implies $(ii-a)$, $(ii-b)$ and Eqn.(\ref{73}) to hold as well.

As a result a general one-particle state $\midd f\rangle \in {\cal F}_1^{\mbox{\footnotesize P}}$ is written as
\beqq \midd f\rangle &=& \sqrt{\frac{1}{2}} \intkt\, \La_{_{P}}^\ga\,_\eta ({\underline k})\,
\La_{_{P}}^\de\,_\ze ({\underline k})\, f^P_\al\,^{\eta\ze} ({\underline k})\, 
\La_{_{P}}^\iota\,_\ga ({\underline k})\, \La_{_{P}}^\ka\,_\de ({\underline k})\,
b^{P\, \dagger \al}\,_{\iota\ka} ({\underline k}) \midd 0\rangle \nonumber \\
&=& \sqrt{\frac{1}{2}} \intkt\, f^{P\, \al}\,_{\ga\de} ({\underline k})\, 
b^{P\, \dagger}_\al\,^{\ga\de} ({\underline k}) \midd 0\rangle \\
&=& \sqrt{\frac{1}{2}} \intkt\, f^{P\, \al}\,_{i j} ({\underline k})\, 
b^{P\, \dagger}_\al\,^{i j} ({\underline k}) \midd 0\rangle =\,\, \midd f^P \rangle \nonumber
\eeqq
with $b^{P\, \al}\,_{\iota\ka} ({\underline k})$, $b^{P\, \dagger}_\be\,^{\eta\zeta} ({\underline k})$ fulfilling the same commutation relations Eqns. (\ref{44}) and (\ref{45}) as $b^\al\,_{\ga\de} ({\underline k}) = \La_{_{P}}^\iota\,_\ga ({\underline k})\, \La_{_{P}}^\ka\,_\de ({\underline k})\, b^{P\, \al}\,_{\iota\ka} ({\underline k})$ and $b^\dagger_\be\,^{\eta\zeta} ({\underline k})$ do due to the covariance of the commutation relations.

Finally the energy operator acts on a one-particle state $\midd f\rangle \in {\cal F}_1^{\mbox{\footnotesize P}}$ as
\beq H \midd f\rangle = \sqrt{2} \intkt\, \om_k\, f^{P\, \al}\,_{i j} ({\underline k})\, 
b^{P\, \dagger}_\al\,^{i j} ({\underline k}) \midd 0\rangle =\,\, H \midd f^P \rangle.
\eeq
The norm of such a state becomes
\beq \langle f \midd f\rangle = \intkt\, f^{P\, *}_\al\,^{i j} ({\underline k})\,  f^{P\, \al}\,_{i j} ({\underline k})
=\,\, \langle f^P \midd f^P\rangle  \geq 0 \nonumber
\eeq
and its energy expectation value
\beq \langle f \midd H \midd f\rangle = \intkt\, \om_k\, f^{P\, *}_\al\,^{i j} ({\underline k})\,  f^{P\, \al}\,_{i j} ({\underline k})
=\,\, \langle f^P \midd H \midd f^P\rangle  \geq 0. \nonumber
\eeq

As a result both the norm and the energy expectation value of a general state $\midd f\rangle\, \in {\cal F}_1^{\mbox{\footnotesize P}}$ depend only on the $i j$-components of the one-particle state vector -- in effect reducing the 24 degrees of freedom we have started with to 4 $\times$ 3 = 12. Note that both the norm and the energy expectation value are positive semidefinite due to the Eqns.(\ref{71}) defining ${\cal F}_1^{\mbox{\footnotesize P}}$.

\section{One-particle subspace of states ${\cal F}_1^{\mbox{\footnotesize LG $\!\&\!$ P}}$ with positive semidefinite norm and energy expectation value}

\paragraph{}
In this section we take the third and crucial step towards the identification of a viable physical Fock space by introducing the one-particle subspace of states ${\cal F}_1^{\mbox{\footnotesize LG $\!\&\!$ P}}$ with positive semidefinite norm and energy expectation value which combines the conditions on the $\al$-index and the $\ga, \de$-indices of the polarization tensor $f_\al\,^{\ga\de} ({\underline k})$ formulated in the two preceding sections.

We start by defining a one-particle state $\midd f\rangle$
\beq \midd f\rangle\, \in {\cal F}_1^{\mbox{\footnotesize LG $\!\&\!$ P}}
\eeq
to be in the one-particle subspace of states ${\cal F}_1^{\mbox{\footnotesize LG $\!\&\!$ P}}$ with positive semidefinite norm and energy expectation value if the three conditions $(i)$, $(ii-a)$ and $(ii-b)$ plus the linearity condition hold
\beqq & & (i)\!: \quad \pa^\al B^{(-)}_\al\,^{\ga\de} (x) \midd f\rangle\, = 0 \nonumber \\
& & \quad {\mbox{or}}\quad\quad k^\al\, f_{\al}\,^{\ga\de} ({\underline k}) = 0  \\
& & (ii-a)\!: \quad \langle f\midd f\rangle\,  \geq 0 \nonumber \\
& & (ii-b)\!: \quad \langle \hat f\midd f\rangle\,  = 0 \nonumber
\eeqq

A general one-particle state $\midd f\rangle\, \in {\cal F}_1^{\mbox{\footnotesize LG $\!\&\!$ P}}$ can always be written as
\beq \quad\quad\quad \midd f\rangle = \sqrt{\frac{1}{2}} \intkt\,  
f^{P\, i}\,_{jk} ({\underline k}) \left( b^\dagger_i\,^{jk} ({\underline k}) 
+ \frac{k^i}{\om_k}\, b^\dagger_0\,^{jk} ({\underline k}) \right) \midd 0\rangle =\,\, \midd f^P\rangle.
\eeq
On such a state the energy operator acts as
\beq H \midd f\rangle = \sqrt{2} \intkt\, \om_k\, 
f^{P\, i}\,_{jk} ({\underline k}) \left( b^\dagger_i\,^{jk} ({\underline k}) 
+ \frac{k^i}{\om_k}\, b^\dagger_0\,^{jk} ({\underline k}) \right) \midd 0\rangle =\,\, H \midd f^P\rangle.
\eeq

Finally the norm of a general one-particle state $\midd f\rangle\, \in {\cal F}_1^{\mbox{\footnotesize LG $\!\&\!$ P}}$  becomes
\beq \langle f \midd f\rangle = \intkt\, f^{P\, *}_i\,^{kl} ({\underline k})\, f^{P\, j}\,_{kl} ({\underline k})
\left( \de^i\,_j - \frac{k^i k_j}{{\underline k}^2} \right) = \langle f^P \midd f^P\rangle \geq 0
\eeq
and the energy expectation value becomes
\beqq \langle f \midd H \midd f\rangle &=& \intkt\, \om_k\, f^{P\, *}_i\,^{kl} ({\underline k})\, f^{P\, j}\,_{kl} ({\underline k})
\left( \de^i\,_j - \frac{k^i k_j}{{\underline k}^2} \right) \nonumber \\
&=& \langle f^P \midd H \midd f^P\rangle \geq 0.
\eeqq

The positive-semidefiniteness becomes obvious if we write
\beq f^{P\, j}\,_{kl} ({\underline k})\, k_j =\,\, \midd {\underline f}^P\,_{kl} ({\underline k}) \midd\,\,
\midd {\underline k} \midd \cos\th({\underline k}),
\eeq
where $\th({\underline k})$ is the angle between $f^{P\, i}\,_{..} ({\underline k})$ and $k_j$. We can now re-write
\beqq \langle f \midd f\rangle &=& \intkt\, \midd {\underline f}^{P\, * kl} ({\underline k}) \midd\,\,
\midd {\underline f}^P\,_{kl} ({\underline k}) \midd \left( 1 - \cos^2 \th({\underline k}) \right) \\
&=&  \intkt\, \midd {\underline f}^{P\, * kl} ({\underline k}) \midd\,\,
\midd {\underline f}^P\,_{kl} ({\underline k}) \midd \sin^2 \th({\underline k})  = \langle f^P \midd f^P\rangle \geq 0
\nonumber \eeqq
as well as
\beqq \langle f \midd H \midd f\rangle &=& \intkt\, \om_k\, \midd {\underline f}^{P\, * kl} ({\underline k}) \midd\,\,
\midd {\underline f}^P\,_{kl} ({\underline k}) \midd \left( 1 - \cos^2 \th({\underline k}) \right) \\
&=&  \intkt\, \om_k\, \midd {\underline f}^{P\, * kl} ({\underline k}) \midd\,\,
\midd {\underline f}^P\,_{kl} ({\underline k}) \midd \sin^2 \th({\underline k})  = \langle f^P \midd H \midd f^P\rangle \geq 0
\nonumber \eeqq

As a result both the norm and the energy expectation value of a general state $\midd f\rangle\, \in {\cal F}_1^{\mbox{\footnotesize LG $\!\&\!$ P}}$ depend only on the transversal parts of the one-particle state vector -- in effect reducing the 24 degrees of freedom we have started with to 2 $\times$ 3 = 6, however only 3 coming with positive-norm, positive-energy and another 3 still coming with negative-norm, negative-energy. Note that the 2 in 2 $\times$ 3 = 6 comes from the transversality condition as discussed in section 6.

\section{Physical one-particle space of states ${\cal F}_1^{\Phys}$ with positive norm and energy expectation value}

\paragraph{}
In this section we take the fourth and final step towards the identification of a physical one-particle space of states ${\cal F}_1^{\Phys}$ with positive norm and energy expectation value by identifying all states $\midd f^P\rangle \in {\cal F}_1^{\mbox{\footnotesize LG $\!\&\!$P}}$ whose difference has norm zero.

As noted in the preceding section a general one-particle state $\midd f^P\rangle\, \in {\cal F}_1^{\mbox{\footnotesize LG $\!\&\!$ P}}$ can be written as
\beq \quad\quad\quad \midd f^P\rangle = \sqrt{\frac{1}{2}} \intkt\,  
f^{P\, i}\,_{jk} ({\underline k}) \left( b^\dagger_i\,^{jk} ({\underline k}) 
+ \frac{k^i}{\om_k}\, b^\dagger_0\,^{jk} ({\underline k}) \right) \midd 0\rangle
\eeq
with norm
\beq \label{91} \langle f^P \midd f^P\rangle = \intkt\, f^{P\, *}_i\,^{kl} ({\underline k})\, f^{P\, j}\,_{kl} ({\underline k})
\left( \de^i\,_j - \frac{k^i k_j}{{\underline k}^2} \right) \geq 0.
\eeq

From Eqn.(\ref{91}) it becomes obvious that longitudinal states with $f^{P\, i}\,_{jk} ({\underline k})$ of the form $f^{P\, i}\,_{jk} ({\underline k})\sim k^i\,h_{jk} ({\underline k})$ are non-zero states with norm zero and energy expectation value zero -- so in some sense equivalent to the zero state, but non-zero.

Separating $\midd f^P\rangle = \,\, \midd f^P_T\rangle\, + \, \midd f^P_L\rangle$ into its transversal and longitudinal parts the longitudinal part
\beq \quad\quad\quad \midd f^P_L\rangle = \sqrt{\frac{1}{2}} \intkt\,  
\frac{k^i k_j}{{\underline k}^2} f^{P\, j}\,_{kh} ({\underline k}) \left( b^\dagger_i\,^{kh} ({\underline k}) 
+ \frac{k^i}{\om_k}\, b^\dagger_0\,^{jk} ({\underline k}) \right) \midd 0\rangle
\eeq
is an example of such a non-zero state with norm zero, $\langle f^P_L \midd f^P_L\rangle = 0$, and energy expectation value zero, $\langle f^P_L \midd H \midd f^P_L\rangle = 0$.

To deal with these non-zero states we next define equivalence classes of states with two states being equivalent
\beqq & & \midd f^P\rangle \sim \,\, \midd g^P\rangle \quad\quad {\mbox{for}}\quad
 \midd f^P\rangle,\,\, \midd g^P\rangle \, \in {\cal F}_1^{\mbox{\footnotesize LG $\!\&\!$ P}} \\
& & \quad\quad {\mbox{if}}\quad 
\langle f^P - g^P\midd f^P - g^P\rangle = 0 \nonumber
\eeqq
if their difference has norm zero \cite{itz}.

The difference of two general states is the state
\beq \midd f^P - g^P\rangle = \sqrt{\frac{1}{2}} \intkt\,  
\left(f^{P\, i}\,_{jk} ({\underline k}) - g^{P\, i}\,_{jk} ({\underline k})\right) \left( b^\dagger_i\,^{jk} ({\underline k}) 
+ \frac{k^i}{\om_k}\, b^\dagger_0\,^{jk} ({\underline k}) \right) \midd 0\rangle
\eeq
Demanding its norm to vanish
\beqq  & & \langle f^P - g^P\midd f^P - g^P\rangle = \intkt\,
\left(f^{P\, *}_i\,^{kl} ({\underline k}) - g^{P\, *}_i\,^{kl} ({\underline k})\right) \nonumber \\
& & \quad\quad \cdot \left(f^{P\, j}\,_{kl} ({\underline k}) - g^{P\, j}\,_{kl} ({\underline k})\right)
\left( \de^i\,_j - \frac{k^i k_j}{{\underline k}^2} \right) {=\!\!\!\!\!^{^!}}\,\,\,\, 0
\eeqq
we find that 
\beq f^{P\, j}_T\,_{kl} ({\underline k}) = g^{P\, j}_T\,_{kl} ({\underline k}),
\eeq
or that the transversal parts  $\midd f^P_T\rangle = \,\, \midd g^P_T\rangle$ of the respective polarization tensors have to be equal.

Put it differently the longitudinal parts of states are "invisible" or equivalent to a zero state
\beq \quad \midd f^P_L\rangle = \sqrt{\frac{1}{2}} \intkt\,  
\frac{k^i k_j}{{\underline k}^2} f^{P\, j}\,_{kh} ({\underline k}) \left( b^\dagger_i\,^{kh} ({\underline k}) 
+ \frac{k^i}{\om_k}\, b^\dagger_0\,^{kh} ({\underline k}) \right) \midd 0\rangle \sim 0
\eeq
whilst a one-particle state is equivalent to its transversal part
\beq \midd f^P\rangle \sim \,\, \midd f^P_T\rangle.
\eeq

Finally we are in a position to define a physical one-particle state as the following equivalent class of states
\beq \midd f^{Phys}\rangle \equiv \left[\, \midd f^P\rangle \right]
= \left\{\, \midd g^P\rangle \in {\cal F}_1^{\mbox{\footnotesize LG $\!\&\!$ P}}\,
\Big\vert\,\, \midd g^P\rangle \sim\,\, \midd f^P\rangle \right\}.
\eeq

There is always one distinguished element in each of the equivalence classes, namely the transversal part of a general state
\beq \quad\quad \midd f^{Phys}_T\rangle = \sqrt{\frac{1}{2}} \intkt\,  
f^{P\, i}_T\,_{kh} ({\underline k}) \left( b^\dagger_i\,^{kh} ({\underline k}) 
+ \frac{k^i}{\om_k}\, b^\dagger_0\,^{kh} ({\underline k}) \right) \midd 0\rangle
\eeq
with 
\beq f^{P\, i}_T\,_{kh} ({\underline k}) =
\left( \de^i\,_j - \frac{k^i k_j}{{\underline k}^2} \right) f^{P\, j}\,_{kh} ({\underline k}).
\eeq

On a general physical one-particle state the energy operator acts as
\beq H \midd f^{Phys}\rangle = \sqrt{2} \intkt\, \om_k\, 
f^{P\, i}\,_{jk} ({\underline k}) \left( b^\dagger_i\,^{jk} ({\underline k}) 
+ \frac{k^i}{\om_k}\, b^\dagger_0\,^{jk} ({\underline k}) \right) \midd 0\rangle,
\eeq
where the polrization tensor $k_i f^{P\, i}\,_{jk} ({\underline k}) = 0$ is transversal.

Finally the norm of a physical one-particle state $\midd f^{Phys}\rangle\, \in {\cal F}_1^{\Phys}$  becomes
\beq \langle f^{Phys} \midd f^{Phys}\rangle = \intkt\, f^{P\, *}_i\,^{kl} ({\underline k})\, f^{P\, i}\,_{kl} ({\underline k}) \geq 0
\eeq
where $k_i f^{P\, i}\,_{jk} ({\underline k}) = 0$, and the energy expectation value becomes
\beqq \langle f \midd H \midd f\rangle = \intkt\, \om_k\, f^{P\, *}_i\,^{kl} ({\underline k})\, f^{P\, i}\,_{kl} ({\underline k}) \geq 0,
\eeqq
where again $k_i f^{P\, i}\,_{jk} ({\underline k}) = 0$.

\section{Covariance of the quantization approach on ${\cal F}_1^{\Phys}$}

\paragraph{}
In this section we demonstrate the covariance of our approach to canonically quantizing the Lorentz gauge fields on the physical one-particle space of states ${\cal F}_1^{\Phys}$ with positive norm and energy expectation value.

Before starting with the demonstration of the covariance we note that our approach allows to consistently represent the creation and annihilation operator algebra given by Eqns.(\ref{44}) and (\ref{45}) on the physical one-particle space of states ${\cal F}_1^{\Phys}$ with positive norm and energy expectation value and -- as we will show in the next section -- on the physical Fock space of states ${\cal F}^{\Phys}$. This is an extension of the Gupta-Bleuler approach to the quantization of free gauge fields for compact gauge groups, where positivity of the norm and energy expectation value of states is assured by the positive definiteness of the Cartan metric on the gauge algebra, and is in itself an interesting result.

Let us turn now to the demonstration of the covariance of our approach to canonically quantize the Lorentz gauge fields. We will split this demonstration into two steps.

The first step starts with a physicist who works in the primed reference frame to write down a general physical one-particle state
\beq \label{105} \quad\quad \midd f'^{Phys}\rangle = \sqrt{\frac{1}{2}} \intkt'\,  
f'^P_i\,^{jk} ({\underline k}') \left( b'^{\dagger\, i}\,_{jk} ({\underline k}') 
+ \frac{k'^i}{\om_{k'}}\, b'^{\dagger\, 0}\,_{jk} ({\underline k}') \right) \midd 0 \rangle.
\eeq

Writing down the state vector above is equivalent to writing down a general one-particle state
\beq \midd f'^{Phys}\rangle = \sqrt{\frac{1}{2}} \intkt'\,
f'^P_\al\,^{\ga\de} ({\underline k}')\, b'^{\dagger\, \al}\,_{\be\ga} ({\underline k}') \midd 0\rangle
\eeq
subject to the conditions 
\beqq & & (i)'\!: \quad \pa'^\al B'^{(-)}_\al\,^{\ga\de} (x') \midd f'\rangle\, = 0 \nonumber \\
& & \quad {\mbox{or}}\quad\quad k'^\al\, f'^P_{\al}\,^{\ga\de} ({\underline k}') = 0  \\
& & (ii-a)'\!: \quad \langle f'^{Phys}\midd f'^{Phys} \rangle\,  \geq 0 \nonumber \\
& & (ii-b)'\!: \quad \langle \hat f'^{Phys}\midd f'^{Phys} \rangle\,  = 0 \nonumber
\eeqq
in the primed frame. Hence we have $f'^P_{\al}\,^{\ga\de} ({\underline k}')$ obeying
\beqq \label{108}
& & f'^P_0\,^{\ga\de} ({\underline k}') = -\frac{k'^i}{\om_{k'}}\, f'^P_i\,^{\ga\de} ({\underline k}') \nonumber \\
& & f'^P_{\al}\,^{0 i} ({\underline k}') = - f'^P_{\al}\,^{i 0} ({\underline k}') = 0. 
\eeqq
Note that the state Eqn.(\ref{105}) does contain a longitudinal part in general.

Next we look at another physicist who works in the unprimed reference frame related to the primed frame by a Lorentz transformation $\La$. This Lorentz transformation relates the unprimed to the primed quantities as
\beqq k'^\al &=& \La^\al\,_\be\, k^\be \nonumber \\
f'^P_{\be}\,^{\eta\ze} \Big({\underline k}' ({\underline k})\Big) \!\!\!\!  &=& \!\!\!\!  
\La_\be\,^\al \La^\eta\,_\ga \La^\ze\,_\de\, f_{\al}\,^{\ga\de} ({\underline k}) \\
b'^{\dagger\, \be}\,_{\eta\ze} \Big({\underline k}' ({\underline k})\Big) \!\!\!\!  &=& \!\!\!\!  
\La^\be\,_\al \La_\eta\,^\ga \La_\ze\,^\de\, b^{\dagger\, \al}\,_{\be\ga} ({\underline k}) \nonumber \\
d {\tilde k}' &=& d {\tilde k} \nonumber
\eeqq
and we find
\beq \midd f'^{Phys}\rangle = \sqrt{\frac{1}{2}} \intkt\,
f_{\al}\,^{\ga\de} ({\underline k})\, b^{\dagger \al}\,_{\be\ga} ({\underline k}) \midd 0\rangle = \,\, \midd f\rangle
\eeq
with the unprimed $b^\dagger$, $b$ obeying the same creation and annihilation operator algebra given by Eqns.(\ref{44}) and (\ref{45}) as do the primed creation and annihilation operators. In addition we have
\beqq \label{111} & & (i)\!: \quad k^\al\, f_{\al}\,^{\ga\de} ({\underline k}) = 0 \nonumber \\
& & (ii-a)\!: \quad \langle f\midd f \rangle\, =  \langle f'^{Phys}\midd f'^{Phys} \rangle\,  \geq 0 \\
& & (ii-b)\!: \quad \langle \hat f\midd f \rangle\,  =  \langle \hat f'^{Phys}\midd f'^{Phys} \rangle\,  = 0. \nonumber
\eeqq
We note that in general the $f_{\al}\,^{\ga\de} ({\underline k})$ are not subject to the unprimed Eqns.(\ref{108})
\beqq
& & f_0\,^{\ga\de} ({\underline k}) = -\frac{k^i}{\om_{k}}\, f_i\,^{\ga\de} ({\underline k}) \nonumber \\
{\mbox{but}} & & f_{\al}\,^{0 i} ({\underline k}) = - f_{\al}\,^{i 0} ({\underline k}) \neq 0 
\eeqq
potentially destroying the covariance of our approach.

So let us turn to the second step. We have demonstrated in section 7 that Eqns.(\ref{111}) ensure the existence of both a unique Lorentz transformation $\La_{_{P}}\left({\underline k}', {\underline k} \right)$ and an $f^P$ such that
\beqq f_\al\,^{\eta\ze} \left({\underline k} ({\underline k}')\right) \!\!\!\!  &=& \!\!\!\!  
\La_{_{P}}^\eta\,_\ga \left({\underline k}', {\underline k} \right)
\La_{_{P}}^\ze\,_\de \left({\underline k}', {\underline k} \right)\, f^P_{\al}\,^{\ga\de} ({\underline k}) \\
b^{\dagger \al}\,_{\eta\ze} \left({\underline k} ({\underline k})'\right) \!\!\!\!  &=& \!\!\!\!  
\La_{_{P}\, \eta}\,^\ga \left({\underline k}', {\underline k} \right)
\La_{_{P}\, \ze}\,^\de \left({\underline k}', {\underline k} \right)\, b^{P\, \dagger \al}\,_{\ga\de} ({\underline k}). \nonumber
\eeqq
so that $f^P_{\al}\,^{\ga\de} ({\underline k})$ is subject to
\beqq
& & f^P_0\,^{\ga\de} ({\underline k}) = -\frac{k^i}{\om_{k}}\, f^P_i\,^{\ga\de} ({\underline k}) \nonumber \\
& & f^P_{\al}\,^{0 i} ({\underline k}) = - f^P_{\al}\,^{i 0} ({\underline k}) = 0. 
\eeqq
Obviously we still have
\beqq & & (i)\!: \quad k^\al\, f^P_{\al}\,^{\ga\de} ({\underline k}) = 0 \nonumber \\
& & (ii-a)\!: \quad \langle f^{Phys}\midd f^{Phys} \rangle\,  = \langle f\midd f \rangle\,  \geq 0 \\
& & (ii-b)\!: \quad \langle \hat f^{Phys}\midd f^{Phys} \rangle\,  = \langle \hat f\midd f \rangle\,  = 0. \nonumber
\eeqq
for the resulting state
\beq \midd f\rangle = \sqrt{\frac{1}{2}} \intkt\,
f^P_\al\,^{\ga\de} ({\underline k})\, b^{P \dagger\, \al}\,_{\be\ga} ({\underline k}) \midd 0\rangle = \,\, \midd f^{Phys}\rangle
\eeq
with $b^{P \dagger}$, $b^P$ obeying the same creation and annihilation operator algebra given by Eqns.(\ref{44}) and (\ref{45}) as do the creation and annihilation operators $b^\dagger$, $b$.

Hence -- leaving aside the superscript $^P$ in the creation and annihilation operators -- we have
\beq \midd f'^{Phys}\rangle = \sqrt{\frac{1}{2}} \intkt\,  
f^P_i\,^{jk} ({\underline k}) \left( b^{\dagger\, i}\,_{jk} ({\underline k}) 
+ \frac{k^i}{\om_{k}}\, b^{\dagger\, 0}\,_{jk} ({\underline k}) \right) \midd 0 \rangle
= \,\, \midd f^{Phys}\rangle
\eeq
which is what we wanted to demonstrate.

In addition it is easy to show that
\beq P'^\al \midd f'^{Phys}\rangle = \La^\al\,_\be\, P^\be \midd f^{Phys}\rangle
\eeq
and
\beq \langle f'^{Phys} \midd f'^{Phys}\rangle = \langle f^{Phys} \midd f^{Phys}\rangle \geq 0
\eeq
as well as
\beq \langle f'^{Phys} \midd P'^\al \midd f'^{Phys}\rangle =
\La^\al\,_\be\, \langle f^{Phys} \midd P^\be \midd f^{Phys}\rangle.
\eeq
Specifically we have
\beqq & & \langle f'^{Phys} \midd H' \midd f'^{Phys}\rangle =
\La^0\,_\be\, \langle f^{Phys} \midd P^\be \midd f^{Phys}\rangle \geq 0 \\
& & \quad\quad\quad\quad \Rightarrow \quad \langle f^{Phys} \midd H \midd f^{Phys}\rangle \geq 0. \nonumber
\eeqq

\section{Physical n-particle space of states ${\cal F}_n^{\Phys}$ and physical Fock space ${\cal F}^{\Phys}$}

\paragraph{}
In this section we extend the construction of the physical one-particle space of states ${\cal F}_1^{\Phys}$ with positive norm and energy expectation value to the physical $n$-particle space of states ${\cal F}_n^{\Phys}$ with positive norm and energy expectation value and finally to the full physical Fock space ${\cal F}^{\Phys}$.

To keep notations simple let us first define the one-particle creation operator $b^\dagger [f^{Phys}]$
\beq b^\dagger [f^{Phys}] \equiv \sqrt{\frac{1}{2}} \intkt\,  
f^P_i\,^{jk} ({\underline k}) \left( b^{\dagger\, i}\,_{jk} ({\underline k}) 
+ \frac{k^i}{\om_{k}}\, b^{\dagger\, 0}\,_{jk} ({\underline k}) \right)
\eeq
the application of which to the vacuum state $\midd 0 \rangle$ generates a physical one-particle state $\midd f^{Phys}\rangle$ -- with an analogous expression for the one-particle annihilation operator $b [f^{Phys}]$. Its application to a general state destroys a physical one-particle state $\midd f^{Phys}\rangle$.

Then the one-particle space of states ${\cal F}_1^{Phys}$ is given by
\beqq {\cal F}_1^{\Phys} &=& \Big\{ \left[  \midd f^{Phys}\rangle \right]
\Big\vert\,\, \midd f^{Phys}\rangle = b^\dagger [f^{Phys}] \midd 0\rangle \,\, {\mbox{with}} \nonumber \\
& & \,\, f^P_0\,^{\ga\de} ({\underline k}) = -\frac{k^i}{\om_{k}}\, f^P_i\,^{\ga\de} ({\underline k}), \\
& & \,\, f^P_{\al}\,^{0 i} ({\underline k}) = - f^P_{\al}\,^{i 0} ({\underline k}) = 0. \nonumber
\Big\}
\eeqq

Analogously the physical two-particle states are given by
\beq \frac{1}{\sqrt{2!}}\, b^\dagger [f_1^{Phys}]\, b^\dagger [f_2^{Phys}] \midd 0\rangle \sim {\mbox{two-particle state}}
\eeq
and the physical $n$-particle states by
\beq \frac{1}{\sqrt{n!}}\, \prod_{i=1}^n\, b^\dagger [f_i^{Phys}] \midd 0\rangle \sim n{\mbox{-particle state}}.
\eeq
It is easy to demonstrate the $n$-particle states above have positive norm and positive energy expectation value.

Finally the following set of $n$-particle states forms a dense set in ${\cal F}^{\Phys}$
\beq {\cal F}_{Basis}^{\Phys} = \left\{
\frac{1}{\sqrt{n!}}\, \prod_{i=1}^n\, b^\dagger [f_i^{Phys}] \midd 0\rangle
\,\, \forall n, \,\, \forall f_i^{Phys} \right\}
\eeq
and serves as a basis of the physical Fock space with states having positive norm and positive energy expectation values.

\section{Conclusions}

\paragraph{}
In this paper we have first derived the non-interacting {\bf SO(1,3)\/} gauge field theory in the Lorentz gauge from the interacting Lorentz gauge field theory of gravitation which is renormalizable by power-counting and equivalent to General Relativity. We then have canoncically quantized the non-interacting theory and step-by-step established the relativistically invariant physical Fock space ${\cal F}^{\Phys}$ for free gauge particles which contains only states with positive norm and energy expectation value.

As an encouraging result we have at this point successfully completed two steps out of our four-step programme aiming to establish a renormalizable quantum theory of gravity -- the first step being to establish a theory for classical gravity which is not equal, but equivalent to GR, and which in return allows for renormalizable actions for the dynamical field, the second to show that a proper canonical quantization of the free gauge fields allows for the definition of positive-norm, positive-energy states and a corresponding relativistically invariant physical Fock space for these fields.

Obviously much remains to be done -- first and foremost the third step of our programme which is to prove the renormalizability of the full quantum theory \cite{chw2} and the final one which is to establish the unitarity of the $S$-matrix on the physical Fock space constructed in this paper. Then there is the question about the $\beta$-functions for both the pure gravitational theory and the gravitational theory coupled to all other fields in the standard model of particle physics.

In addition a deeper understanding is needed of the relativistically invariant mirror Fock space of negative-norm, negative-energy states which can be constructed in exactly the same way as its positive-norm, positive-energy Fock space cousin, especially around the question what protects against transitions from one to the other if the interactions are activated.

Finally as already mentioned in \cite{chw1} the explanation of phenomena such as the accelerated expansion of the universe or the galaxy rotation curves not compatible with the observed matter distribution might after all not be linked to dark energy or dark matter, but rather to a refined theory of gravitation at both the classical and the quantum level -- the free-field limit of which we have established in this paper. Further progress is possible.